\documentclass[njp,10pt]{iopart}
\usepackage{graphicx}
\expandafter\let\csname equation*\endcsname\relax
\expandafter\let\csname endequation*\endcsname\relax
\usepackage{graphicx}
\usepackage{amsmath}
\usepackage{amsfonts}
\usepackage{amssymb}
\usepackage{amstext} 
\usepackage{soul}
\usepackage{dsfont}
\usepackage[utf8]{inputenc}
\usepackage[hidelinks]{hyperref}
\usepackage[caption=false]{subfig}
\usepackage{epstopdf} 
\usepackage{cellspace}
\hypersetup{colorlinks=true, citecolor=blue, urlcolor=blue, linkcolor=blue}
\usepackage{tabularx,ragged2e,booktabs}
\usepackage[dvipsnames]{xcolor}
\usepackage{physics}
\usepackage{ulem}
\usepackage{color}
\usepackage[numbers,sort&compress]{natbib}

\usepackage[left=1in,right=1in,top=1in,bottom=1in,footskip=.25in]{geometry}

\renewcommand\vec{\boldsymbol}

\usepackage{dcolumn}   
\usepackage{bm}        
\usepackage{cellspace}

\renewcommand{\emph}[1]{\textit{#1}}

\renewcommand{\dd}{\mathrm{d}}

\begin{document}

\title{Constraining modified gravity with quantum optomechanics}

\author{Sofia Qvarfort$^{1,2,3,4}$, Dennis R\"atzel$^5$, Stephen Stopyra$^{2,6}$}

\address{$^1$ QOLS, Blackett Laboratory, Imperial College London,  SW7 2AZ London, United Kingdom  \\
$^2$ Department of Physics and Astronomy, University College London, Gower Street, WC1E 6BT London, United Kingdom  \\
$^3$ Nordita, KTH Royal Institute of Technology and Stockholm University, Hannes Alfv\'{e}ns v\"{a}g 12, SE-106 91 Stockholm, Sweden \\
$^4$ Department of Physics, Stockholm University, AlbaNova University Center, SE-106 91 Stockholm, Sweden \\
$^5$ Institut f\"{u}r Physik, Humboldt-Universit\"{a}t zu Berlin, 12489 Berlin, Germany \\
$^6$ The Oskar Klein Centre, Department of Physics, Stockholm University, AlbaNova, SE-10691 Stockholm, Sweden }
\ead{sofia.qvarfort@fysik.su.se, dennis.raetzel@physik.hu-berlin.de, svstopyra@googlemail.com}

\date{\today}

\begin{abstract} 
We derive the best possible bounds that can be placed on Yukawa-- and chameleon--like modifications to the Newtonian gravitational potential with a cavity optomechanical quantum sensor. By modelling the effects on an oscillating source-sphere on the optomechanical system from first-principles, we derive the fundamental sensitivity with which these modifications can be detected in the absence of environmental noise. 
In particular, we take into account the large size of the optomechanical probe compared with the range of the fifth forces that we wish to probe and quantify the resulting screening effect when both the source and probe are spherical. Our results show that optomechanical systems in high vacuum could, in principle, further constrain the parameters of chameleon-like modifications to Newtonian gravity. 
\end{abstract}

\maketitle


\section{Introduction}

General Relativity is one of the most successful theories of nature, but there are compelling reasons to explore modifications to the behaviour of gravity on both large and small scales. Most of the precise predictions of General Relativity have consistently been demonstrated experimentally: among many others these include the perihelion shift of Mercury~\cite{will1990general} and the existence of gravitational waves~\cite{abbott2016observation}. Similarly, the current standard cosmological model, the $\Lambda$ Cold Dark Matter ($\Lambda$CDM) model, is another of General Relativity's success stories. However, in order to match observation, $\Lambda$CDM requires a positive cosmological constant~\cite{Ade:2015rim,Aghanim:2018eyx}. This is backed up by observations of supernovae, which indicate that the Universe's expansion is accelerating~\cite{riess1998observational}. While a natural part of General Relativity, a cosmological constant poses a theoretical challenge to particle physics since the small observed value is inherently sensitive to high-energies, requiring delicate balancing~\cite{Padilla:2015aaa}. Furthermore, many theories of high energy physics that attempt to solve this and other problems -- such as building a consistent quantum theory of gravity -- predict deviations from General Relativity. These theories are collectively known as \emph{modified gravity theories}.

Modified gravity theories, however, typically face a difficult challenge in the form of solar system tests of Newton's laws. Models that differ from General Relativity significantly enough to explain the observed acceleration of the Universe on large scales are typically ruled out by their predicted deviations on smaller scales (solar system and laboratory tests)~\cite{Will:2005va,PhysRevLett.92.121101,2003Natur.425..374B}. There are a large variety of approaches to modified gravity -- see Koyama~\cite{Koyama:2015vza} for a comprehensive review -- but many models attempt to address the problem of solar system tests via a \textit{screening mechanism}~\cite{Joyce:2014kja}. Such mechanisms can be built into modified gravity theories to conceal deviations on solar system scales, without changing the large scale behaviour. An approach considered by many authors is the chameleon mechanism~\cite{Khoury:2003aq,PhysRevD.69.044026,Brax:2004qh}; the basic idea is to add a scalar field that couples directly to gravity in a manner that depends on the local density of matter. In high-density regions, such as inside a galaxy, the effects of modified gravity are screened out, allowing the theory to evade solar system tests. In the low-density void regions between galaxies, however, the effects of modified gravity would be unscreened.

If such a density-dependent gravity mechanism is at play, it ought to be detectable in principle by high-precision laboratory experiments. In particular, the fundamental sensitivity improvements offered by quantum systems are especially promising~\cite{giovannetti2006quantum}. 
At the moment, the detection of modified gravity, and in particular, chameleon fields, has been explored through a diverse variety of methods. Searches with classical systems include theoretical proposals for torsion balance tests of fifth forces~\cite{upadhye2006unv,Mota2006strongly,Mota2007evading,Adelberger2007particle,PhysRevD.78.104021,upadhye2012dark,Upadhye:2012fz}, some of which have already been carried out as experiments~\cite{PhysRevD.70.042004, PhysRevLett.98.021101}. 
Additional proposals suggest that experiments which measure Casimir forces may also be used to constrain chameleon theories~\cite{Mota2007evading,Brax2007detecting,Brax2010tuning,Almasi2015force,Brax2015casimir}.
In atom interferometry, which is already routinely used for quantum sensing, the uniformity of the atoms as well as the additional sensitivity gained from the superposition of flight-paths has led to impressive precision gravimetry sensitivities~\cite{peters2001high, mcguirk2002sensitive, bidel2013compact, hu2013demonstration}. 
Several proposals have explored in depth the possibilities of searching for modified gravity and dark energy with atom interferometry~
\cite{Burrage:2014oza,Burrage:2015lya,elder2016cham,schloegel2016prob,burrage2016proposed,chiow2018multiloop,Hartley:2019wzu}, and some of the most stringent bounds on existing theories have been obtained in this way~\cite{hamilton2015atom,sabulsky2019exp}. 
Further viable routes towards detecting modified gravity include ultra-cold neutron experiments~\cite{Brax:2014zta,Serebrov2011search,Brax2011strongly,Serebrov2014experimental,brax2014testing,Jenke2014gravity,Cronenberg:2016Pt} and neutron interferometry~\cite{Brax2013probing,pokotilovski2013strongly,brax2014testing,lemmel2015neutron,Li2016neutron}. Finally, tests of atomic transition frequencies~\cite{Brax2011atomic,Frugiuele2017constraining}, close examination of vacuum chambers and photo-detectors~\cite{Rybka2010search,upadhye2012design}, as well as tests of the electron magnetic moment~\cite{PhysRevD.97.084050} have also been proposed.

\begin{figure}[t!]
	\includegraphics[width=0.8\linewidth]{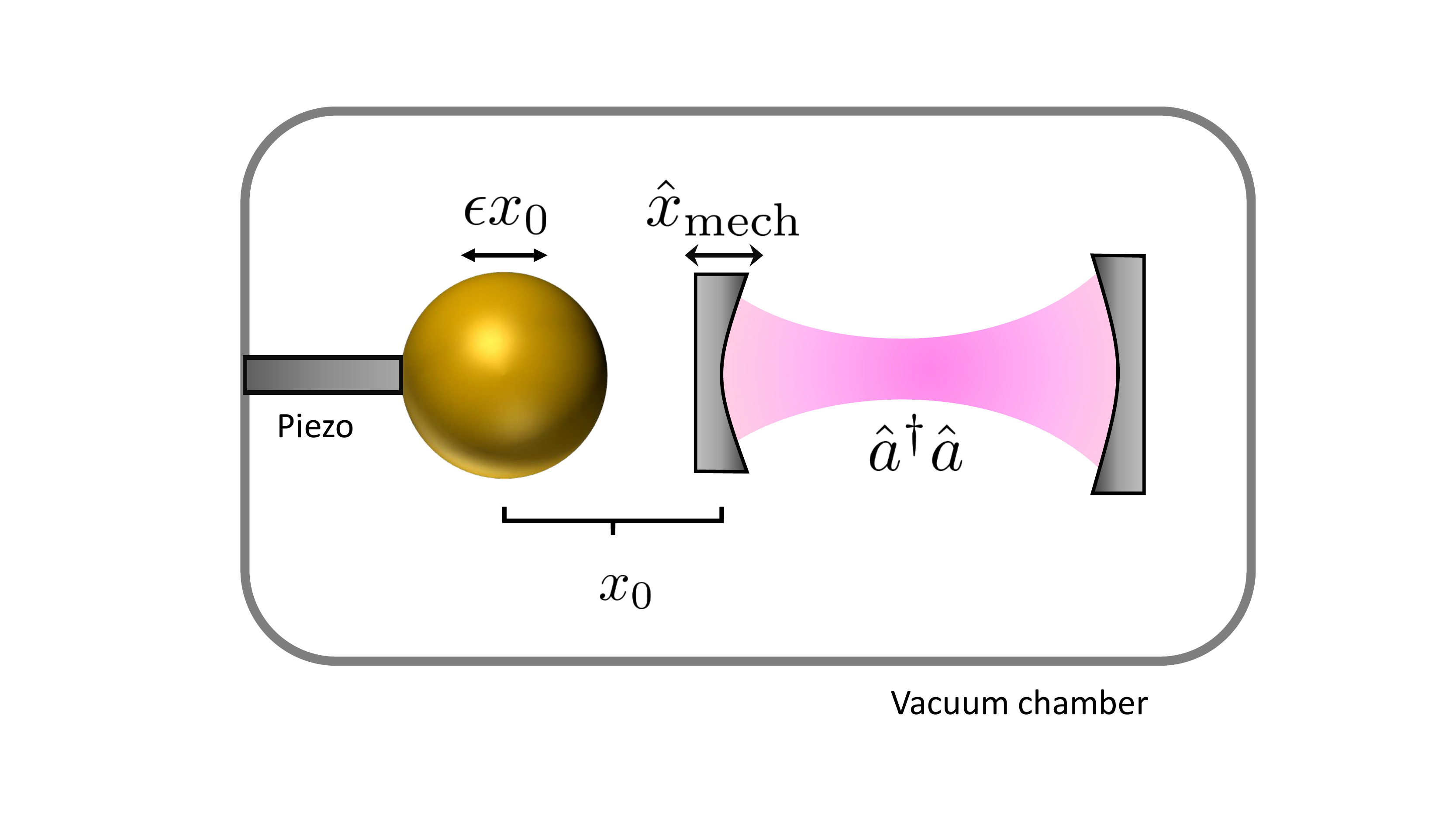}
	\caption{A gold source mass attached to a shear piezo oscillates to create a time-varying gravitational field. The field, which potentially contains deviations from Newtonian gravity, is detected by an optomechanical probe system where the photon number $\hat a^\dag \hat a$ couples to the mechanical position $\hat x_{\mathrm{mech}}$ as $\hat a^\dag \hat a \hat x_{\mathrm{mech}}$, here presented as a moving-end mirror in a Fabry--P\'erot cavity.  The amplitude $\epsilon x_0$ of the source mass oscillation is a fraction of the total distance $x_0$ between the systems. By accounting for the vacuum background density, we may also compute bounds on the parameters of the chameleon screening mechanism.  }\label{fig:cavity}
\end{figure}

An additional approach to detecting the small-scale effects of modified gravity and screening is to take advantage of recent developments in the field of optomechanics, where a small mechanical element is coupled to a laser through radiation-pressure~\cite{bowenbook, aspelmeyer2014cavity}. Optomechanical system encompass a diverse set of platforms which range from microscopic movable  mirrors as part of a Fabry--P\'{e}rot cavity~\cite{favero2009optomechanics}, levitated particles~\cite{barker2010cavity}, clamped membranes~\cite{tsaturyan2017ultracoherent},  liquid Helium~\cite{shkarin2019quantum} and trapped cold atoms~\cite{purdy2010tunable}. When the mechanical element is cooled down to sufficiently low temperatures, it enters into a quantum state that can be manipulated through measurements and optical control techniques. Ground-state cooling has been demonstrated across a number of platforms, including clamped membranes~\cite{chan2011laser,teufel2011sideband} and recently also for levitated systems~\cite{Delic892}. Optomechanical systems show promising potential as both classical and quantum-limited sensors~\cite{arcizet2006high, geraci2010short, hempston2017force}, and recent studies have  proposed their use as gravity sensors~\cite{qvarfort2018gravimetry, armata2017quantum, schneiter2020optimal, qvarfort2020optimal}.  In fact, experimental searches for fifth forces with classical optomechanical setups have already been performed (see e.g.~\cite{PhysRevLett.117.101101,PhysRevD.96.104002}), where the bounds achieved fell within those excluded by atom interferometry. A key question, which we explore in this work, therefore becomes whether an optomechanical sensor in the quantum regime can improve on these bounds. For an overview of searches for new physics with levitated optomechanical systems, see the recent review by Moore \textit{et al}.~\cite{moore2020searching}. The advantage of optomechanical sensors, as opposed to, for example, cold atom interferometry is that the sensitivity of the system can be improved while retaining the compact setup of the experiment. In contrast, improving the sensitivity of atom interferometry primarily relies on increasing the length of the flight-path of the atoms.

The key question we seek to answer in this work is: what fundamental range of parameters of modified gravity theories could ideally be excluded with a quantum optomechanical sensor? To address this question, we consider an idealised system described by a nonlinear, dispersive, optomechanical Hamiltonian which couples the optical and mechanical degrees of freedom through a nonlinear radiation-pressure term. This Hamiltonian is often linearised for a strong coherent input drive, however the fully nonlinear (in the sense of the equations of motion) Hamiltonian is a more fundamental description. 
While all quantum systems are affected by noise, we here assume that the coherence times can be made long enough for the measurement protocol to be carried out. As a result, our analysis explores the bounds in the absence of environmental noise and decoherence.
We then consider the gravitational field that arises when a source mass is placed next to the sensor. 

Since it is often difficult in experiments to distinguish a signal against a constant noise floor, we consider an oscillating source mass, which gives rise to a time-dependent gravitational field. Such a signal can then  be isolated from other common low-frequency $1/f$ noise sources via a Fourier analysis of the data. 
To determine whether our analysis is valid in the case of a chameleon field, we derive the  time-dependent potential that results from the source mass from first principles, where we find that a potential that moves with the mass is the correct choice for non-relativistic velocities. Another key consideration for optomechanical systems is the relatively large size of the optomechanical probe. This has been found to be significant in previous classical experiments with chameleon fields, such as the MICROSCOPE experiment~\cite{pernot2019general,pernot2021constraints}, and we find that it also contributes significantly to the chameleon screening of the fifth force in the envisioned setup of the quantum experiment we consider here (as opposed to, for example, cold atoms, where the screening length of the atomic probes is very small). To take the finite screening length into account, we go beyond the common approximation that the probe radius is small compared to the range of the chameleon field and derive analytic expressions for the modified force seen by the probe. 

Then, using tools from quantum information theory and quantum metrology such as the quantum Fisher information, we are  able to estimate the fundamental sensitivity for detecting deviations from Newtonian gravity. To further improve the sensitivity, we also consider known ways to enhance the optomechanical sensor in the form of squeezed light and a modulated optomechanical coupling~\cite{qvarfort2020optimal}. 

Our main results include the bounds presented in  figure~\ref{fig:exclusion:plot:comparison}, which shows the parameter ranges of modified gravity theories that could potentially be excluded with an ideal optomechanical sensor. The bounds are computed for a specific set of experimental parameters. To facilitate investigations into additional parameter regimes, we have made the code used to compute the bounds available (see the Data Availability Statement). While experiments are unlikely to achieve the predicted sensitivities due to noise and systematic effects, our bounds constitute a fundamental limit for excluding effects beyond Newtonian gravity given the experimental parameters in question. 

This work is structured as follows. In section~\ref{sec:experimental:setup} we present the proposed experimental setup and optomechanical Hamiltonian,  and then we proceed to discuss Yukawa potentials as a modification to the Newtonian gravitational potential in section~\ref{sec:moving:mass:potential}. We consider those sourced by a chameleon field  and provide a first-principles' derivation of the time-dependent potential that results from the mass oscillating around an equilibrium position. We also discuss screening effects inherent to chameleon fields and derive the screening effect that arises from the size of the optomechanical probe. In section~\ref{sec:linearised:potential}, we linearise the modified gravitational potential, and in section~\ref{sec:metrology}, we provide an introduction to quantum metrology and the quantum Fisher information. These tools allow us to present analytic expressions for the fundamental sensitivity of the system, which we do in section~\ref{sec:results}. The work is concluded by a discussion in section~\ref{sec:discussion} and some final remarks in section~\ref{sec:conclusions}.

\section{Optomechanical model and dynamics} \label{sec:experimental:setup}

In this section, we introduce the model of the optomechanical system and show how the effects of a time-varying gravitational field can be included in the dynamics.

\subsection{Experimental setup} 

We envision an experimental setup similar to that used in~\cite{westphal2020measurement}, where an oscillating source mass made of solid gold is placed in a vacuum chamber adjacent to an optomechanical probe (see figure~\ref{fig:cavity}). We have chosen gold because we require the highest possible density in order to detect gravitational effects and maximise the effect of density-dependent screening mechanisms such as chameleon fields\footnote{While there are denser materials, such as depleted Uranium, gold is a stable material that has previously been used for small-mass sensing, see e.g. Ref~\cite{westphal2020measurement}.}. The source mass oscillates back and forth, which can be achieved in a number of different ways~\cite{schmole2017development}. One such implementation is with the help of a shear piezo, which oscillates at a fixed frequency. The optomechanical probe is then allowed to move along the same axis as the oscillating mass. By injecting light into the cavity, the position of the optomechanical coupling is dispersively coupled to the optical field through radiation-pressure. The light then picks up a phase shift conditioned on the displacement of the mechanical mode, which has been influenced by the gravitational force. Therefore, information about the gravitational field is imprinted on the optical state. The light is then collected and measured either as it leaks from the cavity or through a scheme where the cavity is coherently opened to access the full intra-cavity state~\cite{tufarelli2014coherently}.

While the optomechanical interaction can generally be described with the same dynamics for a large range of systems, the force and strength of the coupling differ for each platform. In this work, we begin with a general description of a single interacting mode, but  later specialise towards a spherical mechanical element since it allows for analytical treatments of some modified gravity potentials. 

The optomechanical Hamiltonian, which governs  the dynamics of the optomechanical probe, is given by (in the absence of an external gravitational field):
\begin{equation} \label{eq:basic:Hamiltonian}
\hat H_0 = \hbar \, \omega_{\mathrm{c}} \, \hat N_ a + \hbar \,\omega_{\mathrm{mech}} \, \hat N_b -  \hbar \, k(t)\, \hat N_ a \, \bigl( \hat b^\dag + \hat b \bigr), 
\end{equation}
where $\omega_{\mathrm{c}}$ and $\omega_{\mathrm{mech}}$ are the oscillation frequencies of the optical cavity mode and mechanical mode respectively, with annihilation and creation operators $\hat a ,\hat a ^\dag$ and $\hat b, \hat b^\dag$. We have also defined $\hat N_a = \hat a^\dag \hat a$ and $\hat N_b = \hat b^\dag \hat b$ as the photon and phonon number operators. 

The coupling $ k(t)$ is the (potentially time-dependent) characteristic single-photon interaction strength between the number of photons and the position of the mechanical element. It takes on different forms depending on the optomechanical platform in question. Among the simplest couplings is that for a moving mirror, of mass $m$, that makes up one end of a cavity, $k = (\omega_l/L) \sqrt{\hbar /(2 m\omega_{\mathrm{mech}})}$, where $\omega_l $ is the laser frequency and $L$ is the length of the cavity. In this work, we also consider modulating the coupling in time; it has been previously found that such modulations can be used to enhance the sensitivity of the optomechanical sensor if they can be made to match the oscillation of the external force~\cite{qvarfort2020optimal}. Modulation of the optomechanical coupling can be introduced in different ways depending on the experimental platform in question. For example,  the mechanical frequency of a cantilever can be modified by applying an oscillating electric field~\cite{rugar1991mechanical,szorkovszky2013strong}, and a modulated coupling arises naturally through the micro-motion of a levitated system in a hybrid electro-optical trap~\cite{Millen2015cavity,aranas2016split,fonseca2016nonlinear}.

All quantum systems are affected by noise due to their interaction with the environment. Such an interaction usually results in dissipation and thermalisation, which in turn leads to decoherence of the off-diagonal elements of the quantum state. For cavity optomechanical systems, common sources of noise include photons leaking from the cavity, as well as thermlisation of the mechanical element due to interactions with the surrounding residual gas, or from vibrations from the mount~\cite{aspelmeyer2014cavity}. The nature of the noise is unique to each experimental platform and must be carefully modelled in each case. 

In this work, we are interested in deriving the best-possible sensitivity that an optomechanical system can achieve. We therefore assume that the $Q$-factor of the cavity is high enough that the system stays coherent throughout the duration of our measurement protocols. Recently, $Q$-factors of $10^9$ have been demonstrated in magnetically levitated meso-mechanical systems~\cite{cirio2012quantum}, and linewidths of $81\pm23\,\mu$Hz have been measured~\cite{pontin2020ultranarrow}. We also assume that the system has been cooled to temperatures such that the surrounding environment does not cause the mechanical mode to heat up during the protocol. To reduce unwanted vibrations or gravitational noise, it is also possible to add decoupling stages in the experiments~\cite{pitkin2011gravitational}, such as suspension stages made by fused silica fibres~\cite{penn2001high,cumming2020lowest}.
Under these conditions, it is possible to consider an approximately unitary description of the experiment, which we shall use to derive a fundamental limit of the sensitivity that could in principle be achieved with an optomechanical system. To then describe a realistic experiment, all of  the above effects must be taken into account. We discuss this and other potential future work in section~\ref{sec:discussion}.

When treating the system in a closed and ideal setting, we can model the initial state as a separable state of the light and the mechanical element. For the optical state, we consider injecting squeezed light into the cavity. Squeezed light has been shown to fundamentally enhance the sensitivity to displacements~\cite{giovannetti2006quantum}. By including squeezing here, we generalise our scheme to include these input states. However, we note that in order to improve the sensitivity overall, it is always more beneficial to increase the number of photons rather than squeezing the system. Squeezing also reduces quadrature noise~\cite{mehmet2011squeezed}. 
The state of the mechanical element, on the other hand, is most accurately described as thermal at a non-zero temperature. With these assumptions, the initial state of the system can be written as
\begin{equation}\label{initial:state}
\hat \varrho(0) = \ketbra{\zeta}\otimes \sum_{n=0}^\infty \frac{\tanh^{2n}r_T}{\cosh^2 r_T}\ketbra{n}{n}\;,
\end{equation}
where $\ket{\zeta} = \hat S_\zeta \ket{\mu_{\mathrm{c}}}$ is a squeezed coherent state of the optical field where $\hat S_\zeta = \mathrm{exp} \bigl[ (\zeta^* \hat a^2 - \zeta \hat a^{\dag 2})/2 \bigr]$ and where the coherent state satisfies $\hat a \ket{\mu_{\mathrm{c}}} = \mu_{\mathrm{c}} \ket{\mu_{\mathrm{c}}}$. The squeezing parameter can also be in spherical polar form as $\zeta = r_{\mathrm{sq}} \, e^{i \varphi}$. Squeezed states can be generated through four-wave mixing in an optical cavity~\cite{slusher1985observation} or parametric down-conversion~\cite{wu1986generation}. See also Ref~\cite{andersen201630} for a review of squeezed state generation. The parameter $r_T$ of the thermal state arises from the Bose--Einstein distribution and is defined by $\tanh r_T=\exp[-\frac{\hbar\,\omega_\textrm{mech}}{2\,k_\textrm{B}\,T}]$, where $T$ is the temperature of the system and $k_\mathrm{B}$ is Boltzmann's constant.

\subsection{Modelling the gravitational force} 

In order to compute the sensitivity bounds for detecting modified gravity, we model the effect of the gravitational force from the moving source mass on the optomechanical system as a contribution to the dynamics. When the force is weak, it can be linearised and included as a displacement term in the optomechanical Hamiltonian in equation~\eqref{eq:basic:Hamiltonian}. In this section, we provide a general derivation of this linearised force, while in section~\ref{sec:linearised:potential} we specialise to Yukawa-like and chameleon modifications to the Newtonian potential. The linearisation is necessary to properly describe the quantum dynamics of the setup with the current theoretical machinery; we will however describe the chameleon field in full generality to allow for future work to improve the theoretical description.

We start by assuming that the source mass and the mechanical element of the optomechanical system are constrained to move along the $x$-axis. We let the mechanical element be subject to a harmonic potential centered at $x = 0$ and we label the position of the source mass $x_S(t)$. Then, we assume that there is a small perturbation to the centre-of-mass position of the mechanical element that we call $\delta x$, and, assuming that $|x_S(t)| \gg |\delta x|$ at all times, we write the relative distance between the systems as $x_S(t) - \delta x $. 
Provided that $\delta x$ remains small, we can Taylor expand the system to first order in $\delta x$. Given a generic potential term $V(x_S(t) - \delta x)$, we find,  to first order in $\delta x$:
\begin{equation} \label{eq:expanded:potential}
V(x_S(t) - \delta x ) = V(x_S(t)) - V'(x_S(t))  \, \delta x + \mathcal{O}[(\delta x)^2]. 
\end{equation}
The first term in equation~\eqref{eq:expanded:potential} represents a time-dependent shift of the overall energy, however it does not depend on the position of the optomechanical system. The second term describes the (potentially time-dependent) displacement of the mechanical element with $\delta x$. The second-order term in $\delta x$ leads to a shift in the mechanical frequency that we do not model here, but dynamics of this kind have been previously studied~\cite{qvarfort2020time}. The expansion in equation~\eqref{eq:expanded:potential} is valid as long as $\delta x$ remains small such that the higher-order terms can be neglected. We outline the conditions for this being true in the Discussion (see section~\ref{sec:discussion}). 

We proceed to promote $\delta x$ to an operator $\delta x\rightarrow \hat x_{\mathrm{mech}}$, which can be written in terms of the annihilation and creation operators $\hat b$ and $\hat b^\dag$ of the mechanical element as 
\begin{equation}
\hat x_{\mathrm{mech}} =  x_{\mathrm{zpf}} \, \bigl( \hat b^\dag + \hat b \bigr),
\end{equation}
where $x_{\mathrm{zpf}} = \sqrt{\hbar/(2\,m\omega_{\mathrm{mech}})}$ is the zero-point fluctuation of the mechanical oscillator. It should  be noted here that the dynamics of a nonlinear optomechanical system with a driving term proportional to $\hat x_{\mathrm{mech}}^2$ has been solved, however the inclusion of these effects adds significant complexity the mathematical treatment of the system~\cite{qvarfort2020time}, while it will likely not result in a significant improvement of the sensitivity.

The full optomechanical Hamiltonian including the modified gravitational potential can then be written as
\begin{align} \label{eq:cham:Hamiltonian}
\hat H(t) = \hat H_0 -   V'(x_S(t)) \, x_{\mathrm{zpf}} \, \bigl( \hat b^\dag + \hat b \bigr), 
\end{align}
where $\hat H_0$ is given in equation~\eqref{eq:basic:Hamiltonian} and where the time-dependent modified Newtonian gravitational force is contained in the second term. 

The time-evolution of the system with the Hamiltonian in equation~\eqref{eq:cham:Hamiltonian} can be written as the following time-ordered exponential:
\begin{align}
\hat U(t) &= \overleftarrow{\mathcal{T}} \mathrm{exp}\left[ - \frac{i}{\hbar} \int^t_0 \mathrm{d}t’ \, \hat H(t') \right],
\end{align}
where the time-dependence of the gravitational potential in $\hat H(t)$ requires careful consideration. Such dynamics have been studied previously~\cite{qvarfort2019enhanced, qvarfort2020time} and provide a short overview of the treatment~\ref{app:sensitivity}. Later on in this work, we use the expression for $\hat U(t)$ to derive the sensitivity of the system to modifications of the Newtonian potential, but first, we will study the form of the modifications  in-depth.

\vspace{0.8cm} 

\section{Modified gravitational potential and screening from the source and the probe} \label{sec:moving:mass:potential}

In this section, we discuss an example of how the chameleon mechanism would alter the Newtonian force law on sub-millimetre scales. We write all equations in terms of SI units, but energy units (as used elsewhere in the literature) can be restored by setting $\hbar = c = 1$ throughout.

\subsection{Yukawa modifications to the gravitational force law}

Although there are many ways of modifying Newton's laws on short distances, perhaps one of the best motivated theoretically is to add a Yukawa term to the potential. Yukawa potentials are ubiquitous in scalar field theories, since they are the solution to the sourced (inhomogeneous) Klein--Gordon equation for a massive field in the case of spherical symmetry. As a consequence of Lovelock's theorem~\cite{doi:10.1063/1.1665613}, modifications to General Relativity require either additional degrees of freedom such as a scalar field, or more exotic scenarios such as large extra dimensions, higher derivatives, or non-locality. Consequently, additional scalar fields are common in modified gravity theories. These act like a fifth force, and for a source of mass $M_{S}$ and test particle mass $m$, give rise to a gravitational potential of the form:
\begin{equation} \label{eq:modified:gravitational:potential}
    V(r) = -\frac{G \, M_S m}{r}\left(1+\alpha \, e^{-r/\lambda}\right),
\end{equation}
where $\alpha$ parametrises the intrinsic difference in strength between the Yukawa-like fifth force and gravity, while $\lambda$ parametrises the range of this fifth-force. Note that it is possible to have $\alpha \gg 1$ and still agree with existing constraints, provided the force is sufficiently short range to have evaded tests of gravity on short distances.

For this work, we will consider a chameleon screening mechanism that gives rise to a Yukawa-like force. However, the methods we describe can be broadly applied to many different Yukawa-type modifications of the gravitational field on short distances. In the chameleon mechanism, short distance modifications to the Newtonian force law are \emph{screened} from the reach of solar system tests by the presence of a density-dependent scalar field, known as the chameleon field. In regions of relatively high average density -- such as can be found inside a galaxy -- the chameleon field has a high mass, making it hard to detect at colliders and altering the gravitational force law in such a way as to be consistent with solar-system experiments (this is the `screening' effect). However, in regions of low-density -- such as in cosmological voids -- the field is lighter and the effects of modified gravity unscreened. This allows modified gravity theories to have substantial effects on cosmological scales, while being difficult to detect on galactic or solar-system scales. 

We review the properties of chameleon fields in~\ref{sec:cham:mech}. The net effect of the chameleon scalar field $\phi$ is to modify the effective Newtonian potential affecting a test particle. Specifically, the effective potential at position $\mathbf{X}$ is given by
\begin{equation}
\Phi_{\mathrm{eff}}(\mathbf{X}) = \Phi_N(\mathbf{X}) + \Phi_{\mathrm{C}}(\mathbf{X}) \approx \Phi_N(\mathbf{X}) + \frac{\phi(\mathbf{X})}{M},\label{eq:potential}
\end{equation}
where $\Phi_N$ is the standard Newtonian potential, and $\Phi_C$ is the modification to it arising from the chameleon field. The parameter $M$ (here chosen to be a mass to give the correct units for a potential) determines how strongly the chameleon field affects test particles and arises from the non-minimal coupling of the chameleon field to curvature as discussed in~\ref{sec:cham:mech}. 

In this work, we consider a chameleon model with an effective interaction potential
\begin{equation}
V_{\mathrm{eff}}(\phi) = \frac{\Lambda^{4 + n}}{\phi^n} + \frac{\phi\rho}{M}(\hbar c)^3.
\end{equation}
We explore only the case $n=1$ in this work: other models and choices of $n$ are possible, but we choose this specific example to demonstrate how the method works in principle. This model has two parameters; $\Lambda$, which characterises the energy scale of the chameleon's self-interaction potential; and $M$ which is defined above. 

For $n = 1$ the background value of the field, $\phi_{\mathrm{bg}}$, in an environment of constant  mass density $\rho_{\mathrm{bg}}$ is given by
\begin{equation}
\phi_{\mathrm{bg}} = \sqrt{\frac{M\Lambda^5}{\rho_{\mathrm{bg}}(\hbar c)^3}}\label{eq:phibg}.
\end{equation}
In the centre of the source, the chameleon field reaches its minimum value of $\phi_S$ (which can be obtained by replacing the density $\rho_{\mathrm{bg}}$ in equation~\eqref{eq:phibg} with the source density $\rho_{S}$). The mass of the chameleon field, $m_{\mathrm{bg}}$, is density dependent (see~\ref{sec:cham:mech}) and given by
\begin{equation} \label{eq:mbg}
m_{\mathrm{bg}}c^2 = \left(\frac{4\,\rho_{\mathrm{bg}}^3(\hbar c)^9}{M^3\Lambda^5}\right)^{1/4}. 
\end{equation}
The key question for us is how the field results in a force on the optomechanical sensor. This is what we consider next.

\subsection{Force on the optomechanical sensor}

The effect of a chameleon field is in principle detectable in a high-vacuum environment. In practise, this requires extremely precise acceleration measurements, which an optomechanical system can provide. While the optomechanical probe can come in many diffferent shapes, in this work, for the sake of simplicity, we model both the source mass and the detector probe as spheres. This allows us to compute the sensitivity using the chameleon force between these two spheres. There are therefore two effects to consider: the response of the field in equation~\eqref{eq:potential} to the spherical source, and the response of the probe to that field. Due to the nature of the chameleon field, a non-point-like probe will not simply follow the gradient of equation~\eqref{eq:potential} as would a test particle: instead there is an additional screening effect due to the interactions of the probe itself with the field. 

To derive the force that acts on the sensor, we consider the field inside the vacuum chamber. Burrage \textit{et al}.~\cite{Burrage:2014oza} derived the chameleon field around a spherical source of mass $M_S$ and radius $R_S$ as a function of distance from the centre of the sphere, $r$. They assumed in their derivation that the range of the chameleon force was large compared to the size of the source (that is, $m_{\mathrm{bg}}R_Sc/\hbar \ll 1$). To allow us to consider a broad parameter space, we do not assume that either $m_{\mathrm{bg}}R_Sc/\hbar \ll 1$ or $m_{\mathrm{bg}}R_Pc/\hbar \ll 1$ where the indices $S$ and $P$ denote the source and probe, respectively. In what follows, we go beyond existing studies in this regard by including sources (probes) with non-negligible size compared to the range of the force. 

We use the same asymptotic matching approach as Burrage \textit{et al}.~\cite{Burrage:2014oza} to obtain an expression for the chameleon field around a spherical matter distribution:
\begin{equation} \label{eq:static_field}
\phi(r) = \left\{\begin{matrix}\phi_i, &r < S_i\\
\phi_i + \frac{\hbar c M_i}{8\pi R_i M} \frac{r^3 - 3S_i^2r + 2S_i^3}{rR_i^2}, &S_i < r < R_i\\
\phi_{\mathrm{bg}} - \frac{\hbar c M_i}{4\pi r M(1+m_{\mathrm{bg}}R_ic/\hbar)}\left(1 - \frac{S^3_i}{R_i^3}\right)e^{-m_{\mathrm{bg}}c(r - R_i)/\hbar},&r > R_i\end{matrix}\right\}.
\end{equation}
Here, $i = S, P$ for the source and probe respectively. $\phi_i$ is the equilibrium value of the chameleon field in a material with the source (probe) density: the field will attain this value at some radius $S_i \leq R_i$. There is then a transition layer where the field $\phi_i$ increases to its surface value, before increasing to the equilibrium value in the background density, $\phi_{\mathrm{bg}}$. The range of the force outside the source is controlled by the density-dependent chameleon mass, $m_{\mathrm{bg}}$. The full derivation of equation~\eqref{eq:static_field} is given in~\ref{app:chameleon_field}.

The value of the length scale $S_i$ depends on the source/probe properties, the chameleon model, and environmental properties. For the model we consider here, it is found by solving the following cubic equation:
\begin{equation} \label{eq:solve:for:S}
\frac{S_i^2}{R_i^2} + \frac{2}{3}\left[\frac{1}{1+m_{\mathrm{bg}}R_ic/\hbar} - 1\right]\frac{S_i^3}{R_i^3} = 1- \frac{8\pi M}{3M_i}R_i(\phi_{\mathrm{bg}} - \phi_i)c/\hbar + \frac{2}{3}\left[\frac{1}{1+m_{\mathrm{bg}}R_ic/\hbar} - 1\right].
\end{equation}
In the $m_{\mathrm{bg}}R_ic/\hbar \rightarrow 0$ and $\phi_{\mathrm{bg}} \gg \phi_i$ limits, this reduces to
\begin{equation} \label{eq:S}
S_i = R_i\sqrt{1-\frac{8\pi M}{3M_i}\frac{R_i\phi_{\mathrm{bg}}}{\hbar c}},
\end{equation}
which is the result found by Burrage \textit{et al}.~\cite{Burrage:2014oza}. $S_i$ parametrises the screening effect of the chameleon mechanism for a spherical source/probe: for example, when $S_i$ is much lower than $R_i$, the field is effectively unscreened while for $S_i \approx R_i$ the field is heavily screened. Outside of the source ($r > R_i$), we see from equation~\eqref{eq:static_field} that the scalar field, and thus the modified gravitational potential, has an effective Yukawa form. Thus, a chameleon field of this type would manifest as a Yukawa-like modification to the acceleration of a test particle. Since our proposed experimental setup will involve measuring acceleration outside of the sphere, we only need the $r > R_i$ part of the solution.

When viewed as a Yukawa-type force of the form considered in equation~\eqref{eq:modified:gravitational:potential}, and when the probe itself does not contribute to the screening, the resulting fifth-force strength $\alpha$ and range $\lambda$ are given by 
\begin{align}
&  \alpha_\mathrm{bg}  = 
\frac{2M_{\mathrm{P}}^2}{M^2} \, \xi_S , &&\mbox{and} 
&&\lambda_{\mathrm{bg}} = \frac{\hbar}{m_{\mathrm{bg}}c}, \label{eq:lambda:bg}
\end{align}
where $M_{\mathrm{P}} \approx  \sqrt{\hbar c/(8\pi G)} = 4.341\times 10^{-9}\mathrm{\,kg}$ is the reduced Planck mass (here expressed as a mass rather than an energy), $\alpha_\mathrm{bg}$ depends on the background density through
$\xi_S$, which is given by~\cite{Burrage:2014oza} 
\begin{equation} \label{eq:def:of:xi}
\xi_S = 
\begin{cases}
1, & \rho_S R_S^2 < 3 M \, \phi_{\mathrm{bg}} /(\hbar c) , \\
1 - \frac{S_S^3}{R_S^3}, &\rho_S R^2_S > 3 M \phi_{\mathrm{bg}}/(\hbar c) \,.
\end{cases}
\end{equation}
As long as the optomechanical sensor is approximated as a point-particle, such that $R_P/\lambda_{\mathrm{bg}} \ll 1$, the force felt by the optomechanical probe can therefore be written as 
\begin{align} \label{eq:force:no:screening}
F&= -\frac{G\,M_Sm}{|\mathbf{X}_S|^2}\biggl[1 +  \alpha_{\mathrm{bg}} \left( 1 + \frac{|\mathbf{X}_S(t)|}{\lambda_{\mathrm{bg}}} \right) e^{-|\mathbf{X}_S(t)|/\lambda_{\mathrm{bg}}}\biggr], 
\end{align}
where $\mathbf{X}_S$ is the vector-position of the source. The point-particle approximation is however quite severe, especially for an optomechanical probe, the radius of which can be quite large compared with the range of the force in question. We proceed to consider the screening from the probe in the following section.

\subsection{Chameleon screening from the optomechanical probe} \label{sec:probe:screening}

Compared with the atoms used in alternative approaches to the detection of fifth-force modifications to gravity, such as atom interferometry, the optomechanical  probe can potentially be relatively large compared to the range of fifth forces. This can result in significant contributions to the chameleon field screening. The screening depends strongly on the geometry of the system; in general, numerical methods are needed to compute the full screening~\cite{Burrage:2017shh}. As such, it is difficult to estimate the screening for say a Fabry--P\'{e}rot moving-end mirror; however the problem is simplified when both the source sphere and the probe are spherically symmetric. This is the case when the mechanical element in the optomechanical system is a levitated sphere, made, for example, by silica. 

To estimate the extent of the screening for a spherical optomechanical probe, we consider the force that arises from the movement on the time-dependent mass. See~\ref{app:sec:screening:calculations} for the full calculation. In the limit where the probe radius is much smaller than the distance between the probe and the source sphere $R_P \ll |\mathbf{X}_S(t)|$, we find the following expression for the force: 
\begin{align} \label{eq:force:spheres}
F&= -\frac{G\,M_Sm}{|\mathbf{X}_S|^2}\biggl[1 +  \alpha_{\mathrm{bg},P} \left( 1 + \frac{|\mathbf{X}_S(t)|}{\lambda_{\mathrm{bg}}} \right) e^{-|\mathbf{X}_S(t)|/\lambda_{\mathrm{bg}}}f(R_P/\lambda_{\mathrm{bg}}, |\mathbf{X}_S(t)|/\lambda_{\mathrm{bg}}) \biggr],
\end{align}
where the sensor-dependent fifth-force strength is defined as
\begin{equation}\label{eq:alphabgP}
	\alpha_{\mathrm{bg},P} =  \frac{2M_{\mathrm{P}}^2}{M^2} \,  \xi_S \, \xi_P\,,
\end{equation}
where we have added the subscript '$P$' to denote that screening from the probe is here taken into account. Furthermore,  $\xi_S$ and $\xi_P$ (again labelled $S$ for the source and $P$ probe, respectively), are given in equation~\eqref{eq:def:of:xi}. To compute $\xi_P$, we replace $M_S$, $R_S$ and $\rho_S$ with $M_P$, $R_P$ and $\rho_P$. 
Finally, the function $f$ is a form-factor given by 
\begin{align} \label{eq:form:factor}
f(u,v) = (1 + u) \, e^{-u}\left[\frac{\sinh(u)}{u} - \left(\frac{v}{1 + v} - 2\right)\frac{1}{v}\left(\cosh(u) - \frac{\sinh(u)}{u}\right)\right].
\end{align}
This approaches $1$ in the $x = m_{\mathrm{bg}} R_P c/\hbar = R_P/\lambda_{\mathrm{bg}} \rightarrow 0$ limit, in which case equation~\eqref{eq:force:spheres} reduces to the result of Burrage \textit{et al}.~\cite{Burrage:2014oza} for the force between two spheres. Since spherical probes or source masses generally maximise the screening~\cite{Burrage:2017shh}, equation~\eqref{eq:force:spheres} can be interpreted as a conservative estimate of the screening due to the shielding from the probe. 

Burrage \textit{et al}.~\cite{Burrage:2014oza} make use of the $R_P /\lambda_\mathrm{bg} \rightarrow 0$, since the probe radius in the case of atom interferometry is typically much smaller than $\lambda_{\mathrm{bg}}$. For an optomechanical probe, however, the additional screening introduced by the probe can be substantial, but not, as we shall see, detrimental. In what follows, we compute the sensitivity both with and without the screening from the probe, where the latter corresponds to approximating the probe as a point particle.

\subsection{Potential from a moving source-mass}

In this work, we consider a moving source mass. This brings up a consideration of how the chameleon field responds to the motion of the source mass. For the gravitational field, we know that changes in the potential propagate outwards at the speed of light, and thus the appropriate potential to use is the retarded Newtonian potential. The situation is less clear for the scalar field, however. Since it is massive, it is not immediately obvious information will propagate outwards at the speed of light. To get an idea of its behaviour we need to know the speed, $v_I$, at which information propagates through the scalar field. We show in~\ref{app:scalar_field_evolution} that this is also, in fact, the speed of light. Consequently, the potential at 3D position $\mathbf{X}$ can be approximated by the time dependent form
\begin{align} \label{app:chameleon:potential}
\Phi_{\mathrm{C}}(\mathbf{X},t) = \frac{\phi_{\mathrm{bg}}}{M}-  \frac{GM_S}{|\mathbf{X} - \mathbf{X}_S(t)|}  \alpha_\mathrm{bg}  \, e^{-|\mathbf{X} - \mathbf{X}_S(t)|/\lambda_\mathrm{bg} } .
\end{align}
where $t$ should be replaced with the retarded time given by equation~\eqref{eq:tret}, however, we can ignore this for the non-relativistic speeds and distances considered in this setup. Since both the chameleon field, $\phi$, and the metric, $g_{\mu\nu}$ (which gives rise to the Newtonian potential, $\Phi_N$) are well-defined dynamical quantities, the time-dependence of this potential is well-defined. We note at this point that if quantum corrections are large, the effective speed of information propagation for the scalar field, $v_I$, may differ from $c$~\cite{ellis2007causality}. However, large quantum corrections of this size would mean that we cannot readily use the effective field theory treatment of the chameleon field assumed throughout~\cite{Khoury:2013yya}, so we do not consider this effect here. We can therefore use equation~\eqref{app:chameleon:potential} in the discussion that follows to measure the values of $\alpha$ and $\lambda$, and thus the parameters $\Lambda$ and $M$ of the chameleon field.

\section{Linearised modified Newtonian potential} \label{sec:linearised:potential}

In order to compute the sensitivity of the optomechanical system, we need to include the force on the sensor shown in equation~\eqref{eq:force:spheres} into the dynamics of the optomechanical system. It is possible to obtain the solution numerically, but in order to obtain analytic expressions, we choose to linearise the Yukawa modification of the force for small oscillations of the source-mass. We let the time-dependent distance between the systems $x_S(t)$ be given by:
\begin{equation} \label{eq:time:dependent:distance}
 x_S(t ) = x_0 \, \left( 1 - \epsilon \cos(\omega_0 \, t  + \phi_0) \right) , 
\end{equation}
where $\epsilon$ is a dimensionless oscillation amplitude defined as a fraction of $x_0$, where $\omega_0$ is the oscillation frequency and $\phi_0$ is a phase shift that we specify later in order to maximize the sensitivity. 

In the following two sections, we show the linearisation of the force for a generic Yukawa potential, and for the chameleon force with a large optomechanical probe that contributes to the screening.

\subsection{Linearising the Yukawa potential}

We now linearise the contributions from the Yukawa potential to equation~\eqref{eq:modified:gravitational:potential} for small oscillation amplitudes $\epsilon\ll 1$. We note that, for specific values of $\alpha$ and $\lambda$, higher order contributions to the Newtonian  gravitational force may be larger than the first-order contributions to the Yukawa  force. It is therefore important that, when taking data in an experiment, we determine the origin of the observed values (see the Discussion in section~\ref{sec:discussion}).   
Linearising, we obtain 
\begin{align} \label{eq:linearised:Yukawa:int}
\mathcal{G}_{\mathrm{Yuk}}(t) &\approx   - \frac{G M_S m }{ \, x_S^2(t) }  -
 m g_\mathrm{N} \bigl[ \kappa + \sigma \,\epsilon \,\cos (t  \omega_0 + \phi_0 ) \bigr] \, . 
\end{align}
where $g_\mathrm{N}=G \, M_S/x_0^2$ is the Newtonian gravitational acceleration at the equilibrium distance $x_0$ 
and where we defined the two parameters 
\begin{align} \label{eq:kappa}
	&\kappa =  \alpha \, e^{- x_0/\lambda} \, \left( 1 + \frac{x_0}{\lambda}\right) , 
&& \mbox{and}
	&& \sigma =  \alpha \, e^{- x_0/\lambda} \,  \left(2 + 2 \frac{x_0}{\lambda} + \frac{ x_0^2}{\lambda^2}  \right)   \,,
\end{align}
which quantify the deviation of the constant and the time-dependent part of the  force from the 
Newtonian  one, respectively. We make this distinction because in an experiment, it is often possible to isolate a time-dependent signal from a constant noise floor.  In addition, systematic effects such as the Casimir effect can be effectively screened out in this way (we return to this point in the Discussion in Section~\ref{sec:discussion}). We will therefore focus on estimating $\sigma$ as part of our analysis.

\subsection{Linearising the chameleon potential from a screened spherical probe}
For a spherical optomechanical probe, the force between the probe and the source is given in equation~\eqref{eq:force:spheres}. We note that the form factor shown in equation~\eqref{eq:form:factor}, which arises due to the screening from the optomechanical probe, depends on $x_S(t)$ and is therefore time-dependent. 
To determine the constant and time-dependent contributions, we again assume that the source sphere oscillates around the equilibrium distance $x_0$ according to equation~\eqref{eq:time:dependent:distance}. Noting that $x_S(t)$ only enters into the $y$-terms in equation~\eqref{eq:form:factor}, we find 
\begin{align} \label{eq:linearised:potential:chameleon}
\mathcal{G}_{\mathrm{Cha}}(t) &\approx  - \frac{G M_S m }{ \, x_S^2(t) }  - m g_{\mathrm{N}} \left( \kappa + \sigma \epsilon \cos(\omega_0 \, t + \phi_0) \right), 
\end{align}
where the expressions for $\kappa$ and $\sigma$ now read
\begin{align} \label{eq:kappa:sigma:shielded}
&\kappa =  \alpha_{\mathrm{bg},P} \, e^{-  x_0/ \lambda_{\mathrm{bg}}}  \biggl[  \left( 1 + \frac{x_0}{\lambda_{\mathrm{bg}}}\right)A( R_P/\lambda_{\mathrm{bg}})  +    \left(  1 +2 \frac{\lambda_{\mathrm{bg}}}{x_0}  \right)B(R_P/\lambda_{\mathrm{bg}})  \biggr],
  \\
&\sigma =  \alpha_{\mathrm{bg},P}  \, e^{- x_0/\lambda_{\mathrm{bg}}} \biggl[   \left( 2 + 2 \frac{x_0}{\lambda_{\mathrm{bg}}} + \frac{ x_0^2}{\lambda_{\mathrm{bg}}^2}  \right)A(R_P/\lambda_{\mathrm{bg}}) +   \left( 4 +6  \frac{\lambda_{\mathrm{bg}}}{x_0}  + \frac{x_0}{\lambda_{\mathrm{bg}}}  \right)  B( R_P/\lambda_{\mathrm{bg}}) \biggr], \nonumber
\end{align}
where we have defined:
\begin{align} \label{eq:A:B}
&A(u) = (1 + u) \,  e^{- u} \, \frac{\sinh(u)}{u}, &&\mbox{and} 
&&B(u) = (1 + u) \, e^{- u} \, \left( \cosh(u) - \frac{\sinh(u)}{u}\right). 
\end{align}
The expressions in equation~\eqref{eq:A:B} arise from the form-factor in equation~\eqref{eq:form:factor}.

For the parameter regimes considered in this work, we find that $ R_P/\lambda_{\mathrm{bg}} \ll 1$. This means that the form factors become $A(R_P/\lambda_\mathrm{bg}) = 1$ and $B(R_P/\lambda_{\mathrm{bg}}) = 0$. As a result, $\kappa$ and $\sigma$ simplify to 
\begin{align} \label{eq:kappa:sigma:shielded}
&\kappa =    \alpha_{\mathrm{bg},P}  \, e^{-  x_0/ \lambda_{\mathrm{bg}}} \left( 1 + \frac{x_0}{\lambda_{\mathrm{bg}}}\right), \nonumber
  \\
&\sigma = \alpha_{\mathrm{bg},P} \, e^{- x_0/\lambda_{\mathrm{bg}}}   \left( 2 + 2 \frac{x_0}{\lambda_{\mathrm{bg}}} + \frac{ x_0^2}{\lambda_{\mathrm{bg}}^2}  \right), 
\end{align}
which has the same form as equation~\eqref{eq:kappa}. We are now ready to compute the sensitivities of the optomechanical sensor, but first, we provide a brief introduction to the quantum metrology tools we use for this purpose.

\section{Quantum metrology and ideal bounds} \label{sec:metrology}

In this work, we are interested in the best-possible sensitivity that can be achieved with the optomechanical probe. To determine the sensitivity of the probe, we turn to tools from quantum metrology. 
Specifically, we focus on computing the quantum Fisher information (QFI), which we denote $\mathcal{I}_\theta$, where $\theta$ is the parameter that we wish to estimate. Intuitively the QFI can be seen as a measure of how much the quantum state of the system changes given a specific encoding of $\theta$. The QFI then provides a measure of the change in the state with $\theta$ compared with the case when the state is unaffected. See also Ref~\cite{meyer2021fisher} for an intuitive introduction to the QFI and related concepts in quantum metrology. 

The connection to sensitivity stems from the fact that the QFI provides a lower bound to the variance $\mathrm{Var}(\theta)$ of $\theta$ through the quantum Cram\'{e}r--Rao bound~\cite{cramer1946contribution, rao1992information}:
\begin{equation}
\mathrm{Var} ( \theta) \geq \frac{1}{\mathcal{M} \, \mathcal{I}_\theta } ,
\end{equation}
where $\mathcal{M}$ is the number of measurements or probes used in parallel. The standard deviation of $\theta $ is then given by $\Delta \theta = 1/\sqrt{\mathcal{M}\, \mathcal{I}_\theta} $. 

For unitary dynamics and mixed initial states written in the form of $\hat \varrho = \sum_n \lambda_n \ketbra{\lambda_n}$, the QFI can be cast as~\cite{pang2014,jing2014}:
\begin{align}\label{definition:of:QFI:appendix}
\mathcal{I}_\theta
=&  \, 4\sum_n \lambda_n\left(\bra{\lambda_n}\mathcal{\hat H}_\theta^2\ket{\lambda_n} - \bra{\lambda_n}\mathcal{\hat H}_\theta\ket{\lambda_n}^2 \right) - 8\sum_{n\neq m}\frac{\lambda_n \lambda_m}{\lambda_n+\lambda_m}\left| \bra{\lambda_n}\mathcal{\hat H}_\theta \ket{\lambda_m}\right|^2,
\end{align}
where the operator $\mathcal{\hat H}_\theta$ is defined as $\mathcal{\hat H}_\theta = - i \hat U_\theta^\dag \partial_\theta \hat U_\theta $. Here, $\hat U_\theta$ is the  unitary operator that encodes the parameter $\theta$ into the system.

In our case, $\hat U(\theta)$ is the unitary operator that arises from the Hamiltonian in equation~\eqref{eq:cham:Hamiltonian}, and the effect we wish to estimate is the effect of the Yukawa potential on the probe. Therefore, in order to compute $\mathcal{I}_\theta$, we must first solve the time-evolution of the system, which is often challenging when the signal is time-dependent, as is the case for us here. Some of these challenges can however be addressed by making use of a previously established method for solving the Schrödinger equation using a Lie algebra approach~\cite{wei1963lie}.  Details of this solution were first used to study a purely Newtonian time-dependent gravitational potential~\cite{qvarfort2020optimal} and can be found in~\ref{app:sensitivity}.

Using the expression for $\mathcal{I}_\theta$ in equation~\eqref{definition:of:QFI:appendix}, we can derive a compact expression for the QFI that represent the sensitivity with which modifications to Newtonian gravity can be detected. In our case, we let the parameter $\theta$ of interest be either $\kappa$ or $\sigma$ as defined in equation~\eqref{eq:kappa}.  By then applying the Cram\'er--Rao bound, we can derive the standard deviation for each parameter. We then consider the ratios $\Delta \kappa/\kappa $ or $\Delta \sigma /\sigma$, which describe the relative error of the collective measurements. 

In this work, we say that we can distinguish modifications to the Newtonian potential if the error in $\kappa$ and $\sigma$ is smaller than one, that is, when $\Delta \kappa/\kappa < 1$ or $\Delta \sigma / \sigma < 1$. Note that, to find the sensitivity to the actual values of, for example, $\alpha$ and $\lambda$, we would need a full multi-parameter likelihood analysis, which requires us to go beyond the regular error-propagation formula for the parameter we consider here. Such an analysis is currently beyond the scope of this work. Instead, we focus mainly on detecting $\sigma$, since it is the amplitude of the time-dependent signal. 

Unfortunately, the QFI does not actually reveal the optimal measurement that saturates the quantum Cram\'{e}r--Rao bound. To obtain this information, one must compute the classical Fisher information  for a particular measurement and examine whether it saturates the quantum Fisher information. It is known that, when the optomechanical coupling is constant and takes on specific values, that a homodyne measurement of the optical field is optimal~\cite{qvarfort2018gravimetry,qvarfort2020optimal}. When the optomechanical coupling is modulated at resonance, as is the case here, the optimal measurement is not yet known. The  gravitational interaction between the source and the optomechanical probe results in a phase shift of the optical state.  Therefore, the utility of a homodyne measurements can be expected also for the case of modulated optomechanical coupling, but we leave this specific analysis to future work.
In practise, once the optomechanical probe has interacted with the source, the system is measured to extract information about the gravitational force. Standard measurements that are performed on the optomechanical system include homodyne and heterodyne measurements of the cavity field, as well as photon detection measurements, which can either be resolving (counting the number of photons) or non-resolving (merely detecting the presence of a photon). In a homodyne measurement, the output light from the optomechanical system is brought into interference with a local oscillator light field which comes from the same source as the input light field of the optomechanical system. This is the same measurement principle that is, for example, employed in a Mach-Zehnder interferometer to infer a phase shift on a light field.  Heterodyne measurements, on the other hand, compare the collected light with a different coherent state reference. The usefulness of each measurement depends on the situation at hand. Since we focus on deriving the best-possible sensitivities in this work, we leave it to future work to analyse the sensitivity that can be gained from specific measurements.

\begin{figure*}
\centering
\subfloat[ \label{fig:force:alpha:lambda}]{%
  \includegraphics[width=0.45\linewidth, trim = 0mm 0mm 0mm 0mm]{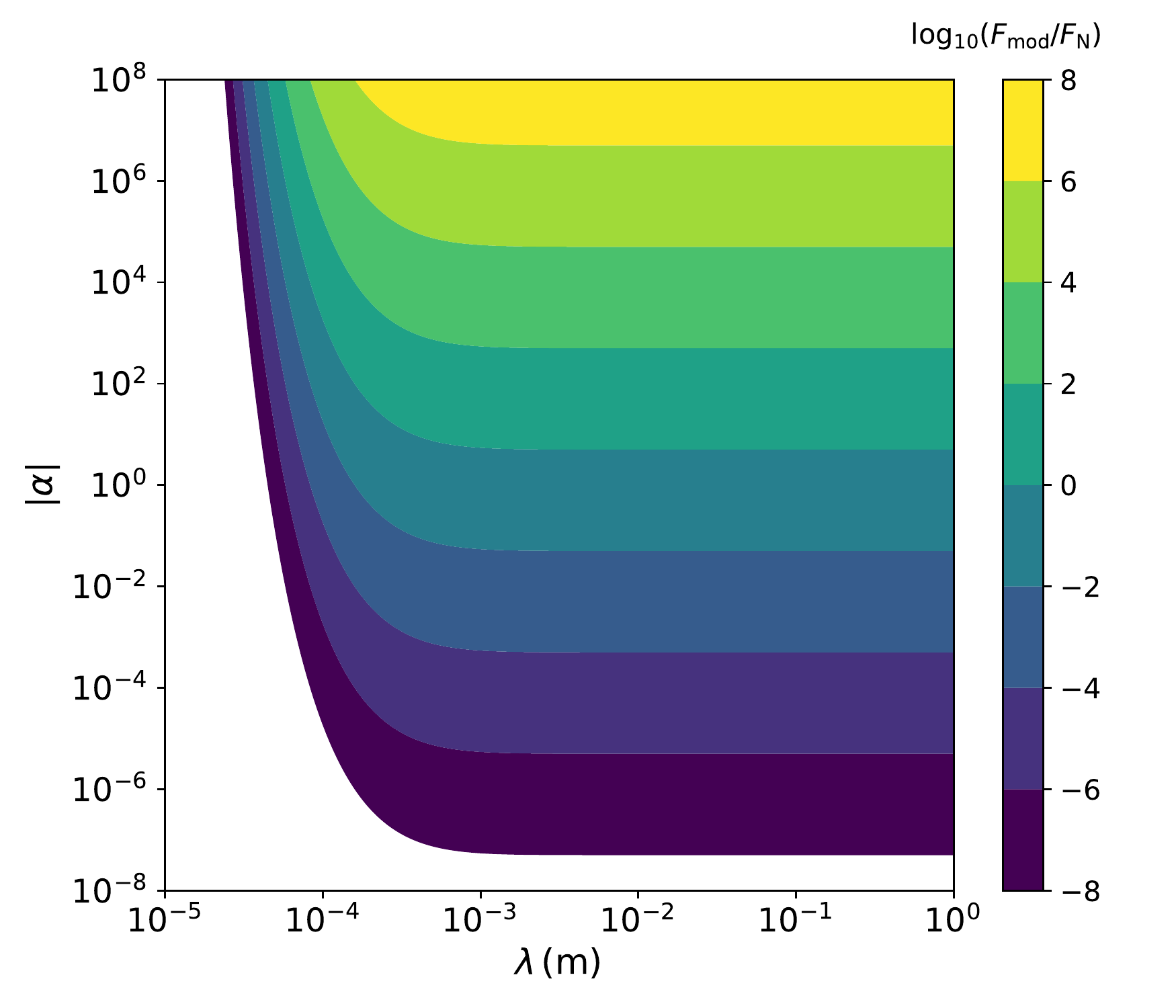}%
} $\qquad$
\subfloat[ \label{fig:force:M:Lambda}]{%
  \includegraphics[width=0.45\linewidth, trim = 0mm 0mm 0mm 0mm]{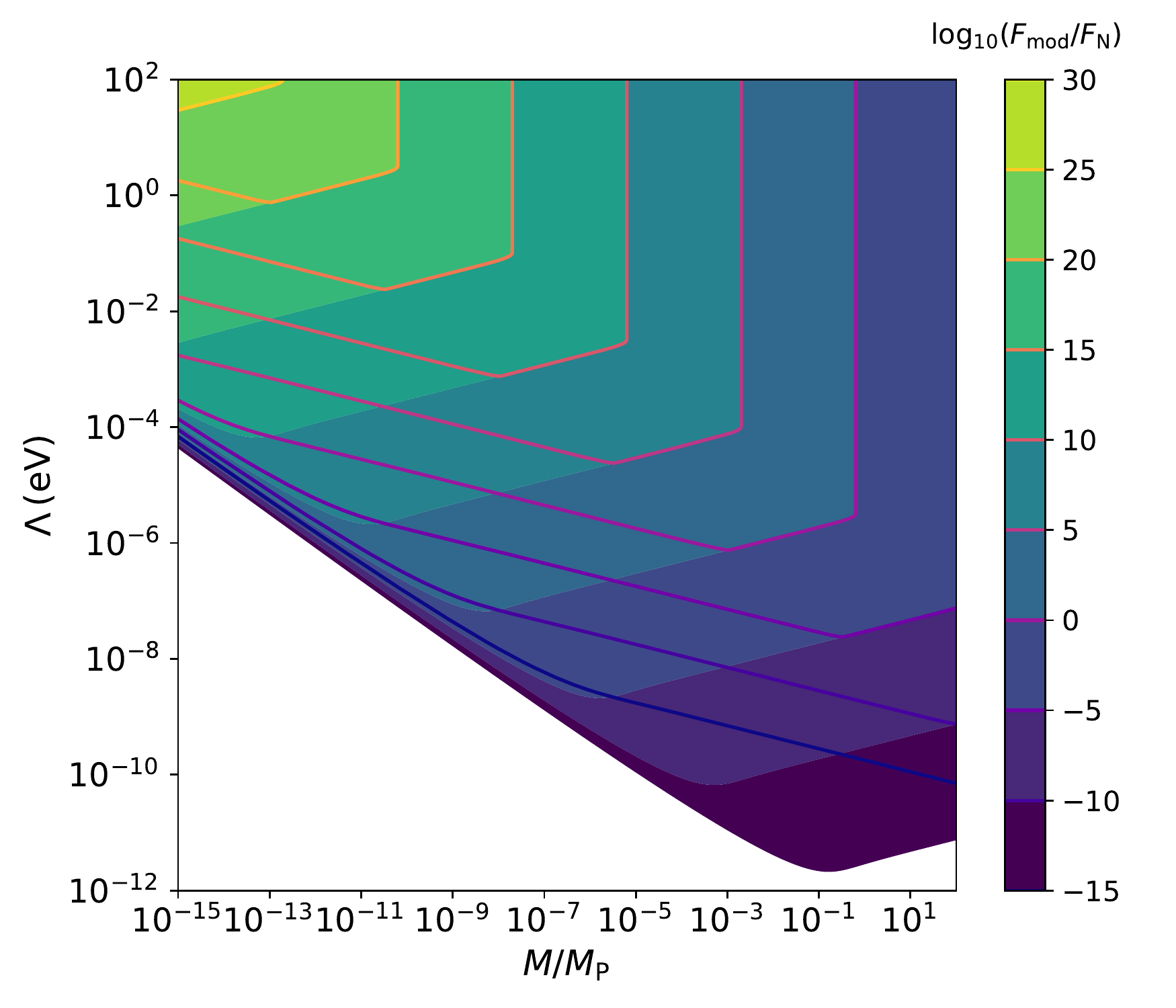}
}\hfill
\caption{Plots of the ratio of the Newtonian force and the modification $ F_{\mathrm{mod}}/F_{\mathrm{N}} = \epsilon \sigma$.  Plot (a) shows the ratio as a function of $\alpha$ and $\lambda$. Plot (b) shows the ratio as a function of $M$ and $\Lambda$ in units of the Planck mass $M_{\mathrm{P}}$ and eV, respectively. The filled-in contours show the radio without screening from the probe. The lines instead show the ratio for when the probe is spherical and contributes to the screening. As a result, the strength of the force is reduced.  The parameters used to make these plots can be found in table~\ref{tab:Values}. The mapping between the $(\alpha,\lambda)$ and $(M,\Lambda)$ spaces is non-trivial. The left hand figure shows the range of force modifications that can potentially detected, which is different to the range of theoretically-interesting chameleon parameters, shown on the right.}
\label{fig:force:plot}
\end{figure*}

\section{Results} \label{sec:results}
We are now ready to compute the sensitivities that can be achieved with an ideal optomechanical sensor for detecting modifications of gravity. Specifically, we consider a region of parameter space to be possible to exclude using the optomechanical sensor when the best precision possible on the parameters $\alpha$ and $\lambda$ (or the chameleon parameters $M$ and $\Lambda$) is sufficient to distinguish them from zero, their values in ordinary General Relativity.

\subsection{Fundamental sensitivities}

We first present some simple expressions for the sensitivities that can be achieved, and we then proceed to compute the  parameter regions that could potentially be excluded with an optomechanical sensor. 
When the source mass oscillates at the same frequency as the optomechanical system, that is, when $\omega_0 =\omega_{\mathrm{mwch}}$, the effects accumulate and cause the position of the optomechanical system to become increasingly displaced.  

Following the outline in~\ref{app:sensitivity}  we find the following expressions for the sensitivities for $\kappa$ and $\sigma$ at time $t \, \omega_{\mathrm{mech}}= 2\pi n $ (see~\cite{qvarfort2020optimal} for a detailed derivation). For large enough temperatures in the mechanical state, such that $r_T\gg1$, the expressions simplify and we find that the sensitivities $\Delta \kappa$ and $\Delta \sigma$ are given by 
\begin{align} 
\Delta \kappa &= \frac{1}{\sqrt{\mathcal{M}} \, g_\mathrm{N} } \frac{1}{\Delta \hat  N_a} \sqrt{\frac{ 2 \hbar \, \omega_{\mathrm{mech}}^5}{m}}  \frac{1}{8  \pi\, n \,  k_0  } , \label{eq:constant:sensitivity:kappa} \\
\Delta \sigma &= \frac{1}{\sqrt{\mathcal{M}} \, g_\mathrm{N} } \frac{1}{\Delta \hat  N_a} \sqrt{\frac{ 2 \hbar \, \omega_{\mathrm{mech}}^5}{m}} \frac{1}{4  \pi\, n \,   k_0 \,\epsilon },  \label{eq:constant:sensitivity:sigma}
\end{align}
where $n$ is an integer, and for an optomechanical coupling  $k(t) \equiv k_0$ and phase $\phi_0 = \pi$, and  where the variance $(\Delta \hat  N_a)^2$ of the photon number is given by~\cite{qvarfort2020optimal}
\begin{align} \label{eq:photon:number:variance}
( \Delta \hat  N_a)^2 &= |\mu_{\mathrm{c}}|^2 e^{ 4 r_{\mathrm{sq}}} + \frac{1}{2} \sinh^2(2\,r_{\mathrm{sq}})  - 2 \, \mathfrak{Re} [ e^{- i \varphi/2} \mu_{\mathrm{c}} ]^2 \sinh(4\,r_{\mathrm{sq}}), 
\end{align}
where $r_{\mathrm{sq}}$ and $\varphi$ are the squeezing amplitude and phase, and where $\mu_{\mathrm{c}}$ is the coherent state amplitude of the optical mode. The expression in equation~\eqref{eq:photon:number:variance} is maximised when $e^{-i\varphi/2 }\mu_{\rm{c}} $ is completely imaginary, which causes the last term of equation~\eqref{eq:photon:number:variance} to vanish. This can be achieved by assuming that $\mu_{\rm{c}}\in \mathbb{R}$ and setting the squeezing phase to $\varphi = \pi/2$. The other parameters in equation~\eqref{eq:photon:number:variance} have been previously defined in the text (see also table~\ref{tab:Values} for a summary).  

The sensitivities can be improved by modulating the optomechanical coupling at the same frequency as the gravitational signal~\cite{qvarfort2020optimal}. 
In this work, we choose a sinusoidal modulation with  $k(t) = k_0 \cos(\omega_k \, t )$, where $k_0$ is the amplitude of the modulation and $\omega_k$ is the modulation frequency. At resonance, when $\omega_k = \omega_{\mathrm{mech}}$, and for the optimal phase choice $\phi_0 = \pi/2$, we find that the sensitivities for measuring $\kappa$ and $\sigma$ become
\begin{align} 
\Delta \kappa^{(\mathrm{mod})} &= \frac{1}{\sqrt{\mathcal{M}} \, g_\mathrm{N} } \frac{1}{\Delta \hat  N_a} \sqrt{\frac{ 2 \hbar \, \omega_{\mathrm{mech}}^5}{m}}  \frac{1}{ 4 \pi\, n \,   k_0  },  \label{eq:modulated:sensitivity:kappa} \\
\Delta \sigma^{(\mathrm{mod})} &= \frac{1}{\sqrt{\mathcal{M}} \, g_\mathrm{N} } \frac{1}{\Delta \hat  N_a} \sqrt{\frac{ 2 \hbar \, \omega_{\mathrm{mech}}^5}{m}} \frac{1}{ 2  \pi^2 \, n^2 \, \,  k_0 \,\epsilon }\label{eq:modulated:sensitivity:sigma} . 
\end{align}
Here, equation~\eqref{eq:modulated:sensitivity:sigma} scales with $n^{-2}$ rather than $n^{-1}$. This enhancement arises from the additional modulation of the optomechanical coupling, and was already noted in the context of time-dependent gravimetry for a purely Newtonian potential~\cite{qvarfort2020optimal}. By now considering the cases where the uncertainty in the parameter is a fraction of the parameter itself, we are able to define the regions in which modifications to Newtonian gravity can be established with certainty.

\begin{table}[h!]
\centering
\caption{ Example parameters used to compute the bounds on modified gravity for a generic optomechanical sensor. We denote the optomechanical (probe) mass by  $m$ so as to not confuse it with the Planck mass $M_{\mathrm{P}}$.  \vspace{0.2cm}}
\begin{tabular}{Sl Sc Sc} \hline \hline
 \textbf{Parameter} & \textbf{Symbol}  &  \textbf{Value} \\ 
\hline 
Source mass & $M_{\mathrm{S}}$ & $10^{-6}\,$kg \\ 
Source mass density & $\rho_{\mathrm{S}}$ & $19.3\times 10^3$\,kg\,m$^{-3}$ \\
Source mass radius & $R_{\mathrm{S}}$ & $2\times 10^{-4}$\,m \\
Equilibrium distance  & $x_0$  & $ 10^{-3}$\,m    \\
Source oscillation amplitude ratio & $\epsilon $ &  0.1  \\ 
Background density & $\rho_{\mathrm{bg}}$  & $8.27\times 10^{-14}$\,kg\,m$^{-3}$ \\ \hline
Optomechanical coupling & $   k_0/(2\pi)$ & $10$\,Hz   \\
Mechanical frequency & $\omega_{\mathrm{mech}}/(2\pi)$ & $100$\,Hz  \\
Probe mass & $m$ & $10^{-14}$\,kg \\ 
Oscillator (probe) mass density (silica) & $\rho_{\mathrm{P}}$ & $1\,538$\,kg\,m$^{-3}$ \\\hline
Coherent state parameter & $|\mu_{\mathrm{c}}|^2$ & $10^6$  \\
Squeezing parameter & $r_{\mathrm{sq}}$ & 1.73 \\
 \hline
Number of measurements & $\mathcal{M}$ & $10^3$ \\
Time of measurement & $\omega_{\mathrm{mech}} \, t = 2\pi n $ & $n = 10$ \\  \hline 
Newtonian gravitational force at equilibrium distance & $ m g_\mathrm{N} $  & $\sim 6.67 \times 10^{-25}$\,N \\\hline
\multicolumn{3}{c}{Sensitivities (constant coupling)} \\\hline
Sensitivity  $\kappa$ & $\Delta \kappa$  & $1.36 \times 10^{-3}$ \\
Sensitivity for constant force &  $m g_{\mathrm{N}} \Delta \kappa$ & $9.08 \times 10^{-28}$\,N \\
Sensitivity  $\sigma$ & $\Delta \sigma$ & $27.1\times 10^{-3}$ \\ 
Sensitivity for res. oscillating force  &  $m g_{\mathrm{N}} \Delta \sigma \epsilon $ & $1.81 \times 10^{-27}$\,N \\\hline
\multicolumn{3}{c}{Sensitivities (resonant coupling)} \\ \hline
Sensitivity $\kappa$ 
 & $\Delta \kappa^{(\mathrm{mod})} $ & $2.71 \times 10^{-3}$  \\
Sensitivity for constant force  &  $m g_{\mathrm{N}} \Delta \kappa^{(\mathrm{mod})}$ & $1.81 \times 10^{-27}$\,N \\
Sensitivity $\sigma$  &$\Delta \sigma^{(\mathrm{mod})} $ & $1.73 \times 10^{-3}$ \\ 
Sensitivity for res. oscillating force &  $m g_{\mathrm{N}} \Delta \sigma^{(\mathrm{mod})} \epsilon $ & $1.15 \times 10^{-28}$\,N\\ \hline\hline
\end{tabular} \label{tab:Values}
\end{table}

\subsection{Experimental parameters}

We assume that the oscillating source mass oscillates at the resonant frequency of the optomechanical system.  We further assume that the source mass is made of solid gold, which has a density of $\rho = 19.3\times10^3$\,kg\,m$^{-3}$. For a mass of $10^{-6}$\,kg ($1$\,mg), this translates into a source mass radius of $R_S = 2.3\times 10^{-4}$\,m.
While this mass is very small compared with those currently used in atom interferometry experiments~\cite{hamilton2015atom}, gravitational fields from masses of slightly larger radii have recently been detected~\cite{westphal2020measurement}. The reason for choosing such a small mass is that the systems can be placed very close together while still achieving a significant oscillation amplitude. This allows us to probe parameter regimes of a short-ranged force. Due to the scaling of the sensitivity as $\Delta \theta \sim x_0^2$, choosing a smaller $x_0$ is always going to be beneficial. We therefore set $x_0  = 10^{-3}$\,m and assume that the oscillation amplitude ratio is $\epsilon = 0.1$. This ensures that, when the source mass oscillates, it does not come into contact with the optomechanical system\footnote{For the choice of such a small source-mass, it might be the case that we must take the mass of the modulation mechanism into account, which would change both the effective mass seen by the optomechanical probe, as well as the screening of the force. A standard piezo stack has a mass of 16\,g, for example.}

For the optomechanical probe, we use the following example parameters: we assume that the effective mass of the optomechanical probe is $ m = 10^{-14}$\,kg, and that the light-matter coupling has an amplitude of $ k_0/(2\pi) = 10$\,Hz. We then assume that the mechanical frequency can be made as low as $\omega_{\mathrm{mch}}/(2\pi) = 100 $\,Hz, which is important since the expressions for $\Delta \kappa$ and $\Delta \sigma$ scale with $\omega_{\mathrm{mech}}^{5/2}$. For the squeezed coherent state, we assume that the coherent state parameter is given by $|\mu_{\mathrm{c}}|^2= 10^6$ and that the phase of the squeezed light can be set to $\varphi = \pi$, which ensures that the photon number variance $(\Delta \hat N_a)^2$ shown in equation~\eqref{eq:photon:number:variance} is maximized. One of the highest squeezing factors that have been achieved to-date is $r_{\mathrm{sq}} = 1.73$~\cite{vahlbruch2016detection}, which is what we choose to include here. We also consider a protocol where we perform $\mathcal{M} = 10^3$ measurements at time $t \, \omega_{\mathrm{mech}} = 20 \pi$, which allows us to improve the sensitivity a bit further.

To derive the bounds on the chameleon parameters $M$ and $\Lambda$, we assume that the optomechanical system can be operated in high vacuum. This also helps in terms of mitigating mechanical noise; in generic oscillators, damping effects are well-understood and largely not present below $10^{-7}$\,mbar~\cite{cole2011phonon}. On the other hand, it can be challenging to confine a levitated optomechanical system at high vacuum~\cite{pontin2020ultranarrow}. Recently, however, several works have demonstrated trapping at $10^{-7}$\,mbar of pressure~\cite{pontin2020ultranarrow, delic2020levitated}, even going as low as $9,2\times 10^{-9}$\,mbar~\cite{magrini2020optimal}. Using these values as our starting point, we note that $ 10^{-9}$\,mbar translates into a molecular background density of $\rho_{\mathrm{bg}} = 8.27\times 10^{-14}$\,kg\,m$^{-3}$. To derive this value, we have used the ideal gas law, which can be rewritten to give $\rho_{\mathrm{bg}} = P m_{N_2}/(k_{\mathrm{B}} T)$. Here, $P$ is the pressure (in Pascal), $k_{\mathrm{B}}$ is Boltzmann's constant, $T$ is the temperature (in Kelvin), and where we have assumed that the vacuum chamber has been vented with hydrogen of molecular mass $m_{H} = 3.3\times10^{-27}$\,kg before being emptied (that is, it was filled with hydrogen gas, such that any residual particles inside the chamber are $H_2$ particles).

All parameters are summarized in table~\ref{tab:Values}. There, we also give values for the Newtonian gravitational force for source and sensor at their respective equilibrium positions, which is approximate equivalent to the time-averaged Newtonian force and
the sensitivities shown in equations~\eqref{eq:constant:sensitivity:kappa},~\eqref{eq:constant:sensitivity:sigma},~\eqref{eq:modulated:sensitivity:kappa}, and~\eqref{eq:modulated:sensitivity:sigma}. We find that for a constant optomechanical coupling, the sensitivities become $\Delta \kappa = 1.36 \times 10^{-3}$ and $\Delta \sigma = 27.1\times 10^{-3}$. For a time-dependent optomechanical coupling modulated sinusoidally at resonance, we find  sensitivities $\Delta \kappa^{(\mathrm{mod})} = 2.71 \times 10^{-3}$ and $\Delta \sigma^{(\mathrm{mod})} = 1.73 \times 10^{-3}$, where $\Delta \kappa^{(\mathrm{mod})}$ is slightly worse than $\Delta \kappa$ and $\Delta \sigma^{(\mathrm{mod})}$ is slightly better than $\Delta \sigma$. In table~\ref{tab:Values}, we also give the corresponding force sensitivities.

To see how strong the modified contributions to the force are compared with just the Newtonian part, we plot the amplitude of the time-dependent modification $F_{\mathrm{mod}} = \frac{G m M_{\mathrm{s}}}{x_0^2}  \epsilon \sigma $ as a fraction of the Newtonian force $F_{\mathrm{N}} = \frac{G m M_{\mathrm{s}}}{x_0^2} $. The result can be found in Figure~\ref{fig:force:plot}, where we have plotted contours for $ F_{\mathrm{mod}}/F_{\mathrm{N}} = \epsilon \sigma$ using the experimental parameters in table~\ref{tab:Values}. Figure~\ref{fig:force:alpha:lambda} shows $F_{\mathrm{mod}}/F_{\mathrm{N}}$ as a function of $\alpha$ and $\lambda$, and Figure~\ref{fig:force:M:Lambda} shows $F_{\mathrm{mod}}/F_{\mathrm{N}}$ as a function of $M$  and $\Lambda$. The filled-in contours in Figure~\ref{fig:force:M:Lambda} correspond to the force shown in equation~\eqref{eq:force:no:screening}, where the screening from the optomechanical probe itself has been ignored. The lines, on the other hand, correspond to the force shown in equation~\eqref{eq:force:spheres} where the screening from a spherical probe has been taken into account.

\subsection{Fundamental bounds for the Yukawa parameters $\alpha$ and $\lambda$}

\begin{figure*}
\centering
\subfloat[ \label{fig:bound:alpha:lambda}]{%
  \includegraphics[width=0.45\linewidth, trim = 10mm 0mm -10mm 0mm]{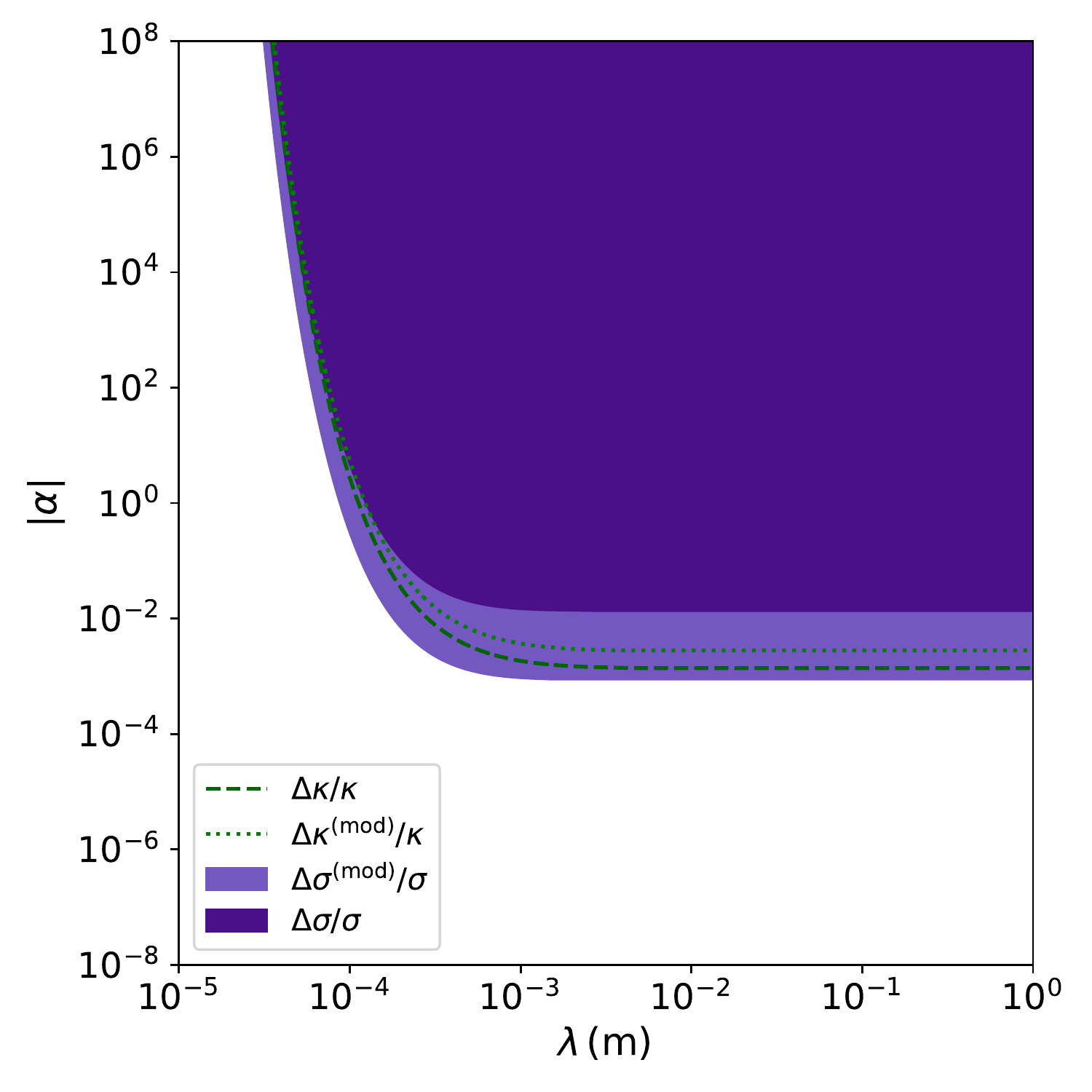}%
} $\qquad$
\subfloat[ \label{fig:bound:M:Lambda}]{%
  \includegraphics[width=0.45\linewidth, trim = 10mm 0mm -10mm 0mm]{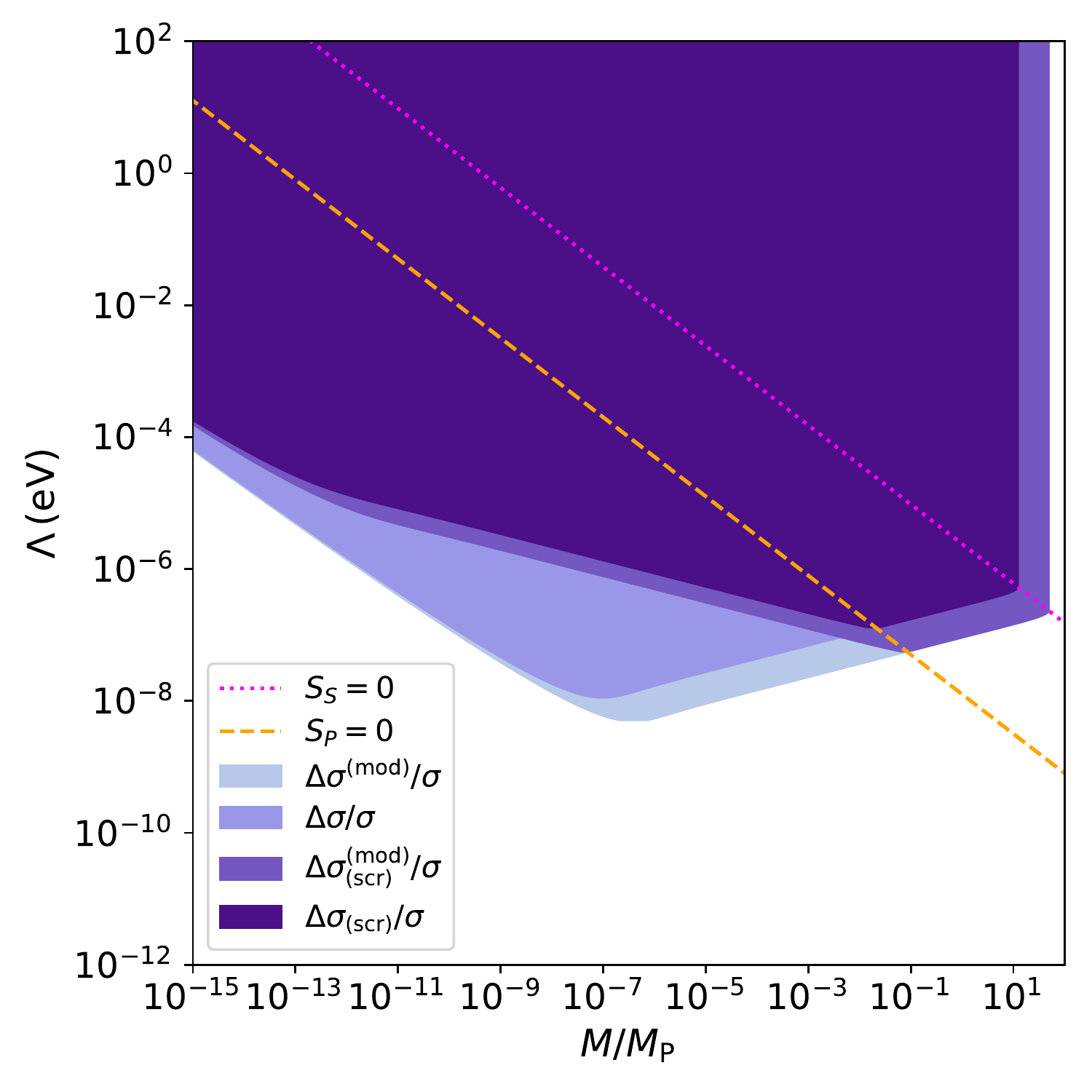}
}\hfill
\caption{Ideal bounds for detecting modifications to  Newtonian gravity with an optomechanical sensor. Each bound shows where the value of the modification is greater than the error bound. The parameters used in both plots are shown in table~\ref{tab:Values}. Plot (a) shows the bounds for the Yukawa parameters $\alpha$ and $\lambda$. The dashed dark green line indicates where $\Delta \kappa/\kappa = 1$, and the dotted lighter green line where $\Delta \kappa^{(\mathrm{res})} / \kappa = 1$. The light purple area shows the parameter regime where $\Delta \sigma^{(\mathrm{res})}/\sigma <1$ and the dark purple area shows where $\Delta \sigma / \sigma <1$. Since $\kappa$ is a constant effect, modulating the optomechanical coupling yields no improvement of the sensitivity.  Plot (b) shows the bounds for the chameleon parameters $M$ in terms of the Planck mass $M_{\mathrm{P}}$ and $\Lambda$ in eV. The bounds include a point-particle approximation of the sensor (the two largest lighter purple areas) and the inclusion of screening from a spherical probe (darker purple lines). The magenta dotted lined shows where the screening length of the source mass is zero $S_S = 0$, below which the screening of the probe starts reducing the sensitivity. Similarly, the orange dashed line shows where the screening length of the probe is zero $S_P = 0$, below which  the screening of the spherical probe reduces the sensitivity. We have refrained from plotting the bounds $\Delta \kappa/\kappa$ and $\Delta \kappa^{(\mathrm{mod})}/\kappa$ here as they roughly follow the outline of the bounds on $\sigma$.}
\label{fig:exclusion:plot}
\end{figure*}

We are now ready to compute the bounds on the parameter ranges that could potentially be tested with a quantum optomechanical system. 
To find the bounds, we consider the ratios $\Delta \kappa/ \kappa$ and $\Delta \sigma/\sigma $ as functions of $\alpha$ and $\lambda$, where $\kappa$ and $\sigma$ were defined in equation~\eqref{eq:kappa} as the modification due to the  gravitational force at the equilibrium distance and the amplitude of the time-dependent contribution. The result can be found in  figure~\ref{fig:bound:alpha:lambda}: the dark green dashed line shows where the relative error satisfies $\Delta \kappa/\kappa = 1$, and the dotted green line shows where $\Delta \kappa^{(\mathrm{mod})}/\kappa = 1$. Since $\kappa$ corresponds to the static modification of the gravitational force, modulating the optomechanical coupling does not improve the sensitivity. We instead focus on the dynamic contribution from $\sigma$. The lighter purple region shows where $\Delta \sigma/\sigma<1 $, and the darker purple region shows where $\Delta \sigma^{(\mathrm{mod})} < 1$. The resonantly modulated optomechanical coupling provides a significant enhancement for $\Delta \sigma$. 

The general features in figure~\ref{fig:bound:alpha:lambda} can be understood by examining the form of $\kappa$ and $\sigma$, which are shown in equation~\eqref{eq:kappa}. When $\lambda \gg x_0$, the exponential can be approximated as $e^{-x_0/\lambda} \sim 1$. This means that $\sigma$ becomes $\sigma \sim 2\alpha$, which is independent $\lambda$ and thereby explains the straight line at $|\alpha| \sim 10^{-3}$. Once $\lambda < x_0$, which corresponds to a short-ranged Yukawa force, the effect can no longer be seen by the optomechanical probe. However, the bounds in figure~\ref{fig:bound:alpha:lambda} could be shifted to the left by decreasing  $x_0$. Care must be taken that the two systems do not touch, which is limited by the source sphere and probe radii, as well as the oscillation amplitude $\epsilon x_0$. For the example parameters used here, the smallest distance between the system is $0.7$\,mm.

\subsection{Fundamental bounds for the chameleon parameters  $M$ and $\Lambda$}

To obtain the bounds on $M$ and $\Lambda$, we rescale $M$ in terms of the reduced Planck mass $M_{\mathrm{P}}$. We then compute the bounds for $M$ and $\Lambda$ by plotting $\Delta \sigma/\sigma$ as a function of $M$ and $\Lambda$ for the following two cases: (i) when the probe is approximated as a point-particle (no probe screening), and (ii) when the screening from the probe is taken into account. The latter we denote by $\Delta \sigma_{(\mathrm{scr})}$ and $\Delta \sigma_{(\mathrm{scr})}^{(\mathrm{mod})}$. 
We compute these quantities by numerically solving equation~\eqref{eq:solve:for:S} for $S_S$ and $S_P$ at each point. The expression for $\sigma$ given in equation~\eqref{eq:kappa:sigma:shielded}. 

The result can be found in figure~\ref{fig:bound:M:Lambda}. Note that we do not plot the bounds for $\Delta\kappa$ and $\Delta\kappa^{(mod)}$ for clarity, and because as static contributions they are more difficult to distinguish from a constant noise floor. The lighter regions show the bounds when the optomechanical probe does not contribute to the screening of the fifth force. This is equivalent to approximating the probe as a point-particle. In contrast, the darker regions show the reduction in sensitivity due to the screening that arises from a spherical optomechanical probe. 

To explain the features of the plot, we draw lines where the screening from the probe $S_P$ and source system $S_S$ vanishes. 
The magenta line shows where $S_S = 0$ and the orange line shows where $S_P = 0$. Above each line, the screening is zero, while below the lines, the screening lengths increase and the modifications to Newtonian gravity can no longer be detected. Finally, the right-most boundary of the dark purple area can be understood as follows: The appearance of $M^{-3}$  in  $m_{\mathrm{bg}}$ (see equation~\eqref{eq:mbg}) ensures that, when $M$ is large compared with the other quantities, $m_{\mathrm{bg}}$ is small. This, in turn, means that the range of the force $\lambda_{\mathrm{bg}}$, as shown in equation~\eqref{eq:lambda:bg} will be large. It then follows that the amplitude $\sigma$ (see equation~\eqref{eq:kappa:sigma:shielded}) will be approximately $\sigma \approx \alpha_{\mathrm{bg}, P}$, where $\alpha_{\mathrm{bg}, P} = 2 M^2/ M_{\mathrm{P}}^2$ from equation~\eqref{eq:alphabgP} (note that $\xi_S = \xi_P = 1$ because we are considering the range of $\Lambda$ above the orange and magenta lines). This means that $\sigma$ is independent of $\Lambda$ and the boundary becomes a vertical line. The point at which the ratio $\Delta \sigma/\sigma = 1$ then occurs is $M/M_{\mathrm{P}} = 48.1$.

\begin{figure*}
\centering
\subfloat[ \label{fig:convex:hull:alpha:lambda}]{%
  \includegraphics[width=0.45\linewidth, trim = 10mm 0mm -10mm 0mm]{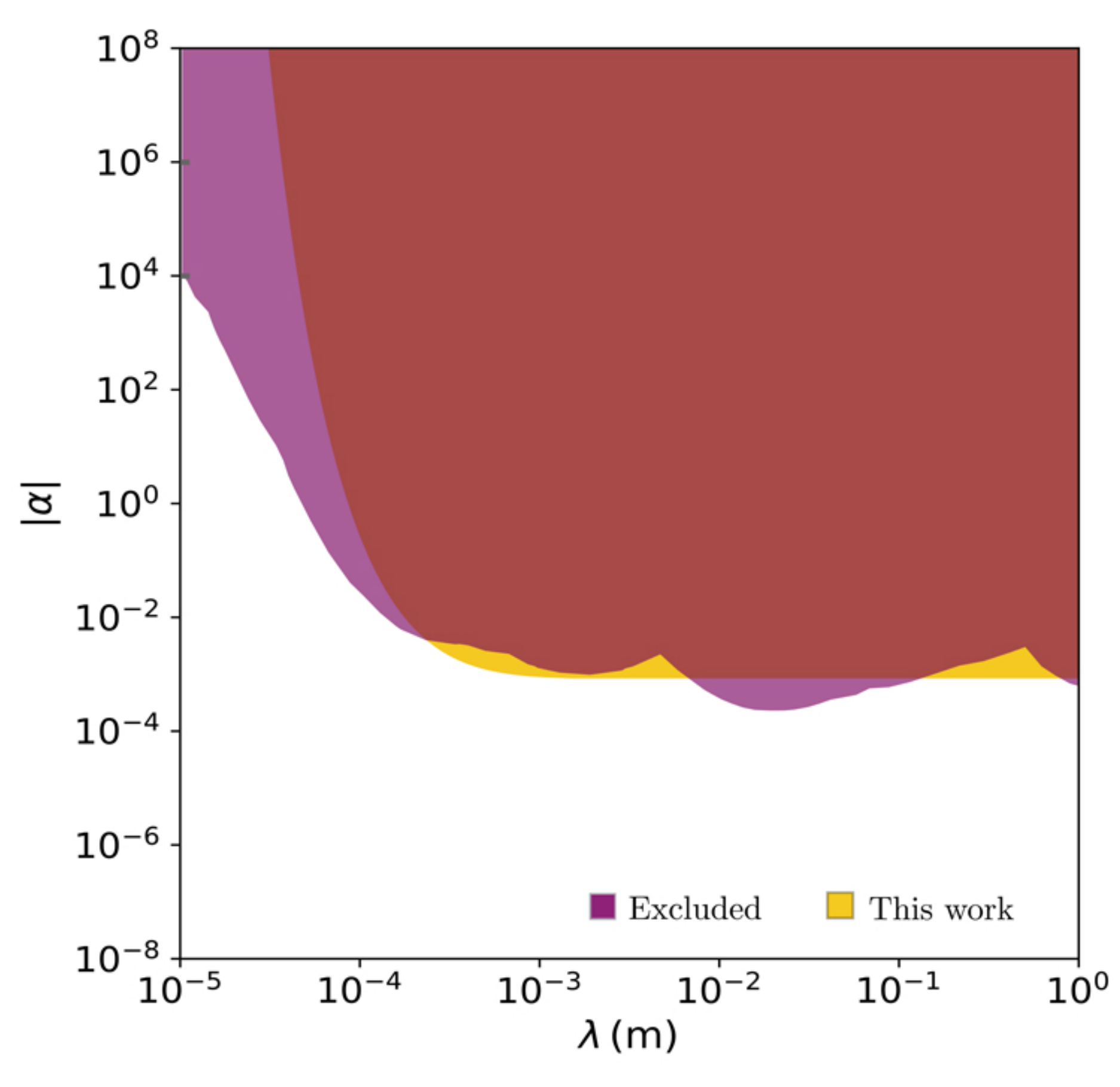}%
} $\qquad$
\subfloat[ \label{fig:convex:hull:M:Lambda}]{%
  \includegraphics[width=0.45\linewidth, trim = 10mm 0mm -10mm -0mm]{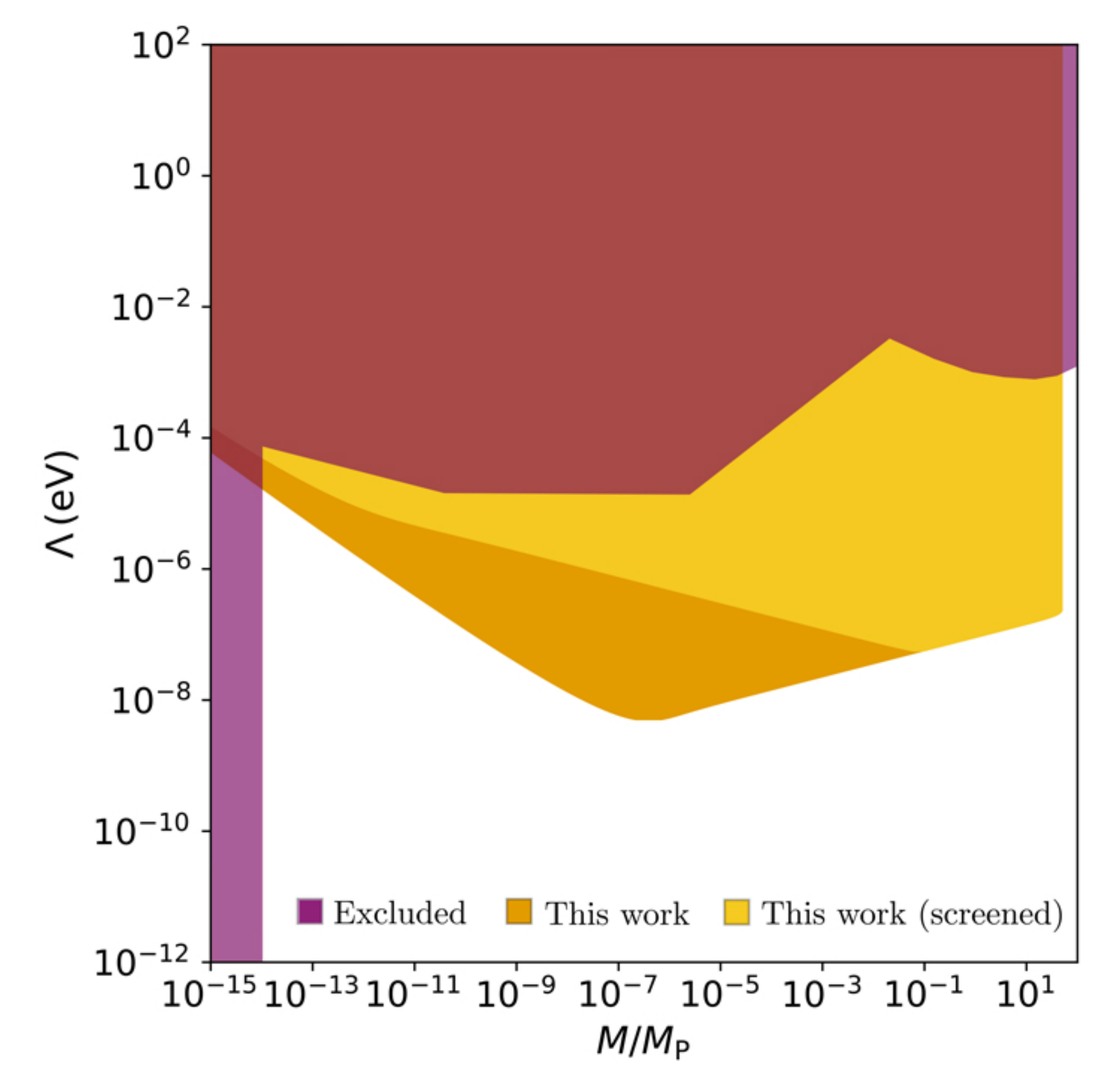}
}\hfill
\caption{Comparison between predictions (this work) and known experimental bounds (pink region). Both plots show the convex hull (yellow) of the bounds derived in this work in figure~\ref{fig:exclusion:plot}. Plot (a) shows the bounds in terms of the Yukawa parameters $\alpha$ and $\lambda$, while Plot (b) shows the bounds in terms of the chameleon screening parameters $M$ and $\Lambda$. Plot (b) also includes the bounds (yellow) for when the optomechanical probe contributes to the screening of the chameleon field. The pink areas represent the experimentally excluded regions based on figure~8 of~\cite{Murata_2015}  and recent results presented in~\cite{PhysRevLett.124.051301} (see figure~6). (b) shows bounds in terms of $M$ and $\Lambda$, which are the mass and energy-scale for the chameleon screening mechanism. The experimentally excluded regions are based on those reported in Ref~\cite{burrage2018tests}. }
\label{fig:exclusion:plot:comparison}
\end{figure*}

\subsection{Relation to existing experimental bounds}
To see how the theoretical bounds in figure~\ref{fig:exclusion:plot} relate to known experimental bounds on Newtonian gravity, we plot the convex hull of the shaded areas in figures~\ref{fig:bound:alpha:lambda} and~\ref{fig:bound:M:Lambda} against the bounds presented in Refs.~\cite{Murata_2015,PhysRevLett.124.051301,burrage2018tests}.By comparing with experimental results, we are able to demonstrate where optomechanical systems could help further constrain known bounds according to the results in this work. We emphasise however that this comparison is highly hypothetical, since experimental challenges such as noise, long-term stability, and integration over many runs of the experiment have not been included in our analysis. Much more work is required before it is known exactly how the optomechanical probe compares with other platforms (see section~\ref{sec:discussion}). 

The bounds can be found in figure~\ref{fig:exclusion:plot:comparison}, where figure~\ref{fig:convex:hull:alpha:lambda} shows the bounds in terms of $\alpha$ and $\lambda$, and where figure~\ref{fig:convex:hull:M:Lambda} shows the bounds in terms of $M$ and $\Lambda$. The yellow regions show the convex hull of the bounds derived in this work, and the purple region shows the combined parameter spaces that have been experimentally excluded. The orange area in figure~\ref{fig:convex:hull:M:Lambda} shows the excluded region for when the optomechanical probe is approximated as a point-particle, i.e.~the chameleon screening due to the finite size of the probe is neglected. 

Our results indicate that, for the values used in this work, even the ideal realisation of a nonlinear optomechanical sensor achieves similar bounds on $\alpha$ and $\lambda$ to those already reported in the literature. The decoherence, dissipation and thermalisation effects not accounted for in this description are likely to further reduce the sensitivity. This suggests that the sensitivity of the system must be improved further, should we wish to probe the hitherto unexplored regions in figure~\ref{fig:convex:hull:alpha:lambda}. From inspecting  equations~\eqref{eq:constant:sensitivity:kappa},~\eqref{eq:constant:sensitivity:sigma},~\eqref{eq:modulated:sensitivity:kappa}, and~\eqref{eq:modulated:sensitivity:sigma}, we note that the strongest dependence is with the mechanical frequency $\omega_{\rm{m}}$. Thus the lower $\omega_{\rm{m}}$, the better the sensitivity. Another strategy would be to increase the strength of the light--matter coupling $k_0$, however this is a long-standing challenge for many experimental platforms. More effective perhaps would be to decrease the separation distance $x_0$ between the probe and source systems, which would allow the optomechanical sensor to explore a larger range of $\lambda$, in particular smaller $\lambda$, since the Yukawa potential will not be as suppressed there. However, as the sensor is moved closer to the source sphere, the Casimir effect is expected to strongly contribute to the resulting acceleration (see below). On the other hand, our results according to  figure~\ref{fig:convex:hull:M:Lambda} indicate that optomechanical systems could be used to probe some hitherto unexplored regions of the chameleon parameters $M$ and $\Lambda$. The advances here likely depend on the quality of the background vacuum. 

\section{Discussion} \label{sec:discussion}
In this section, we discuss the challenges that must be overcome when considering an experiment of this nature. They include systematics and noise that affect the experiment, as well as forces that arise from the Casimir effect.

\subsection{Examining the conditions for linearising the force}

In order to definitely rule out modifications to the Newtonian potential, we must experimentally determine if the observed data deviates from that predicted by Newtonian gravity. Doing so requires extensive knowledge of the full dynamics of the system, including higher-order contributions from the Newtonian potential that we have  neglected in our main analysis. 
With this in mind, we examine the derivation of the linearised gravitational potential (see the expansion in equation~\eqref{eq:expanded:potential}) to determine when this linearisation breaks down. We assumed that the perturbation  $\delta x$ to the position of the optomechanical element is small compared with $x_S(t)$ (the distance from the probe to the source mass) at all times.  However, depending on the intended precision of the measurement of the force, Newtonian gravitational terms of second order in $\delta x$ may become relevant, that is, terms of the form $\propto\bigl( \hat b^\dag + \hat b\bigr)^2$. These terms can be included into the full dynamical analysis, which has been done in~\cite{bruschi2020time}. We leave performing the same analysis for modified gravity to future work.

Moreover, the radiation pressure found in an optomechanical setup has the explicit effect of displacing the mechanical element. When the light-matter coupling is modulated at mechanical resonance, the maximum position increases linearly as a function of time~\cite{qvarfort2020optimal}. Once this displacement grows too large, the approximation under which the optomechanical Hamiltonian in equation~\eqref{eq:basic:Hamiltonian} was derived is no longer valid (see e.g. Ref~\cite{law1995interaction} for details of how the optomechanical Hamiltonian is derived). A method for dealing with a displacement driven by radiation pressure would be attempting to cancel the expected radiation pressure by manually introducing a time-dependent linear potential $\sim (\hat b^\dag + \hat b)$ into the dynamics~\cite{qvarfort2020optimal}. In this way, the displacement from the light-radiation pressure is cancelled, while the phase from the gravitational interaction is still imparted on the optical state. The drawback of this method is that it most likely introduces additional noise into the experimental setup from the linear driving term. We do however leave the full quantum metrology analysis to future work.

\subsection{Limitations due to the Casimir effect}

Due to the relative weakness of gravity compared with electromagnetic force, the latter are likely to dominate any experimental setting. Therefore, any stray electromagnetic effects must be controlled very precisely in order to detect deviations from Newtonian gravity. One of the most important effects that has to be taken into account is the Casimir force~\cite{casimir1948influence},  which becomes significant when the distance between the probe and the source mass is small. To estimate the effect of the Casimir force, we use an analytic formula given in~\cite{Bimonte2018beyond} (based on the results of~\cite{Bimonte2012exact}) for the force due to the Casimir effect between two homogeneous perfectly conducting spheres at a distance much larger than their radii. The model of two perfectly conducting spheres is unlikely to accurately describe the experimental realisation of optomechanical setup described in this article, both in terms of geometry and material. Therefore, we will use this case to give only a first estimate of the effect and discuss how to suppress it.

 We consider the Drude boundary condition model for isolated conductors (see~\cite{Bimonte2018beyond} for details). For the distance between the probe and the source $x_0-R_S-R_P$ being much larger than the thermal wavelength, i.e. $ x_0-R_S-R_P \gg \lambda_T = \hbar c/(2\pi k_B T)$ (where the thermal wavelength is about $1\mathrm{\mu m}$ at room temperature) the classical thermal contribution to the Casimir force dominates, which leads to the expressions
\begin{equation}\label{eq:Casimir:effect:Dr}
F_C  \approx 18k_{\mathrm{B}} T \frac{R_S^3 R_P^3}{(x_0-R_S-R_P)^7}, 
\end{equation}
where $R_P$ and $R_S$ are the radii of the probe and the source, respectively.
At room temperature and for the parameters given in table~\ref{tab:Values}, equation~\eqref{eq:Casimir:effect:Dr} leads to an acceleration of the order of $ 9\times 10^{-13}\,\mathrm{m\,s^{-2}}$, experienced by the probe mass, while the gravitational acceleration induced by the source mass is of the order of $6\times 10^{-11}\,\mathrm{m\,s^{-2}}$. Casimir forces are therefore of order  $10^{-2}$ smaller than the main gravitational component. The size of the fifth force corrections we consider here are largely controlled by the two parameters $\sigma$ and $\kappa$ as shown in equation~\eqref{eq:linearised:potential:chameleon}. At peak sensitivity, when $x_0\sim \lambda$, this means that the ratio of the Casimir force to the Newtonian gravitational force should be compared to $\alpha$. We see from figure~\ref{fig:bound:alpha:lambda} that this ratio can be as low as $10^{-3}$ at the edge of the detectable region; the corrections are even smaller at lower $\lambda$. Furthermore, since the Casimir force grows very strongly with the inverse distance of the source and probe mass, the Casimir force quickly overshadows the fifth force contributions by many orders of magnitude when the source-probe distance is decreased to achieve better sensitivities. This shows that the Casimir effect is a relevant systematic that has to be controlled, that is, either precisely quantified or reduced. One way to reduce the force is to lower the temperature of the setup.

Another option to suppress the Casimir effect is to place a material in-between the source mass and the sensor that acts as a shield to the Casimir effect~\cite{chiaverini2003new,munday_measured_2009}. The Casimir force of the shield will be stationary while the un-shielded gravitational acceleration will be time-dependent, and therefore, clearly distinguishable~\cite{schmole2016micromechanical}. This approach is, however, limited by the size of the shield.
For example, in levitated optomechanics, the screening scheme can be naturally realized by placing the source mass behind one of the cavity end mirrors such that the mirror serves as a shield. However, in the case of detecting modifications due to a chameleon field, the presence of the mirror might introduce additional screening effects that need to be accounted for. The Casimir effect may also be reduced by modulating or compensating for the Casimir force with radiation pressure~\cite{banishev_modulation_2012}, nano-structuring of the source and probe surfaces~\cite{intravaia2013strong}, or an optical modulation of the charge density~\cite{chen_control_2007}.

Further analysis of the impact of a shield, or other techniques for accounting for the impact of the Casimir force, will require detailed numerical modelling. For example, Pernot-Borr et al.~\cite{pernot2019general} considered the impact of cylindrical walls on the screening of a source, finding that it can depend strongly on the thickness of the wall used for screening. Since we here consider the fundamental limits of an optomechanical setup, we leave a numerical analysis of the impact of different approaches to future work.

\subsection{Improvements to the sensitivity}

There are a number of ways in which the sensitivity of the optomechanical system can be further improved. In this work, we considered spherical source masses and probes in order to analytically derive the screening from the probe, however, choosing a different shaped source may improve the bounds that could be achieved. For example, a source mass in the form of a slab much larger than the probe system would mitigate gradient contributions from the Newtonian part of the potential, since the gravitational force from an infinite plane is constant. Furthermore, it was shown in Ref~\cite{Burrage:2017shh} that symmetric source masses tend to be much more strongly screened (and thus have smaller detectable effects) then asymmetric sources. Therefore, we would expect to obtain more favourable precision bounds than those presented in this work by considering asymmetric sources. An interesting prospect also arises from the fact that the optomechanical probe itself can also be asymmetric, e.g. in the shape of a levitated rod~\cite{kuhn2017optically}, which offers an additional avenue compared with, for example, atomic systems. However, these non-spherical cases bring with them additional challenges. The approximation used in equation~\eqref{eq:static_field} assumes a spherical source (probe), and approximates the nonlinear solution of the chameleon field equation with an analytic expression derived by asymptotic matching. To accurately obtain measurements with a non-spherical setup would require precise numerical modelling of the chameleon field around a (non-spherical) source and probe, such as done in Ref~\cite{Burrage:2017shh}. The precise effect on the sensitivity is left to future work. 

As a final note, we mention that the nonlinear radiation-pressure term in the Hamiltonian in equation~\eqref{eq:basic:Hamiltonian} appears in many different contexts, of which not all fall under the category of optomechanics (such as for example electromechanical setups~\cite{tsang2010cavity}). Our results therefore apply to these systems as well. We therefore have a large range of systems to choose from when it comes to optimising the geometry and resulting sensitivity.

\subsection{Future work towards an experimental proposal}

The sensitivities calculated in this work give us an indication of the \textit{resolution} of the force that the optomechancial probe can achieve in principle. That is, we learn the magnitude of the gravitational force that can be detected. In practice, however, we must then determine whether this force is simply the Newtonian force, or whether it is due to the Newtonian force and an additional force that arises from the modification. With a good-enough resolution, such a modification can be detected even if the Newtonian force is much stronger than the modification. 
There are several methods by which the modification can be detected. The first is to very carefully model the influence of the Newtonian force on the optomechanical dynamics and data that is collected through e.g. a homodyne measurement. If a deviation in the collected data is then seen, steps should be taken to rule out any other source. Another way is to carefully change the equilibrium separation distance $x_0$  between the source sphere and the optomechanical probe. Since the modifications considered in this work changes quite drastically due to the inclusion of  the exponential term in equation~\eqref{eq:modified:gravitational:potential}, it should be possible to detect such a exponential change in the data. Both of these methods here can be theoretically explored in future work. 

 Our results can be used to evaluate the fundamental ability of a quantum optomechanical system to probe a particular parameter regime of modified gravity theories. A realistic optomechanical system, however, will be affected by a number of systematics and noise sources, including optical dissipation from photons leaking from the cavity, mechanical thermal noise, Brownian motion noise, damping effects, and noise from the trapping or clamping mechanism, as well as radiation back-action noise and shot-noise.  Yet additional noise sources include external gravitational noise and environmental vibrations (see e.g.~\cite{schmole2016micromechanical,schmole2017development} for a discussion of a related experimental setup).  Generally, such noise sources have spectral contributions at the resonant frequency of the sensor and are enhanced as well as the signal from the source mass that we wish to detect. Therefore, in practice, it may be favourable to consider an off-resonant sensing scheme, such as those discussed in Refs~\cite{schmole2016micromechanical,schmole2017development}.  We also note that such additional noise sources will be particularly dominant when the mechanical frequency is low, however we see from equation~\eqref{eq:constant:sensitivity:kappa} and~\eqref{eq:constant:sensitivity:sigma} that a low mechanical frequency is a necessary requirement if we wish to achieve a high sensitivity. We also note that it is not clear how the sensitivity gained from e.g. modulating the optomechanical coupling changes when the $Q$-factors of the cavity and the oscillators are considered. 

To model the noise and systematics mentioned in the previous paragraph, a plausible next step beyond this work involves linearising the optomechanical dynamics around a strong coherent input-state~\cite{aspelmeyer2014cavity}. With the help of phase-space methods~\cite{serafini2017quantum}, it is then possible to include most of the systematics and noise terms mentioned above into the dynamics. In addition, a homodyne measurement could be modelled using input-output theory for the optical mode. One can then examine the susceptibility of the mode and determine the noise levels required for these effects to be detectable~\cite{motazedifard2021ultraprecision}. An important question that must be addressed is the laser power required to maximise the sensitivity. 
Since the linearisation gives rise to equations of motion that differ from those used here, it is difficult to predict what the resulting bounds on modified gravity theories will look like compared with those presented here. Most likely, the presence of noise and absence of non-Gaussian resources (which arise from the nonlinear coupling) means that the prediction for the sensitivity is reduced. 

To instead extend the analysis in this work even further in the nonlinear optomechanical regime, we must include noise in the solution of the dynamics for the nonlinear Hamiltonian in equation~\eqref{eq:basic:Hamiltonian}. However, since the resulting nonlinear Langevin equations are generally much more difficult to solve (although certain solutions in the weak-coupling limit and for systems with weak optical decoherence exist~\cite{rabl2011photon, nunnenkamp2011single}), we expect this to be challenging. A preliminary step towards modeling Markovian optical decoherence affecting the intra-cavity state was recently taken~\cite{qvarfort2020master}, and mechanical thermal noise has been modelled using a range of methods~\cite{bassi2005towards, bernad2006quest}. For a strongly coupled system, however, optical and mechanical noise cannot be treated separately, and must instead be considered together~\cite{hu2015quantum,betzholz2020breakdown}. To our knowledge, fundamental quantum metrology bounds in the noisy nonlinear regime have not yet been considered. 

Another aspect that needs to be modelled is the additional screening that  arises from the inclusion of a shield to block out Casimir forces. In addition, for a levitated optomechanical sphere, a mirror must be placed between the optomechanical probe and the source, which also contributes to the screening (but which can, at the same time, act as the Casimir shield). To carry out a full analysis of the screening, the geometry of the vacuum chamber, along with the trapping mechanism of the optomechanical system and the Casimir shield, must be carefully modelled. It is then possible to exactly predict the magnitude of the modified force that the optomechanical probe can detect.

\section{Conclusions} \label{sec:conclusions}

In this work, we derived the best-possible bounds for detecting modified gravity with a quantum optomechanical sensor. We modelled the effects of a force from an oscillating source mass on the optomechanical probe and estimated the sensitivity of the system by computing the quantum Fisher information. In particular, we considered the additional screening that arises due to the relatively large size of the optomechanical probe. 
Our results show that optomechanical sensors could, in principle, be used to improve on existing experimental bounds for the chameleon screening mechanism, although more work is needed to evaluate the prospects for using experimental optomechanical systems as probes for modified gravity.

\section*{Data availability statement}
The code used to compute the screening and sensitivity to chameleon fields can be found in the following online GitHub repository: \href{https://github.com/sqvarfort/modified-gravity-optomech}{https://github.com/sqvarfort/modified-gravity-optomech}.

\section*{Acknowledgments}
We thank Markus Rademacher, Niall Moroney, David Edward Bruschi, Doug Plato,  Alessio Serafini, Daniel Braun, Michael R. Vanner,  Peter F. Barker, Witlef Wieczorek, Clare Burrage, and Hendrik Ulbricht for helpful comments and discussions. S.Q. was supported in part by an Engineering and Physical Sciences Research Council (EPSRC) Doctoral Prize Fellowship,  the Wallenberg Initiative on Networks and Quantum Information (WINQ), and the Marie Skłodowska-Curie Action IF programme “Nonlinear optomechanics for verification, utility, and sensing” (NOVUS) -- Grant- Number 101027183. D.R. would like to thank the Humboldt Foundation for supporting his work with their Feodor Lynen Research Fellowship and acknowledges funding by the Marie Skłodowska-Curie Action IF programme -- Project-Name “Phononic Quantum Sensors for Gravity” (PhoQuS-G) -- Grant-Number 832250. The work of S.S. was supported by the G\"oran Gustafsson Foundation for Research in Natural Sciences and Medicine, by the Royal Society, and partially supported by the UCL Cosmoparticle Initiative and the European Research Council (ERC) under the European Community’s Seventh Framework Programme (FP7/2007-2013)/ERC grant agreement number 306478-CosmicDawn.

\appendix

\section{The chameleon mechanism}
 \label{sec:cham:mech}

In this appendix we briefly review the derivation of the chameleon mechanism and how it gives rise to a fifth-force; the reader is directed to Refs~\cite{Khoury:2003aq,PhysRevD.69.044026,Brax:2004qh} for further details. Throughout this appendix we will use energy units ($\hbar = c = 1$) for notational simplicity. The basic idea of the chameleon screening mechanism is to screen the effects of additional degrees of freedom in a modified gravity model (typically light scalar fields), by making their mass dependent on the local density. This results in a scalar field whose mass is large inside the solar system where the average density is high and is thus difficult to create in collider experiments, but has a lighter mass in the intergalactic medium where the density of matter is lower. Typically, this is achieved using a scalar field whose action is of the form:
\begin{align}
S =&   \, S_m( \psi_{(m)} , \Omega^{-2} (\phi) g_{\mu \nu})\nonumber \\
&+\int \mathrm{d}^4 x \sqrt{-g} \left[\frac{1}{16\pi G} R - \frac{1}{2} \nabla_\mu \phi\nabla^\mu \phi - V(\phi) \right], \label{eq:chamAction}
\end{align}
where $g_{\mu \nu}$ is the spacetime curvature, $R$ is the Ricci tensor,
$V(\phi)$ is the chameleon potential, and $S_{(m)}$ is the matter action. 
Various choices of screening mechanism are possible with this action, but the Chameleon mechanism corresponds the following choice of the functions $V(\phi)$ (the interaction potential) and $\Omega(\phi)$ (which represents direct, non-minimal coupling between the Chameleon field and gravity):
\begin{align}
    V(\phi) &= \frac{\Lambda^{4+n}}{\phi^n}\label{eq:ChamPot},\\
    \Omega(\phi) &= 1 - \frac{\phi}{M}.\label{eq:ChamOmega}
\end{align}
In this work, we will consider the case $n=1$ for simplicity. One way to understand the effect of the $\Omega(\phi)$ term is to regard matter as coupling to the so called Jordan frame metric, $\tilde{g}_{\mu\nu} = \Omega^{-2}(\phi)g_{\mu\nu}$, while the Chameleon field sees a different metric, $g_{\mu\nu}$, which suggests quanta of the scalar field, if we were somehow able to isolate them, would be observed to fall differently to normal matter (violating the equivalence principle). One can either regard $\tilde{g}_{\mu\nu}$ as the ``real'' metric, in which case $\phi$ has an unusual direct coupling to gravity, or $g_{\mu\nu}$, in which case all particles have a special coupling to $\phi$ via the function $\Omega(\phi)$ appearing wherever the metric does in the matter action, $S_m$. Ultimately what matters, however, is how objects will be observed to move in the presence of this scalar field. Formally, we can obtain this from the geodesic equation for the 4-vector position $X^{\mu}=(t,\mathbf{X})$, since normal matter sees the metric $\tilde{g}_{\mu\nu}$:
\begin{equation}
    \frac{\dd^2X^{\rho}}{\dd\lambda^2} + \tilde{\Gamma}^{\rho}_{\mu\nu}\frac{\dd X^{\mu}}{\dd\lambda}\frac{\dd X^{\nu}}{\dd\lambda} = 0.\label{eq:geodesicJordan}
\end{equation}

If we regard $g_{\mu\nu}$ as the true metric, then the effect of the chameleon field is to add what appears to be a fifth force, since when we take the Newtonian limit we can re-write $\tilde{\Gamma}^{\rho}_{\mu\nu}$ in equation(\ref{eq:geodesicJordan}) in terms of the Newtonian potential, $\Phi_N$ and $\Omega$ to obtain:
\begin{equation}
\frac{\dd^2X^k}{\dd\lambda^2} = -\partial^{k}\Phi + \partial^{k}\log\Omega(\phi(X)) = -\partial^{k}(\Phi_N+ \Phi_{C})\label{eq:newtonCorrection},
\end{equation}
where $\Phi_C = -\log\Omega(\phi(X))$ is an effective fifth-force potential. The strength of this fifth force is characterised by $\Omega$, but to compute its effects we need to know how the scalar field couples to matter. Varying equation~\eqref{eq:chamAction} with respect to $\phi$ we obtain:
\begin{equation}
    \nabla_{\mu}\nabla^{\mu}\phi - V'(\phi) - \frac{\dd\log\Omega(\phi)}{\dd\phi}g^{\mu\nu}T_{\mu\nu} = 0,\label{eq:chamKG}
\end{equation}
where $T_{\mu\nu}$ is the Hilbert Stress energy tensor. For non-relativistic matter, the scalar field is sourced by the local matter density, with $g^{\mu\nu}T_{\mu\nu} = -\rho$, giving
\begin{equation}
    \nabla_{\mu}\nabla^{\mu}\phi - V'(\phi) + \frac{\dd \log \Omega}{\dd \phi}\rho = 0.\label{eq:chamKGNonRel}
\end{equation}
This is equivalent to the scalar field interacting via the potential:
\begin{equation}
    V_{\mathrm{eff}}(\phi) = V(\phi) - \log\Omega(\phi)\rho.\label{eq:Veff}
\end{equation}
If we use the form equation~\eqref{eq:ChamOmega}, then for cases where $\phi \ll M$, we can approximate $\log\Omega(\phi)$ as:
\begin{equation}
\log\Omega(\phi) \approx -\frac{\phi}{M}.
\end{equation}
Under this approximation, the effective scalar field potential becomes
\begin{equation}
V_{\mathrm{eff}}(\phi) = V(\phi) + \frac{\phi\rho}{M}.\label{eq:VeffDef}
\end{equation}
and the Chameleon fifth force potential is:
\begin{equation}
\Phi_{\mathrm{cham}}(X) = -\log\Omega(\phi(X))\approx \frac{\phi(X)}{M}.
\end{equation}
For the purposes of this work, we will consider the $n=1$ chameleon field. In a region with constant mass density $\rho_{\mathrm{bg}}$, this means that the Chameleon rests at the vacuum value:
\begin{equation}
    \phi_{\mathrm{bg}} =\sqrt{\frac{M\Lambda^5}{\rho_{\mathrm{bg}}}},\label{eq:phbg}
\end{equation}
for which fluctuations of the field have mass:
\begin{equation}
     m^2_{\mathrm{bg}}(\rho_{\mathrm{bg}}) = V''_{\mathrm{eff}}(\phi_{\mathrm{bg}}) = 2 \, \sqrt{\frac{\rho_{\mathrm{bg}}^3}{M^3\Lambda^5}}.\label{app:eq:mbg}
\end{equation}
As expected, the mass of field fluctuations increases with the background density, which means in areas of comparative high density such as inside the solar system\footnote{Compared to the average density inside a cosmological void.}, the mass is large and the scalar field difficult to excite and detect.

\subsection{Chameleon Field From a Spherical Source}
\label{app:chameleon_field}

The chameleon field in the vicinity of a spherical source of mass $M_S$ and radius $R_S$ can be computed by solving the Klein-Gordon equation,
\begin{equation}
\frac{\dd^2\phi}{\dd r^2} + \frac{2}{r}\frac{\dd\phi}{\dd r} - V'_{\mathrm{eff}}(\phi) = 0.\label{eq:kg}
\end{equation}

This is\ a non-linear equation, but an approximate solution was found by Burrage \textit{et al}.~\cite{Burrage:2014oza} in the limit $m_{\mathrm{bg}}R_S \ll 0$. This is valid for the atoms considered there, but in our case we may need to consider larger sources. We therefore repeat the derivation of Burrage \textit{et al}.~\cite{Burrage:2014oza} for the case of arbitrary $m_{\mathrm{bg}}R_S$. The fundamental strategy uses the method of asymptotic matching~\cite{bender2013advanced} to derive an approximate solution over for the full domain of the differential equation, by smoothly matching together solutions valid in different domains.

In the strongly perturbing case ($\rho_SR_S^2 > 3M\phi_{\mathrm{bg}}/(\hbar c)$), the solution reaches its equilibrium value, $\phi_S$, for density $\rho_S$ at some radius $S \leq R_S$. We denote the region with $r < S$ the interior region, or region I. The solution there can thus be approximated as constant
\begin{equation}
\phi_I(r) = \phi_S.
\end{equation}
There is then a transition layer (region II) between $S < r < R_S$ where the solution rapidly shifts towards the background value, $\phi_{\mathrm{bg}}$. Since the density ratio between the source-sphere and the external vacuum is high, we will find $\phi_{\mathrm{bg}} \gg \phi_S$ as a result of equation~(\ref{eq:phbg}). In the transition layer ($S < r < R_S$), the field will begin to increase, eventually reaching a regime where $\phi \gg \phi_S$. Because we can re-write equation~(\ref{eq:VeffDef}) as

\begin{equation}
V_{\mathrm{eff}}(\phi) = \frac{\rho\phi}{M}\left[\frac{\phi_S^2}{\phi^2} + 1\right],\label{eq:veffPhiS}
\end{equation}
then once $\phi \gg \phi_S$ the density-dependent term dominates the potential. Under such conditions, $V'_{\mathrm{eff}}(\phi) \approx \rho/M$ and we can solve equation~(\ref{eq:kg}) analytically:
\begin{equation}
\phi_{II}(r) = \frac{M_S}{8\pi MR_S}\frac{r^2}{R_S^2} + \frac{C}{r} + D.\label{eq:II}
\end{equation}
Finally, far away from the source sphere, $\phi$ is close to its background value, $\phi_{\mathrm{bg}}$, and we can approximate the potential as quadratic: $V_{\mathrm{eff}}(\phi) \approx m_{\mathrm{bg}}^2(\phi - \phi_{\mathrm{bg}})^2/2$. The solution there takes the form
\begin{equation}
\phi_{III}(r) = \phi_{\mathrm{bg}} + \frac{E}{r}e^{-m_{\mathrm{bg}}r} + \frac{F}{r}e^{+m_{\mathrm{bg}}r}.
\end{equation}
Here, region III is defined as $r > R_{S}$. Note that although $\phi \gg \phi_S$ outside the sphere, equation~(\ref{eq:II}) does not apply because the density outside the sphere is now $\rho_{\mathrm{bg}}$ which is typically much smaller than $\rho_S$: the density dependent term in the potential is thus no longer dominant outside the source. We note, however, that solutions $\phi_{II}$ and $\phi_{III}$ are technically only valid in the vicinity of $r\sim R_S$ and $r \gg R_S$ respectively. However, we can approximate the behaviour of the fully-non-linear solution for all $r$ by matching these asymptotic solutions at $S$ and $R_S$, which imposes four constraints to ensure smoothness of the asymptotically matched solution: $\phi_{I}(S) = \phi_{II}(S), \phi_{I}'(S) = \phi_{II}'(S), \phi_{II}(R_S) = \phi_{III}(R_S), \phi_{III}'(R_S) = \phi_{III}'(R_S)$. We also note that we require $F = 0$ to have a solution approaching $\phi_{\mathrm{bg}}$ as $r\rightarrow\infty$, which means that there are four unknowns, $C, D, E, $ and the radius $S$. We solve for these four unknowns, finding
\begin{align}
C =& \, \frac{M_S}{4\pi M}\frac{S^3}{R_S^3} \label{eq:C}, \\
D =&\,  \phi_S - \frac{3M_S S^2}{8\pi M R_S^3} \label{eq:D}, \\
E =&  - \frac{M_S}{4\pi M(1 + m_{\mathrm{bg}}R_S)}\left(1 - \frac{S^3}{R_S^3}\right)e^{m_{\mathrm{bg}R_S}}\label{eq:E},
\end{align}
where $S$ satisfies 
\begin{equation}
\frac{S_S^2}{R_S^2} + \frac{2}{3}\left[\frac{1}{1+m_{\mathrm{bg}}R_S} - 1\right]\frac{S_S^3}{R_S^3} = 1- \frac{8\pi M}{3M_S}R_S(\phi_{\mathrm{bg}} - \phi_S) + \frac{2}{3}\left[\frac{1}{1+m_{\mathrm{bg}}R_S} - 1\right].\label{eq:Scubic}
\end{equation}
Taken together, Eqs~(\ref{eq:C})--(\ref{eq:E}) and~(\ref{eq:Scubic}) imply equation~(\ref{eq:static_field}), and reduce to the Burrage \textit{et al}.~\cite{Burrage:2014oza} result in the case $m_{\mathrm{bg}}R_S \rightarrow 0$ limit. As a cubic equation for $S$, it is of limited use to express $S$ in closed form, and for the purposes of this work we solve equation~(\ref{eq:Scubic}) numerically. This can run into catastrophic cancellation problems, due to the finite floating-point precision, when $S\approx R_S$ (that is, the heavily screened regime), since we need to compute $1 - S^3/R_S^3$. Hence, when the numerical solution gives $S/R_S$ close to 1, we switch over to an analytic approximation obtained by substituting $S/R_S = 1 + \epsilon$ and solving for $\epsilon$ to first order. This gives
\begin{equation}
\epsilon = -\frac{4\pi M}{3M_S}R_S(\phi_{\mathrm{bg}} - \phi_S)(1 + m_{\mathrm{bg}}R_S),
\end{equation}
which implies
\begin{equation}
1 - \frac{S^3}{R_S^3} \approx \frac{4\pi M}{M_S}R_S(\phi_{\mathrm{bg}} - \phi_S)(1 + m_{\mathrm{bg}}R_S).\label{eq:SCubicTaylor}
\end{equation}
We see that equation~(\ref{eq:SCubicTaylor}) agrees with the Taylor expansion of equation~(\ref{eq:S}) in the $m_{\mathrm{bg}}R_S \rightarrow 0$ limit.

\section{Time dependence of a chameleon field \label{sec:app:moving_source}}

The main result of this work is that an oscillating optomechnanical system can be used to detect the presence of chameleon fields. However, throughout we have made the assumption that such a field responds essentially instantaneously to the motion of the mass that sources it. For purely gravitational effects this assumption is justified since information about the position of the source mass propagates outwards at the speed of light. However, the situation is less clear for a scalar field sourced by a moving mass, since the field is massive: there is no a priori guarantee that information can propagate outwards at the speed of light. To address this, we derive the propagator for a chameleon field sourced by a point mass and demonstrate that the resulting retarded chameleon potential can be treated as if information propagates instantaneously, for non-relativistic oscillating masses.

\subsection{Time dependence of the gravitational potential}
First, we will derive the response of the gravitational field to a small mass, neglecting back-reaction and gravitational wave emission as negligible. In the linearised limit which allows us to make contact with Newtonian gravity, the metric perturbation $h_{\mu\nu}$ around $\eta_{\mu\nu}$ satisfies:
\begin{align}
    \partial_{\sigma}\partial_{\nu}{h^{\sigma}}_{\mu}+\partial_{\sigma}\partial_{\mu}{h^{\sigma}}_{\nu} - \partial_{\mu}\partial_\nu h - \partial_{\lambda}\partial^{\lambda}h_{\mu\nu}\nonumber\\- \eta_{\mu\nu}\partial_{\rho}\partial_{\lambda}h^{\rho\lambda} + \eta_{\mu\nu}\partial_{\lambda}\partial^{\lambda}h = 16\pi GT_{\mu\nu},
\end{align}
where we are using the $-+++$ metric convention. We choose the Lorenz gauge, defined by the condition $\partial_{\mu}h^{\mu\nu} = \frac{1}{2}\partial^{\nu}h$, which simplifies this to:
\begin{equation}
    \partial_{\lambda}\partial^{\lambda}\bar{h}_{\mu\nu} = -16\pi GT_{\mu\nu},\label{eq:trace_rev_einstein}
\end{equation}
where $\bar{h}_{\mu\nu} = h_{\mu\nu} - \frac{1}{2}h\eta_{\mu\nu}$ is the \emph{trace-reversed} perturbation. A generic expression for the Hilbert stress energy tensor is:
\begin{equation}
    T_{\mu\nu} = -\frac{2}{\sqrt{-\det g}}\frac{\delta S_m}{\delta g^{\mu\nu}},
\end{equation}
where $S_m$ is the matter action. For a point particle of mass $M_S$, the appropriate matter action is:
\begin{equation}
    S_m = M_S\int\dd\tau\sqrt{-g_{\mu\nu}\dot{q}^\mu\dot{q}^{\nu}},
\end{equation}
where $q^{\mu}(\tau)$ describes the particle trajectory and dots denote differentiation with respect to proper time, $\tau$ (the variation with respect to $q^{\mu}$ yields the geodesic equation, verifying that this is indeed the action we seek). This means that the stress-energy tensor at position $\mathbf{X}$ for a point particle following trajectory $\mathbf{q}(\tau)$ is:
\begin{equation}
    T_{\mu\nu} = \frac{M_S\dot{q}_{\mu}\dot{q}_{\nu}\delta^{(3)}(\mathbf{X} - \mathbf{q}(\tau))}{\sqrt{-\det g}\sqrt{-g_{\alpha\beta}\dot{q}^{\alpha}\dot{q}^{\beta}}}.
\end{equation}
For low (non-relativistic) velocities in a Minkowski background this means $T_{00} = M_S\delta^{(3)}(\mathbf{X} - \mathbf{q}(\tau)) = \rho$, as we would expect, with all other components zero. Note that we are ignoring any back-reaction from this moving particle, which would give only higher order corrections. Taking the trace of equation~\eqref{eq:trace_rev_einstein} yields:
\begin{equation}
    \partial_{\lambda}\partial^{\lambda}h = -16\pi G\rho, \label{eq:wave}
\end{equation}
which is the same equation satisfied by $\bar{h}_{00}$, and so we have $h = \bar{h}_{00}$ (note the sign -- $\eta^{\mu\nu}\bar{h}_{\mu\nu} = -h$, but $\eta^{\mu\nu}T_{\mu\nu} = - \rho$). Meanwhile, $\partial_{\lambda}\partial^{\lambda}\bar{h}_{ij} = 0$ as there are no spatial parts of $T_{\mu\nu}$. This describes propagating gravitational waves, which we neglect, and can be safely set to zero. Thus, we need only solve equation~\eqref{eq:wave}, which has a known solution in terms of a retarded potential:
\begin{equation}
    h = 4G\int\dd^3\mathbf{X'}\frac{\rho(t-\frac{|\mathbf{X} -\mathbf{X'}|}{c},\mathbf{X'})}{|\mathbf{X} - \mathbf{X'}|}.
\end{equation}
For the point source with density $\rho(\mathbf{X},t) = M_S\delta^{(3)}(\mathbf{X} - \mathbf{q(t)})$, this means:
\begin{equation}
    h(\mathbf{X},t) = \frac{4GM_S}{|\mathbf{X} - \mathbf{q}(t_{\mathrm{ret}})|}.
\end{equation}
where $t_{\mathrm{ret}}$ is the retarded time that solves:
\begin{equation}
    t_{\mathrm{ret}} = t-\frac{|\mathbf{X} - \mathbf{q}(t_{\mathrm{ret}})|}{c}.\label{eq:tret}
\end{equation}
Note, comparing with the standard perturbative parameterisation of the metric:
\begin{align}
\dd s^2 =& -\dd t^2(1 + 2\Phi) + w_i(\dd x^i\dd t + \dd t \dd x^i)\nonumber\\
& + [(1-2\Psi)\delta_{ij} + 2s_{ij}]\dd x^i\dd x^j,
\end{align}
we see that $\Phi = -h/4$, or in other words:
\begin{equation}
    \Phi(\mathbf{X},t) = -\frac{GM_S}{|\mathbf{X} - \mathbf{q}(t_{\mathrm{ret}})|},
\end{equation}
which is the retarded gravitational potential for a point source of mass $M$, as we would expect. Note that this is well defined (in the Lorenz gauge), as it is not derived from energy considerations but from the equation of motion of the metric perturbation.
\subsection{Time dependence of the chameleon field}
\label{app:scalar_field_evolution}
The evolution of the scalar field is significantly more complicated, due to the fact that it satisfies a highly non-linear equation of motion, equation~\eqref{eq:chamKG}. However, provided the source mass is not large the we can consider small deviations from the background value $\phi_{\mathrm{bg}}$ and linearise the equation:
\begin{equation}
    \partial_{\lambda}\partial^{\lambda}\Delta\phi-\frac{m_{\mathrm{bg}}^2c^2}{\hbar^2}\Delta\phi = \frac{\Delta\rho(\mathbf{X},t)}{M},\label{eq:cham_linear}
\end{equation}
where $\Delta\phi = \phi - \phi_{\mathrm{bg}}$ and $\Delta\rho$ is the deviation from $\rho_{\mathrm{bg}}$ that sources the field deviation. This is the Klein--Gordon equation with mass $m_{\mathrm{bg}}$, but with a source on the RHS. To solve this, we need to make use of the retarded propagator of the Klein--Gordon equation~\cite{scharf2014finite}:
\begin{align}
    G_{\mathrm{ret}}(X,Y) =& \frac{\Theta(X^0 - Y^0)}{2\pi}\delta(\tau^2(x,Y))\label{eq:retProp}\\
    &- \Theta(X^0 - Y^0)\Theta(\tau^2(X,Y))\frac{m_{\mathrm{bg}}cJ_1(m_{\mathrm{bg}}c\tau(X,Y)/\hbar)}{4\pi\hbar\tau(X,Y)}\nonumber ,\\
    \tau(X,Y) =& \sqrt{c^2(X^0-Y^0)^2 - (\textbf{X-Y})^2}.\label{eq:tauxy}
\end{align}
Here, $J_1$ is a Bessel function of the first kind. Note that we choose the retarded, rather than advanced or Feynman propagator here in order to ensure that the field responds causally to the movements of the source. The general solution of equation~\eqref{eq:cham_linear} is
\begin{equation}
    \Delta\phi(X) = \int\dd^4X'\frac{\Delta\rho(x')}{M}G_{\mathrm{ret}}(X,X').
\end{equation}
For our point particle, the density deviation is
\begin{equation}
    \Delta\rho(\mathbf{X},t) = M_S\delta^{(3)}(\mathbf{X}-\mathbf{q}(t)).
\end{equation}
The corresponding solution is thus:
\begin{align}
    \phi(\mathbf{X},t)& = \phi_{\mathrm{bg}}+ \frac{M_Sc}{M}\int\dd t'\left[\frac{\theta(t-t')}{2\pi}\delta(c^2[t-t']^2 - |\mathbf{X} - \mathbf{q}(t')|^2) \right.\nonumber \\
    &-\theta(t-t')\theta(c^2[t-t']^2 - [\mathbf{X} - \mathbf{q}(t')]^2)\left.\frac{m_{\mathrm{bg}}cJ_1(m_{\mathrm{bg}}c\sqrt{c^2(t-t')^2 - |\mathbf{X} - \mathbf{q}(t')|^2}/\hbar)}{4\pi\hbar\sqrt{c^2(t-t')^2 - |\mathbf{X} - \mathbf{q}(t')|^2}}\right].\label{eq:retardedIntegral}
\end{align}
We use the formula:
\begin{equation}
    \delta(f(X))=\sum_i\frac{\delta(X - X_i)}{|f'(X_i)|},
\end{equation}
where $X_i$ are solutions of $f(X_i) = 0$. In the $\mathbf{q}$ constant case, it is easy to see that there are two solutions, $t' = t \pm |\mathbf{q} - \mathbf{X}|$, but only the negative solution matters, due to the $\theta(t - t')$ term (this is the causal effect of the retarded propagator, and ensures that we only integrate over contributions from the past of the time $t$ we are looking at). In the case where $\mathbf{q}$ is time dependent, solving $(t-t')^2 - |\mathbf{X} - \mathbf{q}(t')|^2 = 0$ is less trivial, but still results in a unique retarded time, $t_{\mathrm{ret}}$, exactly the same quantity that is well known from electrodynamics. To see that it is unique, consider that $t_{\mathrm{ret}}$ is by definition the time at which light arriving at an observer at time $t$ was emitted by the source. Assume there are two such times, $t_1,t_2$. We have (temporarily putting back the factors of $c$ for clarity) $c(t - t_i) = |\mathbf{q}(t_i) - \mathbf{X}|$, and can subtract these two equations from each other to obtain $c(t_2 - t_1) = |\mathbf{q}(t_1) - \mathbf{X}| - |\mathbf{q}(t_2) - \mathbf{X}|$. This implies that the distance of $\mathbf{q}$ from $\mathbf{X}$ has changed at the speed of light - not possible unless the source itself is moving at the speed of light. Thus, for sub-luminal sources, $t_{\mathrm{ret}}$ is unique\footnote{Actually, there is still a solution for $t_{\mathrm{adv}} > t$, but this is eliminated due to the causal $\theta(t - t')$ function.}. Hence:
\begin{align}
    \delta(c^2[t-t']^2 - |\mathbf{X} - \mathbf{q}(t')|^2) &=\frac{\delta(t' - t_{\mathrm{ret}})}{2(c^2(t_{\mathrm{ret}} - t) + (\mathbf{X} - \mathbf{q}(t_{\mathrm{ret}}))\cdot \mathbf{v}(t_{\mathrm{ret}}))} \nonumber \\
&= -\frac{\delta(t' - t_{\mathrm{ret}})}{2c|\mathbf{X} - \mathbf{q}(t_{\mathrm{ret}})|(1 - \frac{(\mathbf{X} - \mathbf{q}(t_{\mathrm{ret}}))}{|\mathbf{X} - \mathbf{q}(t_{\mathrm{ret}})|}\cdot \mathbf{v}(t_{\mathrm{ret}})/c)},
\end{align}
where we have used equation~\eqref{eq:tret} and $\mathbf{v}(t)$ is the velocity of the source. Thus, the first part of the integral reduces to:
\begin{equation}
    I_1 = -\frac{M_S}{4\pi M|\mathbf{X} - \mathbf{q}(t_{\mathrm{ret}})|(1 - \frac{(\mathbf{X} - \mathbf{q}(t_{\mathrm{ret}}))\cdot \mathbf{v}(t_{\mathrm{ret}})}{|\mathbf{X} - \mathbf{q}(t_{\mathrm{ret}})|c})}.\label{eq:I1}
\end{equation}
This is indeed exactly the same as the expression found in electrodynamics, where the propagating field (the photon) is massless, and thus the second term in equation~\eqref{eq:retardedIntegral} is not present. In our case, however, we have to deal with the massive part of the integral too:
\begin{align}
    I_2 =& -\frac{cM_S}{4\pi M}\int\dd t'\theta(t - t')\theta(c^2[t - t']^2 - |\mathbf{X} - \mathbf{q}(t')|^2)\bigg{(}\nonumber\left.\frac{m_{\mathrm{bg}}cJ_1(m_{\mathrm{bg}}c\sqrt{c^2(t - t')^2 - |\mathbf{X} - \mathbf{q}(t')|^2}/\hbar)}{\hbar\sqrt{c^2(t - t')^2 - |\mathbf{X} - \mathbf{q}(t')|^2}}\right).\label{eq:massiveIntegral}
\end{align}
In this case, the effect of the two Heaviside step functions is to force us to integrate over the past, up to the retarded time:
\begin{equation}
    I_2 = -\frac{cM_S}{4\pi M}\int_{-\infty}^{t_{\mathrm{ret}}}\dd t'\frac{m_{\mathrm{bg}}cJ_1(m_{\mathrm{bg}}c\sqrt{c^2(t - t')^2 - |\mathbf{X} - \mathbf{q}(t')|^2}/\hbar)}{\hbar\sqrt{c^2(t - t')^2 - |\mathbf{X} - \mathbf{q}(t')|^2}}.
\end{equation}
Now, make the substitution:
\begin{align}
    u =& \frac{m_{\mathrm{bg}}c}{\hbar}\sqrt{c^2(t - t')^2 - |\mathbf{X} - \mathbf{q}(t')|^2},\\
    t'(u) =& t - \frac{\hbar}{c^2m_{\mathrm{bg}}}\sqrt{u^2 + \frac{m_{\mathrm{bg}}^2c^2}{\hbar^2}|\mathbf{X} - \mathbf{q}(t'(u))|^2}, \\
    \dd u =& \frac{m_{\mathrm{bg}}c[c^2(t' - t) + (\mathbf{X} - \mathbf{q}(t'))\cdot \mathbf{v}(t')]}{\hbar\sqrt{c^2(t - t')^2 - |\mathbf{X} - \mathbf{q}(t')|^2}}\dd t',
\end{align}
to obtain:
\begin{equation}
    I_2 = +\frac{m_{\mathrm{bg}}cM_S}{4\pi M\hbar}\int_{0}^{\infty}\dd u \frac{J_1(u)}{\sqrt{u^2 + m_{\mathrm{bg}}^2c^2|\mathbf{X} - \mathbf{q}(t'(u))|^2/\hbar^2} - [\mathbf{X} - \mathbf{q}(t'(u))]\cdot \mathbf{v}(t'(u))m_{\mathrm{bg}}/\hbar}.
\end{equation}
This would not seem to offer a significant simplification, unless we adopt the low-velocity approximation, $|\mathbf{v_0}|/c \ll 1$. The second term in the denominator is $O(v/c)$, so we can in general neglect it (note that $|\mathbf{X} - \mathbf{q}(t')|/c$ is not in general small as $\mathbf{X}$ can be arbitrarily far from the source). We can also expand $|\mathbf{X} - \mathbf{q}(t'(u))|^2$ as a power series in $u$ around the retarded time, $t_{\mathrm{ret}} = t'(0)$:
\begin{align}
    |\mathbf{X} - \mathbf{q}(t'(u))|^2 &= |\mathbf{X} - \mathbf{q}(t_{\mathrm{ret}})|^2 + \left.\frac{\dd|\mathbf{X} - \mathbf{q}(t')|^2}{\dd t'}\right|_{t = t_{\mathrm{ret}}}\frac{\dd t'}{\dd u}u+ \frac{\dd^2t'}{\dd u^2}\left.\frac{\dd |\mathbf{X} - \mathbf{q}(t')|^2}{\dd t'}\right|_{t' = t_{\mathrm{ret}}}\frac{u^2}{2} \nonumber \\
    &\quad + \left(\frac{\dd t'}{\dd u}\right)^2 \left.\frac{\dd^2|\mathbf{X} - \mathbf{q}(t')|^2}{\dd t'^2}\right|_{t' = t_{\mathrm{ret}}}\frac{u^2}{2} + O(u^3).
\end{align}
We find:
\begin{align}
    \frac{\dd|\mathbf{X} - \mathbf{q}(t')|^2}{\dd t'} &= -2(\mathbf{X} - \mathbf{q}(t'))\cdot \mathbf{v}(t'), \frac{\dd^2|\mathbf{X} - \mathbf{q}(t')|^2}{\dd t'^2} = 2|\mathbf{v}(t')|^2-2(\mathbf{X} - \mathbf{q}(t'))\cdot \frac{\dd \mathbf{v}'}{\dd t'}, \\
    \frac{\dd t'}{\dd u} &= \frac{\hbar\sqrt{c^2(t - t')^2 - |\mathbf{X} - \mathbf{q}(t')|^2}}{m_{\mathrm{bg}}c[c^2(t' - t) + (\mathbf{X} - \mathbf{q}(t'))\cdot \mathbf{v}(t')]}, \\
    \frac{\dd^2t'}{\dd u^2} &= \frac{\hbar}{m_{\mathrm{bg}}c\sqrt{c^2(t - t')^2 - |\mathbf{X} - \mathbf{q}(t')|^2}}\left(\frac{\dd t'}{\dd u}\right)\nonumber\\ 
 &\quad - \frac{\hbar\sqrt{c^2(t - t')^2 - |\mathbf{X} - \mathbf{q}(t')|^2}}{m_{\mathrm{bg}}c\left[c^2(t' - t)+(\mathbf{X} - \mathbf{q}(t'))\cdot\mathbf{v}(t')\right]^2}\times \left[c^2 - |\mathbf{v}(t')|^2 + (\mathbf{X} - \mathbf{q}(t'))\cdot \frac{\dd \mathbf{v}(t')}{\dd t'}\right]\frac{\dd t'}{\dd u}.
\end{align}
And evaluated at $u = 0$ (or $t' = t_{\mathrm{ret}}$), this gives:
\begin{align}
    \left.\frac{\dd|\mathbf{X} - \mathbf{q}(t')|^2}{\dd t'}\right|_{t' = t_{\mathrm{ret}}} &= -2(\mathbf{X} - \mathbf{q}(t_{\mathrm{ret}}))\cdot \mathbf{v}(t_{\mathrm{ret}}),\\
    \left.\frac{\dd^2|\mathbf{X} - \mathbf{q}(t')|^2}{\dd t'^2}\right|_{t' = t_{\mathrm{ret}}} &= +2|\mathbf{v}(t_{\mathrm{ret}})|^2-2(\mathbf{X} - \mathbf{q}(t_{\mathrm{ret}}))\cdot \left.\frac{\dd \mathbf{v}'}{\dd t'}\right|_{t' = t_{\mathrm{ret}}},\\
    \left.\frac{\dd t'}{\dd u}\right|_{u = 0} &= 0, \\ \left.\frac{\dd^2 t'}{\dd u^2}\right|_{u = 0} &= \frac{\hbar^2}{m_{\mathrm{bg}}^2c^2\left[c^2(t_{\mathrm{ret}} - t) + (\mathbf{X} - \mathbf{q}(t_{\mathrm{ret}})\cdot\mathbf{v_0}(t_{\mathrm{ret}}))\right]}.
\end{align}
Hence:
\begin{align}
    &|\mathbf{X} - \mathbf{q}(t'(u))|^2 = |\mathbf{X} - \mathbf{q}(t_{\mathrm{ret}})|^2 - \frac{\hbar^2(\mathbf{X} - \mathbf{q}(t_{\mathrm{ret}}))\cdot\mathbf{v}(t_{\mathrm{ret}})}{m_{\mathrm{bg}}^2c^2\left[c^2(t_{\mathrm{ret}} - t) + (\mathbf{X} - \mathbf{q}(t_{\mathrm{ret}})\cdot\mathbf{v}(t_{\mathrm{ret}}))\right]}u^2\nonumber + O(u^3). 
\end{align}
We substitute in equation~\eqref{eq:tret} to obtain 
\begin{align}
&|\mathbf{X} - \mathbf{q}(t'(u))|^2 = |\mathbf{X} - \mathbf{q}(t_{\mathrm{ret}})|^2 + \frac{\hbar^2\frac{(\mathbf{X} - \mathbf{q}(t_{\mathrm{ret}}))}{|\mathbf{X} - \mathbf{q}(t_{\mathrm{ret}})|}\cdot\mathbf{v}(t_{\mathrm{ret}})/c}{m_{\mathrm{bg}}^2c^2\left[1 - \frac{(\mathbf{X} - \mathbf{q}(t_{\mathrm{ret}}))}{|\mathbf{X} - \mathbf{q}(t_{\mathrm{ret}})|}\cdot\mathbf{v}(t_{\mathrm{ret}})/c\right]}u^2\nonumber + O(u^3). 
\end{align}
 The $u^2$ term is proportional to $v/c$ once SI units are restored and so we can ignore it in the non-relativistic limit. Generally speaking, higher order terms in the expansion about $u = 0$ will also have terms proportional to $v/c \ll 1$, so we neglect them\footnote{One can argue that this must be true for the solution to reduce to the static Yukawa potential in the $v\rightarrow c$ limit.}. This reduces the integral to:
\begin{align}
    I_2 &= +\frac{M_Sm_{\mathrm{bg}}c}{4\pi M\hbar}\int_{0}^{\infty}\dd u\frac{J_1(u)}{\sqrt{u^2 + m_{\mathrm{bg}}^2c^2|\mathbf{X} - \mathbf{q}(t_{\mathrm{ret}})|^2/\hbar^2}}\nonumber\\
    &= +\frac{M_S}{4\pi M|\mathbf{X} - \mathbf{q}(t_{\mathrm{ret}})|}\left(1 - e^{-m_{\mathrm{bg}}c|\mathbf{X} - \mathbf{q}(t_{\mathrm{ret}})|/\hbar}\right).
\end{align}
Combining this with equation~\eqref{eq:I1}, again in the $v \ll c$ limit, gives:
\begin{equation}
    \Delta\phi(\mathbf{X,t}) =  - \frac{M_S}{4\pi M|\mathbf{X} - \mathbf{q}(t_{\mathrm{ret}}(\mathbf{X},t))|}e^{-m_{\mathrm{bg}}|\mathbf{X} - \mathbf{q}(t_{\mathrm{ret}}(\mathbf{X},t))|}.
\end{equation}
This is the expected Yukawa potential, only with the retarded time for the position of the source. It is worth noting that the fact that the force carrier (in this case the scalar bosonic chameleon field excitations) is massive does not appear to affect the retarded time, which describes information propagating through the field at the speed of light, even though the bosons themselves do not. The fact that the retarded time appears, both here and in electrodynamics, is not because the force carriers themselves (photons in the case of electrodynamics) travel at the speed of light - it is due to the causal structure of space-time itself.

\subsection{Screening by the optmomechanical Probe} \label{app:sec:screening:calculations}

Burrage \textit{et al}.~\cite{Burrage:2014oza} derived an expression for the force between two extended spheres due to a chameleon field. However, their approach considered the forces between individual atoms, for which the range of the force could largely be ignored (corresponding to the $m_{\mathrm{bg}}r \ll 1$ limit). With the larger sensor devices we consider in this work this is not necessarily applicable. Furthermore, since in this case we envision an oscillating source, we need to be certain that time-dependent effects do not come into play. For this reason, we re-derive the force between two spheres without the static spheres and $m_{\mathrm{bg}}r \ll 1$ assumption. The derivation closely follows that of Burrage \textit{et al}.~\cite{Burrage:2014oza}, with the assumption of a spherical probe to provide a simplification (more accurate modelling of the optomechanical probe will in principle be necessary for performing the actual experiment).

As in Burrage \textit{et al}.~\cite{Burrage:2014oza}, we denote the source sphere $A$ and the test sphere ($B$) for which we are computing the force (as a model for the optomechanical probe). The force is determined by the rate of change of the momentum resulting from the energy momentum flux across the surface of ball $B$:
\begin{equation}
F_i = \dot{P}_i  = -\int_{\partial B}\tau^j_in_j\dd S,
\end{equation}
where $\tau^j_i$ are the spatial components of the energy momentum tensor (including gravity), $\partial B$ is the surface of ball $B$, $n^j$ the surface normal vector and $\dd S$ the surface area element. The gravitational and matter contributions to the energy momentum tensor follow as in Burrage \textit{et al}.~\cite{Burrage:2014oza}, with the main change being to the chameleon field term:
\begin{equation}
T^{(\phi)j}_i = -\nabla_{i}\phi\nabla^{j}\phi+ \delta_i^j\left(\frac{1}{2}\nabla_{\mu}\phi\nabla^{\mu}\phi + V(\phi)\right).
\end{equation}
The chameleon field at position $\mathbf{X}$ and time $t$ with $\mathbf{X}$ centred on ball $B$ is given by
\begin{equation}
\phi(\mathbf{X},t) = \phi_{\mathrm{bg}} + \phi_A(\mathbf{X},t) + \phi_B(\mathbf{X}) = \phi_{\mathrm{bg}} - \frac{\xi_AM_A}{4\pi M(1 + m_{\mathrm{bg}R_A})}\frac{e^{-m_{\mathrm{bg}}(|\mathbf{X} - \mathbf{X}_A(t)|-R_A)}}{|\mathbf{X} - \mathbf{X}_A(t)|} - \frac{\xi_BM_B}{4\pi M(1 + m_{\mathrm{bg}R_B})}\frac{e^{-m_{\mathrm{bg}}(r-R_B)}}{r},
\end{equation}
where $r = |\mathbf{X}|$ and where the prefactors $\xi_A$ and $\xi_B$ are given by 
\begin{equation}
\xi_i = 
\begin{cases}
1, & \rho_i R_i^2 < 3 M \, \phi_{\mathrm{bg}}, \\
1 - \frac{S_i^3}{R_i^3}, &\rho_i R^2_i > 3 M \phi_{\mathrm{bg}}.
\end{cases}
\end{equation}
Note that that we have assumed the linear regime, where the fields from the two chameleon sources add. The gradient of the contribution $\phi_B(\mathbf{x})$ from ball $B$ is given by
\begin{equation}
\nabla_i\phi_B = \frac{\xi_BM_B}{4\pi M}\frac{1}{r^2}\frac{(1+m_{\mathrm{bg}}r)}{(1 + m_{\mathrm{bg}}R_B)}e^{-m_{\mathrm{bg}}(r - R_B)}\nabla_ir = \frac{\xi_BM_B}{4\pi M}\frac{x_i}{r^3}\frac{(1+m_{\mathrm{bg}}r)}{(1 + m_{\mathrm{bg}R_B})}e^{-m_{\mathrm{bg}}(r - R_B)}.
\end{equation}
The chameleon stress-energy tensor is therefore given by
\begin{align}
T_i^{(\phi)j} =& -\left(\partial_i\phi_A+ \frac{\xi_BM_B}{4\pi M}\frac{xX_i}{r^3}\frac{(1+m_{\mathrm{bg}}r)}{(1 + m_{\mathrm{bg}R_B})}e^{-m_{\mathrm{bg}}(r - R_B)}\right)\left(\partial^j\phi_A+ \frac{\xi_BM_B}{4\pi M}\frac{X^j}{r^3}\frac{(1+m_{\mathrm{bg}}r)}{(1 + m_{\mathrm{bg}R_B})}e^{-m_{\mathrm{bg}}(r - R_B)}\right) \nonumber \\& + \delta_i^j\biggl[\frac{1}{2}\left(\partial_k\phi_A+ \frac{\xi_BM_B}{4\pi M}\frac{X_k}{r^3}\frac{(1+m_{\mathrm{bg}}r)}{(1 + m_{\mathrm{bg}R_B})}e^{-m_{\mathrm{bg}}(r - R_B)}\right)\left(\partial^k\phi_A+ \frac{\xi_BM_B}{4\pi M}\frac{X^k}{r^3}\frac{(1+m_{\mathrm{bg}}r)}{(1 + m_{\mathrm{bg}R_B})}e^{-m_{\mathrm{bg}}(r - R_B)}\right) \nonumber \\
&\qquad\qquad- \frac{1}{2}(\partial_t(\phi_A + \phi_B))^2\biggr],
\end{align}
whereas Burrage \textit{et al}.~\cite{Burrage:2014oza}, we ignore the contribution from the potential, $V(\phi)$. Note that Burrage \textit{et al}.~\cite{Burrage:2014oza} assume the $m_{\mathrm{bg}}r \ll 1$ limit, which we generalise here to arbitrarily large $r$. To do this, we will make only the assumption that $|\mathbf{X}|/|\mathbf{X}_A| \ll 1$ on the surface of ball $B$, that is, that the radius of ball $B$ is much smaller than the distance to ball $A$. The means we can expand
\begin{equation}
|\mathbf{X} - \mathbf{X}_A| \approx |\mathbf{X}_A|\left(1 - \frac{\mathbf{X}\cdot \mathbf{X}_A}{|\mathbf{X}_A|^2} + O(|\mathbf{X}|^2/|\mathbf{X}_A|^2)\right).
\end{equation}
We can then expand the derivative of the field from sphere $A$ as
\begin{align}
\partial_i\phi_A = & \frac{\xi_AM_A}{4\pi M|\mathbf{X}_A|^2}\frac{(1 + |\mathbf{X}_A|m_{\mathrm{bg}})}{(1 + m_{\mathrm{bg}}R_A)}\exp\left(-m_{\mathrm{bg}}(|\mathbf{X}_A| - R_A) + \frac{m_{\mathrm{bg}}\mathbf{X}\cdot\mathbf{X}_A}{|\mathbf{X}_A|}+O(|\mathbf{X}|^2/|\mathbf{X}_A|^2)\right) \times \nonumber \\
&\left[-\frac{X_{Ai}}{|\mathbf{X}_A|} + X^j\left(\frac{\delta_{ij}}{|\mathbf{X}_A|} - \frac{3X_{Ai}X_{Aj}}{|\mathbf{X}_A|^3} + \frac{m_{\mathrm{bg}}X_{Ai}X_{Aj}}{(1 + |\mathbf{X}_A|m_{\mathrm{bg}})|\mathbf{X}_A|^2}\right) + O(|\mathbf{X}|^2/|\mathbf{X}_A|^2)\right],
\end{align}
where we have deliberately not expanded the exponential in $|\mathbf{X}|/|\mathbf{X}_A|$. This is because we cannot generally assume that $m_{\mathrm{bg}}|\mathbf{X}| << 1$ (though we will examine this limit later). 

We are seeking to perform an integral of the form
\begin{equation}
-\int_{\partial B}\dd S T_i^{(\phi)j}n_j.
\end{equation}
First, use the fact that $\partial_t\phi_B =0$ (we are in the frame of reference of $B$, so all time dependence is in the motion of ball $A$). The time derivative of the field is
\begin{equation}
\partial_t\phi_A = \frac{\xi_A M_A}{4\pi M}\frac{(\mathbf{X} - \mathbf{X}_A)\cdot(-\mathbf{v}_{A})}{|\mathbf{X} - \mathbf{X}_A|^3}\frac{(1 + m_{\mathrm{bg}}|\mathbf{X} - \mathbf{X}_A|)}{(1 + m_{\mathrm{bg}}R_A)}e^{-m_{\mathrm{bg}}(|\mathbf{X} - \mathbf{X}_A| - R_A)} = -v_A^i\partial_i\phi_A,
\end{equation}
which can be written in terms of the spatial derivative because the time dependence only appears through $\mathbf{X}_A$, which always appears together with $\mathbf{X}$ in the form $|\mathbf{X} - \mathbf{X}_A|$. This allows us to simplify the expression for the stress-energy tensor:
\begin{equation}
T_i^{(\phi)j} \approx \left(-\delta_i^k\delta^{jl} + \frac{\delta_i^j\delta^{kl}}{2}\right)(\partial_k\phi_A+ \partial_k\phi_B)(\partial_l\phi_A+ \partial_l\phi_B) - \frac{\delta_i^j}{2}v_A^kv_A^l\partial_k\phi_A\partial_l\phi_A.
\end{equation}
We therefore find that the integral splits into three parts:
\begin{align}
-\int_{\partial B}\dd S T_i^{(\phi)j}n_j =& I_{AA} + I_{AB} + I_{BB},\\
I_{AA} =& \left(\delta_i^{(k}\delta^{l)j} - \frac{\delta_i^j}{2}[\delta^{kl} - v_A^kv_A^l]\right)\frac{1}{R_B}\int_{\partial B}\dd S X_j\partial_k\phi_A\partial_l\phi_A,\\
I_{AB} =& \left(\delta_i^{(k}\delta^{l)j} - \frac{\delta_i^j}{2}\delta^{kl}\right)\frac{2}{R_B}\int_{\partial B}\dd S X_j\partial_k\phi_A\partial_l\phi_B,\\
I_{BB} =& \left(\delta_i^{(k}\delta^{l)j} - \frac{\delta_i^j}{2}\delta^{kl}\right)\frac{1}{R_B}\int_{\partial B}\dd S X_j\partial_k\phi_B\partial_l\phi_B.
\end{align}
We can write
\begin{equation}
\partial_i\phi_B|_{|\mathbf{X}| = R_B} = \frac{\xi_BM_B}{4\pi M}\frac{X_i}{R_B^3} \equiv Q_BX_i,
\end{equation}
where we define
\begin{equation}
Q_B = \frac{\xi_BM_B}{4\pi M}\frac{1}{R_B^3}.
\end{equation}
Similarly:
\begin{align}
\partial_i\phi_A =& Q_A\left(-X_{Ai} + X^jA_{ij}\right)\frac{(1 + m_{\mathrm{bg}}|\mathbf{X}_A|)}{(1 + m_{\mathrm{bg}}R_A)}e^{-m_{\mathrm{bg}}(|\mathbf{X}_A| - R_A)}\exp\left( \frac{m_{\mathrm{bg}}\mathbf{X}\cdot\mathbf{X}_A}{|\mathbf{X}_A|}\right),\\
Q_A =& \frac{\xi_AM_A}{4\pi M}\frac{1}{|\mathbf{X}_A|^3},\\
A_{ij} =& \left[\delta_{ij} - \frac{3X_{Ai}X_{Aj}}{|\mathrm{X}_A|^2} + \frac{m_{\mathrm{bg}}X_{Ai}X_{Aj}}{(1 + m_{\mathrm{bg}}|\mathbf{X}_A|)|\mathbf{X}_A|}\right].
\end{align}
This means we can write the three integrals $I_{AA},I_{AB}, I_{BB}$ as
\begin{align}
I_{AA} =& \left(\delta_i^{(k}\delta^{l)j} - \frac{\delta_i^j}{2}[\delta^{kl} - v_A^kv_A^l]\right)\frac{Q_A^2}{R_B}\bigl[X_{Ak}X_{Al}B_j(2m_{\mathrm{bg}})  - (X_{Ak}A_{lm} + X_{Al}A_{km})B_{jm}(2m_{\mathrm{bg}}) \nonumber \\
&\qquad\qquad\qquad\qquad\qquad\qquad+ A_{km}A_{ln}B_{jmn}(2m_{\mathrm{bg}})\bigr]\frac{(1 + m_{\mathrm{bg}}|\mathbf{X}_A|)^2}{(1 + m_{\mathrm{bg}}R_A)^2}e^{-2m_{\mathrm{bg}}(|\mathbf{X}_A| - R_A)},\\
I_{AB} =&  \left(\delta_i^{(k}\delta^{l)j} - \frac{\delta_i^j}{2}\delta^{kl}\right)\frac{2Q_AQ_B}{R_B}(-X_{Ak}B_{il}(m_{\mathrm{bg}}) + A_{km}B_{jml}(m_{\mathrm{bg}}))\frac{(1 + m_{\mathrm{bg}}|\mathbf{X}_A|)}{(1 + m_{\mathrm{bg}}R_A)}e^{-m_{\mathrm{bg}}(|\mathbf{X}_A| - R_A)},\\
I_{BB} =& \left(\delta_i^{(k}\delta^{l)j} - \frac{\delta_i^j}{2}\delta^{kl}\right)\frac{Q_B^2}{R_B}\int\dd S X_jX_kX_l,
\end{align}
where we have expressed everything in terms of the following integrals:
\begin{align}
B_i(m) =& \int\dd S X_ie^{m \mathbf{X}\cdot\mathbf{X}_A/|\mathbf{X}_A|},\label{eq:Bi}\\
B_{ij}(m) =& \int\dd S X_iX_je^{m \mathbf{X}\cdot\mathbf{X}_A/|\mathbf{X}_A|},\label{eq:Bij}\\
B_{ijk}(m) =& \int\dd S X_iX_jX_ke^{m \mathbf{X}\cdot\mathbf{X}_A/|\mathbf{X}_A|}.\label{eq:Bijk}
\end{align}

In fact, one can immediately see that $I_{BB} = 0$, because the integral of a cubic polynomial vanishes over a sphere:
\begin{equation}
\int_{\partial B}X_iX_jX_k\dd S = 0.
\end{equation}
This follows from symmetry since the integrand is odd when $\mathbf{X} \rightarrow -\mathbf{X}$ and we integrate over the whole sphere (one can also verify this directly by evaluating it in spherical polar co-ordinates).

To evaluate the remaining integrals in equations (\ref{eq:Bi})-(\ref{eq:Bijk}), we switch to polar co-ordinates. Without loss of generality, we can arrange our co-ordinates at any time to be such that $\mathbf{X}_A$ is along the $z$ axis, which simplifies the calculation. For $B_i(m)$ we obtain
\begin{equation}
B_i(m) = \int X_i e^{mz}\dd S.
\end{equation}
We define our polar co-ordinates as
\begin{align}
X =& r\sin\theta\cos\psi,\nonumber \\
Y =& r\sin\theta\sin\psi,\nonumber \\
Z =& r\cos\theta.
\end{align}
We can see immediately that $B_1 = B_2 = 0$ since
\begin{equation}
B_1\propto \int_{0}^{2\pi}\dd\psi \cos\psi = 0, B_2\propto \int_{0}^{2\pi}\dd\psi\sin\psi = 0.
\end{equation}
We therefore only need to evaluate $B_3$:
\begin{equation}
B_3(m) = 2\pi R_B^3\int_{0}^{\pi}\dd\theta \sin\theta\cos\theta e^{mR_B\cos\theta} = \frac{4\pi R_B}{m^2}(mR_B\cosh(mR_B) - \sinh(mR_B)).
\end{equation}
We can summarise this as
\begin{equation}
B_i(m) = \delta_{i3}\frac{4\pi R_B}{m^2}(mR_B\cosh(mR) - \sinh(mR)),
\end{equation}
which generalises, for an arbitrary axis choice, to
\begin{equation}
B_i(m) = \frac{X_{Ai}}{|\mathbf{X}_A|}\frac{4\pi R_B}{m^2}(mR_B\cosh(mR_B) - \sinh(mR_B)).
\end{equation}
Next, we evaluate $B_{ij}(m)$. We can make similar arguments here to conclude that $B_{13} = B_{23} = 0$ (for the same reason that $B_1 = B_2 = 0$). Furthermore:
\begin{equation}
B_{12}\propto \int_{0}^{2\pi}\sin\psi\cos\psi = 0.
\end{equation}
Hence, all the cross-terms vanish, so this tensor is diagonal. The diagonals are:
\begin{align}
B_{11} =& R_B^4\int_{0}^{2\pi}\dd\psi\cos^2\psi\int_{0}^{\pi}\sin^3\theta e^{mR_B\cos\theta} = \frac{\pi R_B}{m^3}(4mR\cosh(mR_B) - 4\sinh(mR_B)), \label{eq:B11}\\
B_{22} =& R_B^4\int_{0}^{2\pi}\dd\psi\sin^2\psi\int_{0}^{\pi}\sin^3\theta e^{mR_B\cos\theta} = \frac{\pi R_B}{m^3}(4mR_B\cosh(mR_B) - 4\sinh(mR_B)), \label{eq:B22}\\
B_{33} =& 2\pi R_B^4\int_{0}^{\pi}\dd\theta \cos^2\theta\sin\theta e^{mR_B\cos\theta} = \frac{\pi R}{m^3}(-8mR_B\cosh(mR_B)+ 4(2 + m^2R_B^2)\sinh(mR_B)).\label{eq:B33}
\end{align}
This can be summarised as
\begin{align}
B_{ij}(m) &= \delta_{ij}\frac{\pi R_B}{m^3}(4mR_B\cosh(mR_B) - 4\sinh(mR_B)) \nonumber \\
&\qquad + \delta_{i3}\delta_{j3} \frac{\pi R_B}{m^3}(-12mR_B\cosh(mR_B)+ 4(3 + m^2R_B^2)\sinh(mR_B)),
\end{align}
or, for a generic axis choice:
\begin{align}
B_{ij}(m) &= \delta_{ij}\frac{\pi R_B}{m^3}(4mR_B\cosh(mR_B) - 4\sinh(mR_B)) \nonumber \\
&\qquad+ \frac{X_{Ai}X_{Aj}}{|\mathbf{X}_A|^2} \frac{\pi R_B}{m^3}(-12mR_B\cosh(mR_B)+ 4(3 + m^2R_B^2)\sinh(mR_B)). 
\end{align}
Finally, we can compute $B_{ijk}(m)$. First, note that this is a rank-3 symmetric tensor. Rank $m$ symmetric tensors in $d$ dimensions have $(m + d -1)!/(d-1)!m!$ independent components, which for $d=3$ gives $(m+1)(m+2)/2$. This means we have to compute 10 independent components in total. However, as before most are actually 0. Again, similar arguments to before imply $B_{331} = B_{221} =  B_{321} = 0$. We also find that:
\begin{align}
B_{111}\propto & \int_{0}^{2\pi}\dd\psi\cos^3\psi = 0,\nonumber \\
B_{222}\propto & \int_{0}^{2\pi}\dd\psi\sin^3\psi = 0,\nonumber \\
B_{112} \propto & \int_{0}^{2\pi}\dd\psi \cos^2\psi\sin\psi = 0,\nonumber \\
B_{221} \propto & \int_{0}^{2\pi}\dd\psi \sin^2\psi\cos\psi =0.
\end{align}
There are, in fact, only three non-zero components:
\begin{align}
B_{333} =& 2\pi R_B^5\int_{0}^{\pi}\dd\theta \cos^3\theta\sin\theta e^{mR_B\cos\theta} = \frac{\pi R_B}{m^4}(4mR_B(6 + m^2R_B^2)\cosh(mR_B) - 12(2 + m^2R_B^2)\sinh(mR_B)).\label{eq:B333}\\
B_{311} =& R_B^5\int_{0}^{2\pi}\dd\psi\cos^2\psi \int_{0}^{\pi}\dd\theta \cos\theta\sin^3\theta e^{mR_B\cos\theta} = \frac{\pi R_B}{m^4}(-12mR_B\cosh(mR_B)+ 4(3 + m^2R_B^2)\sinh(mR_B)).\label{eq:B311}\\
B_{322} =& R_B^5\int_{0}^{2\pi}\dd\psi\sin^2\psi \int_{0}^{\pi}\dd\theta \cos\theta\sin^3\theta e^{mR_B\cos\theta} = \frac{\pi R_B}{m^4}(-12mR_B\cosh(mR_B)+ 4(3 + m^2R_B^2)\sinh(mR_B)).\label{eq:B322}
\end{align}
To summarise this, we note that only terms with at least one of $i,j,k$ equal to 3 are non-zero. The other two indices must then be a diagonal matrix. This means that we can write:
\begin{align}
B_{ijk} =& \delta_{i3}\left[\delta_{jk}\frac{\pi R_B}{m^4}(-12mR_B\cosh(mR_B) + 4(3 + m^2R_B^2)\sinh(mR_B)) \nonumber \right.\\
&\left. + \delta_{j3}\delta_{k3}\frac{\pi R_B}{m^4}(4mR_B(9 + m^2R_B^2)\cosh(mR_B) - 4(9 + 4m^2R_B^2)\sinh(mR_B))\right] + (\mbox{all permutations}).
\end{align}
Or, using symmetrised index notation:
\begin{align}
B_{ijk} =& \frac{X_{A(i}}{|\mathbf{X}_A|}\left[\delta_{jk)}\frac{\pi R_B}{m^4}(-12mR_B\cosh(mR_B) + 4(3 + m^2R_B^2)\sinh(mR_B)) \nonumber \right.\\
&\left. + \frac{X_{Ai}X_{Aj)}}{|\mathbf{X}|_A^2}\frac{\pi R_B}{m^4}(4mR_B(9 + m^2R_B^2)\cosh(mR_B) - 4(9 + 4m^2R_B^2)\sinh(mR_B))\right],
\end{align}
where for an arbitrary tensor, circular brackets around indices indicate the symmetrised indices:
\begin{equation}
M_{(i_1\ldots i_n)} \equiv \frac{1}{n!}\sum_{\mathrm{permutations}}M_{i_{p1}\ldots i_{pn}},
\end{equation}
where $p_1\ldots p_n$ run over all permutations of $1\ldots n$. For example:
\begin{align}
M_{(ij)} \equiv & \frac{1}{2}(M_{ij} + M_{ji}), \\
M_{(ijk)} \equiv & \frac{1}{6}(M_{ijk} + M_{ikj} + M_{jki} + M_{jik} + M_{kij} + M_{kji}).
\end{align}
We can now proceed to substitute these into the expressions for $I_{AA}$ and $I_{AB}$. First, we consider $I_{AA}$, since this term is argued to be zero by Burrage \textit{et al}.~\cite{Burrage:2014oza}	in the $m_{\mathrm{bg}}r \ll 0$ limit. First, we can ignore the velocity dependent terms, since we work in the non-relativistic limit where $|\mathbf{v}_A| \ll 0$ (we are using units where $c = 1$). To gain some further understanding of the behaviour, let us consider the $R_B\rightarrow 0$ limit:
\begin{align}
B_i(m)\approx & \frac{4\pi X_{Ai}}{3|\mathbf{X}_A|m^3}(mR_B)^4 + O((mR_B)^6), \\
B_{ij}(m) \approx & \delta_{ij} \frac{4\pi}{3m^4}(mR_B)^4 + O((mR_B)^6),\\
B_{ijk}(m) \approx & \frac{X_{A(i}}{|\mathbf{X}_A|}\left[\delta_{jk)}\frac{4\pi}{15m^5}(mR_B)^6 + \frac{X_{Aj}X_{Ak)}}{|\mathbf{X}_A|^2}\frac{8\pi}{15m^5}(mR_B)^6 + O((mR_B)^8)\right].
\end{align}
This gives
\begin{align}
I_{AA} \approx & \left(\delta_i^{(k}\delta^{l)j} - \frac{\delta_i^j}{2}\delta^{kl}\right) \frac{Q_A^2}{R_B}\left[\frac{\pi X_{Ak}X_{Al}X_{Aj}}{6|\mathbf{X}_A|m_{\mathrm{bg}}^3}(2m_{\mathrm{bg}}R_B)^4 - \frac{\pi X_{Ak}A_{lj}}{6m_{\mathrm{bg}}^4}(2m_{\mathrm{bg}}R_{B})^4 + O((2m_{\mathrm{bg}}R_B)^6)\right]\nonumber \\
\times &\frac{(1 + m_{\mathrm{bg}}|\mathbf{X}_A|)^2}{(1 + m_{\mathrm{bg}}R_A)^2}e^{-2m_{\mathrm{bg}}(|\mathbf{X}_A| - R_A)}, \nonumber \\
 =& \left(\delta_i^{(k}\delta^{l)j} - \frac{\delta_i^j}{2}\delta^{kl}\right) \frac{Q_A\xi_AM_A(1 + m_{\mathrm{bg}}|\mathbf{X}_A|)}{4\pi M}e^{-m_{\mathrm{bg}}|\mathbf{X}_A|}\left[\frac{8\pi m_{\mathrm{bg}} X_{Ak}X_{Al}X_{Aj}}{3|\mathbf{X}_A|}\left(\frac{R_B}{|\mathbf{X}_A|}\right)^3 - \frac{8\pi X_{Ak}A_{lj}}{6}\left(\frac{R_{B}}{|\mathbf{X}_A|}\right)^3\right]\nonumber\\
 &\times \frac{(1 + m_{\mathrm{bg}}|\mathbf{X}_A|)^2}{(1 + m_{\mathrm{bg}}R_A)^2}e^{-2m_{\mathrm{bg}}(|\mathbf{X}_A| - R_A)}, \\
I_{AB} \approx & \left(\delta_i^{(k}\delta^{l)j} - \frac{\delta_i^j}{2}\delta^{kl}\right) \frac{2Q_AQ_B}{R_B}\left[-\frac{4\pi X_{Ak}}{3m_{\mathrm{bg}}^4}\delta_{il}(m_{\mathrm{bg}}R_B)^4 + O((m_{\mathrm{bg}}R_B)^6)\right]\frac{(1 + m_{\mathrm{bg}}|\mathbf{X}_A|)}{(1 + m_{\mathrm{bg}}R_A)}e^{-m_{\mathrm{bg}}(|\mathbf{X}_A| - R_A)}\nonumber   \\
= & \left(\delta_i^{(k}\delta^{l)j} - \frac{\delta_i^j}{2}\delta^{kl}\right) Q_A \frac{\xi_BM_B}{4\pi M}(1 + m_{\mathrm{bg}}R_B) e^{-m_{\mathrm{bg}}R_B}\left[-\frac{8\pi X_{Ak}}{3}\delta_{il}\right]\frac{(1 + m_{\mathrm{bg}}|\mathbf{X}_A|)}{(1 + m_{\mathrm{bg}}R_A)}e^{-m_{\mathrm{bg}}(|\mathbf{x}_A| - R_A)}.
\end{align}
This suggests that $I_{AA}$ is subdominant in this limit, provided:
\begin{equation}
\frac{Q_A}{Q_B}\frac{(1 + m_{\mathrm{bg}}|\mathbf{X}_A|)}{(1 + m_{\mathrm{bg}}R_A)}e^{-m_{\mathrm{bg}}(|\mathbf{X}_A| - R_A)} \ll \frac{R_B}{|\mathbf{X}_A|}. \label{eq:ratio_cond}
\end{equation}
Then this means that we can essentially ignore the $I_{AA}$ term in the $R_B\rightarrow 0$ limit, which agrees with the calculation of Burrage \textit{et al}.~\cite{Burrage:2014oza}. Furthermore, provided we stay in the limit in equation~\eqref{eq:ratio_cond}, then $I_{AA}$ will always be suppressed relative to $I_{AB}$. Generally speaking, this may not be the case, however, in which case we would have to include the $I_{AA}$ term. For now, let us compute the force without this term, as just include the $I_{AB}$ contribution:
\begin{equation}
F_i = \left(\delta_i^{(k}\delta^{l)j} - \frac{\delta_i^j}{2}\delta^{kl}\right)\frac{2Q_AQ_B}{R_B}(-X_{Ak}B_{jl}(m_{\mathrm{bg}}) + A_{km}B_{jml}(m_{\mathrm{bg}}))\frac{(1 + m_{\mathrm{bg}}|\mathbf{X}_A|)}{(1 + m_{\mathrm{bg}}R_A)}e^{-m_{\mathrm{bg}}(|\mathbf{X}_A| - R_A)}.
\end{equation}
Let us evaluate each piece in turn:
\begin{align}
\left(\delta_i^{(k}\delta^{l)j} - \frac{\delta_i^j}{2}\delta^{kl}\right)X_{Ak}B_{jl}(m_{\mathrm{bg}}) = & \frac{1}{2}(\delta_i^{k}\delta^{jl} + \delta_i^j\delta^{kl} - \delta_{i}^j\delta^{kl})X_{Ak}B_{jl} \nonumber \\
=& \frac{X_{Ai}}{2}B_{jj} \nonumber \\
= & X_{Ai}\frac{2\pi R_B^3}{m_{\mathrm{bg}}}\sinh(m_{\mathrm{bg}}R_b),
\end{align}
where in the last line we simply add the non-zero diagonals, equations (\ref{eq:B11}) - (\ref{eq:B33}). The next piece is slightly more complicated:
\begin{equation}
\left(\delta_i^{(k}\delta^{l)j} - \frac{\delta_i^j}{2}\delta^{kl}\right)A_{km}B_{jml}(m_{\mathrm{bg}}) = \frac{1}{2}A_{im}B_{mjj}.
\end{equation}
Consider $m = 1,2,3$ in turn, and use co-ordinates where $\mathbf{X}_A$ is along the $z$ axis for simplicity. We find that $B_{1jj} = B_{111} + B_{122} + B_{133} = 0, B_{2jj} = B_{211} + B_{222} + B_{233} = 0$ and the only non-zero component is $B_{3jj} = B_{311} + B_{322} + B_{333}$, which is obtained by summing equations (\ref{eq:B311}) - (\ref{eq:B333}). Together this implies:
\begin{equation}
B_{m33} = \delta_{m3} \frac{4\pi R_B^2}{m_{\mathrm{bg}}^3}(m_{\mathrm{bg}}R_B\cosh(m_{\mathrm{bg}}R_B) - \sinh(m_{\mathrm{bg}}R_B)). 
\end{equation}
In the same co-ordinates, we find
\begin{equation}
A_{i3} = \delta_{i3}\left(\frac{m_{\mathrm{bg}}|\mathbf{X}_A|}{1+m_{\mathrm{bg}}|\mathbf{X}_A|} - 2\right),
\end{equation}
which implies:
\begin{equation}
\left(\delta_i^{(k}\delta^{l)j} - \frac{\delta_i^j}{2}\delta^{kl}\right)A_{km}B_{jml}(m_{\mathrm{bg}}) = \delta_{i3}\left(\frac{m_{\mathrm{bg}}|\mathbf{X}_A|}{1+m_{\mathrm{bg}}|\mathbf{X}_A|} - 2\right)\frac{2\pi R_B^3}{m_{\mathrm{bg}}^2}(m_{\mathrm{bg}}R_B\cosh(m_{\mathrm{bg}}R_B) - \sinh(m_{\mathrm{bg}}R_B)).
\end{equation}
We conclude that
\begin{align}
F_i = & Q_AQ_B\left[-X_{Ai}\frac{4\pi R_B^2}{m_{\mathrm{bg}}}\sinh(m_{\mathrm{bg}}R_B) + \frac{X_{Ai}}{|\mathbf{X}_A|}\left(\frac{m_{\mathrm{bg}}|\mathbf{X}_A|}{1+m_{\mathrm{bg}}|\mathbf{X}_A|} - 2\right) \frac{4\pi R_B^2}{m_{\mathrm{bg}}^2}(m_{\mathrm{bg}}R_B\cosh(m_{\mathrm{bg}}R_B) - \sinh(m_{\mathrm{bg}}R_B)\right] \nonumber \\
&\times \frac{(1 + m_{\mathrm{bg}}|\mathbf{X}_A|)}{(1 + m_{\mathrm{bg}}R_A)}e^{-m_{\mathrm{bg}}(|\mathbf{X}_A| - R_A)} \nonumber \\
 = & \frac{\xi_A\xi_BM_AM_B}{16\pi^2 M^2|\mathbf{X}_A|^3R_B^3}\frac{(1 + |\mathbf{X}_A|m_{\mathrm{bg}})}{(1 + m_{\mathrm{bg}R_A})}e^{-m_{\mathrm{bg}}(|\mathbf{X}_A| - R_A)} \times \nonumber \\
 & \left[-X_{Ai}\frac{4\pi R_B^2}{m_{\mathrm{bg}}}\sinh(m_{\mathrm{bg}}R_B) + \frac{X_{Ai}}{|\mathbf{X}_A|}\left(\frac{m_{\mathrm{bg}}|\mathbf{X}_A|}{1+m_{\mathrm{bg}}|\mathbf{X}_A|} - 2\right) \frac{4\pi R_B^2}{m_{\mathrm{bg}}^2}(m_{\mathrm{bg}}R_B\cosh(m_{\mathrm{bg}}R_B) - \sinh(m_{\mathrm{bg}}R_B)\right] \nonumber \\
 =& -\frac{G\xi_A\xi_BM_AM_BM_{\mathrm{P}}^2}{2\pi M^2|\mathbf{X}_A|^2}\frac{(1 + |\mathbf{X}_A|m_{\mathrm{bg}})}{(1 + m_{\mathrm{bg}R_A})}e^{-m_{\mathrm{bg}}(|\mathbf{X}_A| - R_A)} \frac{X_{Ai}}{|\mathbf{X}_A|}\times \nonumber \\
 & \left[\frac{4\pi}{m_{\mathrm{bg}}R_B}\sinh(m_{\mathrm{bg}}R_B) - \left(\frac{m_{\mathrm{bg}}|\mathbf{X}_A|}{1+m_{\mathrm{bg}}|\mathbf{X}_A|} - 2\right) \frac{4\pi }{m_{\mathrm{bg}}^2|\mathbf{X}_A|R_B}(m_{\mathrm{bg}}R_B\cosh(m_{\mathrm{bg}}R_B) - \sinh(m_{\mathrm{bg}}R_B)\right] \nonumber \\
  =& -\frac{GM_AM_B}{|\mathbf{X}_A|^2}2\xi_A\xi_B\left(\frac{M_{\mathrm{P}}}{M}\right)^2\frac{(1 + |\mathbf{X}_A|m_{\mathrm{bg}})}{(1 + m_{\mathrm{bg}R_A})}e^{-m_{\mathrm{bg}}(|\mathbf{X}_A| - R_A)} \frac{X_{Ai}}{|\mathbf{X}_A|}\times \nonumber \\
 & \left[\frac{1}{m_{\mathrm{bg}}R_B}\sinh(m_{\mathrm{bg}}R_B) - \left(\frac{m_{\mathrm{bg}}|\mathbf{X}_A|}{1+m_{\mathrm{bg}}|\mathbf{X}_A|} - 2\right) \frac{ 1}{m_{\mathrm{bg}}^2|\mathbf{X}_A|R_B}(m_{\mathrm{bg}}R_B\cosh(m_{\mathrm{bg}}R_B) - \sinh(m_{\mathrm{bg}}R_B)\right]. \nonumber \\
\end{align}
Including the gravitational force, and moving to co-ordinates centred on ball $A$, we find:
\begin{equation}
F(r) = -\frac{GM_AM_B}{r^2}\left[1 + 2\xi_A\xi_B\left(\frac{M_{\mathrm{P}}}{M}\right)^2\frac{(1 + |\mathbf{X}_A|m_{\mathrm{bg}})}{(1 + m_{\mathrm{bg}R_A})}e^{-m_{\mathrm{bg}}(|\mathbf{X}_A| - R_A)}f(m_{\mathrm{bg}}R_B,m_{\mathrm{bg}}|\mathbf{X}_A|) \right],
\end{equation}
where we have a form-factor function that modifies the force, given by
\begin{equation}
f(u,y) = (1 + u)e^{-u}\left[\frac{\sinh(u)}{u} - \left(\frac{v}{1 + v} - 2\right)\frac{1}{v}\left(\cosh(u) - \frac{\sinh(u)}{u}\right)\right].
\end{equation}
This force has several peculiar features. Firstly, there appears to be a double-counting of the distance from the centre of ball $B$ to its surface ($R_B$) featuring in the exponential suppression with distance. Note, however, that this additional exponential arises from the screening affect of the probe itself, that is, its lack of response to the chameleon field when the probe is itself large. Secondly, we note that if we computed the force of ball $B$ on ball $A$, we would \emph{not} obtain a symmetric force of opposite sign. In other words, this force appears to violate Newton's third law, which on the face of it suggest that momentum is not conserved. However, this is misleading, because the derivation takes into account the stress energy tensor not only of the ball itself ($\tau^{(m)\mu\nu}$) but of the chameleon field ($\tau^{(\phi)\mu\nu}$). Momentum should still be conserved if the momentum of both the balls \emph{and} the field are included in the calculation. This differs from the Newtonian force derived by Burrage \textit{et al}.~\cite{Burrage:2014oza} since in the $m_{\mathrm{bg}}R_B \ll 1$ limit the force is inverse-square, meaning that Newton's shell theorem applies; a sphere should exert the same inverse-square force as an equivalent point mass. But this does not apply to Yukawa potentials, implying that the chameleon force does not obey the strong equivalence principle (in addition to the explicit violation of the weak equivalence principle due to the force being density-dependent). In any case, the form factor $f$ is typically close to 1 (it is equal to 1 in the $m_{\mathrm{bg}}R_B \ll 1$ limit), so can in most cases be neglected. 

\section{Derivation of the sensitivity} \label{app:sensitivity}

The dynamics that arise for a moving source-mass was solved in Refs~\cite{qvarfort2020time}. Resonant gravimetry and enhancements from modulated optomechanical couplings were considered in~\cite{qvarfort2020optimal}. We here outline the solution and refer the reader to~\cite{qvarfort2020optimal} for the full derivation. 

\subsection{Solution of the dynamics}

The time-evolution that arises from the Hamiltonian in equation~\eqref{eq:cham:Hamiltonian} can be written as 
\begin{align}\label{U}
\hat U(t)= \, &e^{-i\,F_{\hat{N}_a}\,\hat{N}_a}\,e^{-i\,F_{\hat{N}^2_a}\,\hat{N}^2_a}\,e^{-i\,F_{\hat{B}_+}\,\hat{B}_+}\,e^{-i\,F_{\hat{N}_a\,\hat{B}_+}\,\hat{N}_a\,\hat{B}_+}\,e^{-i\,F_{\hat{B}_-}\,\hat{B}_-}\,e^{-i\,F_{\hat{N}_a\,\hat{B}_-}\,\hat{N}_a\,\hat{B}_-},
\end{align}
where we have transformed into a frame that is rotating with the free optical evolution $\hat a^\dag \hat a $, and where the operators are given by 
\begin{align}\label{basis:operator:Lie:algebra}
 \hat{N}^2_a &= (\hat a^\dagger \hat a)^2, \nonumber \\
	\hat{N}_a &= \hat a^\dagger \hat a, & 
	\hat{N}_b &= \hat b^\dagger \hat b, & \nonumber \\
	\hat{B}_+ &=  \hat b^\dagger +\hat b, &
	\hat{B}_- &= i\,(\hat b^\dagger -\hat b), &
	 &  \nonumber\\
	\hat{N}_a\,\hat{B}_+ &= \hat{N}_a\,(\hat b^{\dagger}+\hat b), &
	\hat{N}_a\,\hat{B}_- &= \hat{N}_a\,i\,(\hat b^{\dagger}-\hat b). &
	 & 
\end{align}
 By differentiating $\hat U(t)$ and equating the result with the Hamiltonian in equation~\eqref{eq:cham:Hamiltonian}, we can use the linear independence of the Hilbert space to find the following coefficients~\cite{qvarfort2020time}
\begin{align}\label{sub:algebra:decoupling:solution}
F_{\hat{N}_a}&= 2 \frac{x_{\mathrm{zpf}}}{ \hbar} \,\int_0^t\,\mathrm{d}t'\,V'(x_S(t’))\,\sin(\omega_{\mathrm{mech}}t')\int_0^{t'}\mathrm{d}t''\,g(t'')\,\cos(\omega_{\mathrm{mech}}t'')\, \nonumber \\
&\quad + 2  \frac{x_{\mathrm{zpf}}}{ \hbar}\,\int^t_0\,\mathrm{d}t' \,g(t')\, \sin (\omega_{\mathrm{mech}}t') \, \int^{t'}_0 \,\mathrm{d}t''\, V'(x_S(t^{\prime \prime}))\, \cos(\omega_{\mathrm{mech}}t'') \, ,  \nonumber\\
F_{\hat{N}^2_a}&= -2 \,\int_0^t\,\mathrm{d}t'\,g(t')\,\sin(\omega_{\mathrm{mech}}t')\int_0^{t'}\mathrm{d}t''\,g(t'')\,\cos(\omega_{\mathrm{mech}}t'') \, , \nonumber\\
F_{\hat{B}_+}&=  \frac{x_{\mathrm{zpf}}}{ \hbar}\int_0^t\,\mathrm{d}t'\,V'(x_S(t'))\,\cos(\omega_{\mathrm{mech}}t') \, , \nonumber\\
F_{\hat{B}_-}&= \frac{x_{\mathrm{zpf}}}{ \hbar}\int_0^t\,\mathrm{d}t'\,V'(x_S(t')) \,\sin(\omega_{\mathrm{mech}}t') \, , \nonumber\\
F_{\hat{N}_a\,\hat{B}_+}&=- \int_0^t\,\mathrm{d}t'\,g(t')\,\cos(\omega_{\mathrm{mech}}t') \, , \nonumber\\
F_{\hat{N}_a\,\hat{B}_-}&=- \int_0^t\,\mathrm{d}t'\,g(t')\,\sin(\omega_{\mathrm{mech}}t'),
\end{align}
where $V'(x_S(t)) $ is a generic potential given in equation~\eqref{eq:cham:Hamiltonian}. 

\subsection{Derivation of the quantum Fisher information} 

In the case where we are estimating a linear mechanical displacement, $\hat{\mathcal{H}}_\theta$ can be decomposed into~\cite{schneiter2020optimal}
\begin{equation}
\hat{\mathcal{H}}_\theta = B \hat N_a  + C_+ \hat B_+  + C_- \hat B_-, 
\end{equation}
where the coefficients are given by  
\begin{align}
B&=-\partial_\theta F_{\hat N_a} - 2 \, 
F_{\hat N_a \, \hat B_-}\partial_\theta F_{\hat B_+}
 ,\ \nonumber \\
C_+&= -\partial_\theta F_{\hat B_+}, \nonumber \\
C_-&= -\partial_\theta F_{\hat B_-}.
\end{align}
For an initially coherent state in the optical field and a thermal state of the mechanical element, as that shown in equation~\eqref{initial:state}, we have $\lambda_n = \tanh^{2n}(r_T)/\cosh^2(r_T)$ and $\ket{\lambda_n} = \ket{\zeta}\otimes \ket{n}$. 
The QFI can then be written as the following expression given the initially coherent state of the optical mode and thermal state of the mechanical element shown in equation~\eqref{initial:state}
\begin{align}\label{qfi:estimate:d1}
\mathcal{I}_\theta &= 4\,B^2\,|\mu_\textrm{c}|^2+4\,\frac{\left(C_+^2+C_-^2\right)}{\cosh(2r_T)}.
\end{align}
We can then obtain the expressions in equations~\eqref{eq:constant:sensitivity:kappa},~\eqref{eq:constant:sensitivity:sigma}
and~\eqref{eq:modulated:sensitivity:kappa},~\eqref{eq:modulated:sensitivity:sigma} by dropping the second term, which we can assume to be much smaller than the first term, and taking the derivative of $F_{\hat N_a}$ and $F_{\hat B_+}$ for $\kappa$ and $\sigma$ while assuming $k(t) = k_0$ and $k(t) = k_0\cos(\omega_m t)$, respectively. 

We emphasise here that the reason that the time dependent Newtonian gravitational acceleration does not appear in the result is due to the linearity of the derivative. Ultimately, a sensing scheme of this form has a certain resolution, which we are able to compute from these results. In practise, the data must still be analysed in order to distinguish between the Newtonian and the modified gravitational force, which we discuss in section~\ref{sec:discussion}.

\footnotesize
\bibliographystyle{iopart-num}
\bibliography{draft_bibliography}

\providecommand{\newblock}{}
\begin{thebibliography}{100}
\expandafter\ifx\csname url\endcsname\relax
  \def\url#1{{\tt #1}}\fi
\expandafter\ifx\csname urlprefix\endcsname\relax\def\urlprefix{URL }\fi
\providecommand{\eprint}[2][]{\url{#2}}

\bibitem{will1990general}
Will C~M 1990 {\em Science\/} {\bf 250} 770--776
  \urlprefix\url{10.1126/science.250.4982.770}

\bibitem{abbott2016observation}
Abbott B~P, Abbott R, Abbott T, Abernathy M, Acernese F, Ackley K, Adams C,
  Adams T, Addesso P, Adhikari R {\em et~al.\/} 2016 {\em Physical Review
  Letters\/} {\bf 116} 061102
  \urlprefix\url{https://doi.org/10.1103/PhysRevLett.116.061102}

\bibitem{Ade:2015rim}
Ade P~A~R {\em et~al.\/} (Planck) 2016 {\em Astronomy \& Astrophysics\/} {\bf
  594} A14 \urlprefix\url{https://doi.org/10.1051/0004-6361/201525814}

\bibitem{Aghanim:2018eyx}
Aghanim N, Akrami Y, Ashdown M, Aumont J, Baccigalupi C, Ballardini M, Banday
  A, Barreiro R, Bartolo N, Basak S {\em et~al.\/} 2020 {\em Astronomy \&
  Astrophysics\/} {\bf 641} A6
  \urlprefix\url{https://doi.org/10.1051/0004-6361/201833910}

\bibitem{riess1998observational}
Riess A~G, Filippenko A~V, Challis P, Clocchiatti A, Diercks A, Garnavich P~M,
  Gilliland R~L, Hogan C~J, Jha S, Kirshner R~P {\em et~al.\/} 1998 {\em The
  Astronomical Journal\/} {\bf 116} 1009
  \urlprefix\url{https://doi.org/10.1086/300499}

\bibitem{Padilla:2015aaa}
Padilla A 2015 {\em arXiv preprint arXiv:1502.05296\/} (\textit{Preprint}
  \eprint{1502.05296})

\bibitem{Will:2005va}
Will C~M 2006 {\em Living Reviews in Relativity\/} {\bf 9} 3 (\textit{Preprint}
  \eprint{gr-qc/0510072}) \urlprefix\url{https://doi.org/10.12942/lrr-2006-3}

\bibitem{PhysRevLett.92.121101}
Shapiro S~S, Davis J~L, Lebach D~E and Gregory J~S 2004 {\em Physical Review
  Letters\/} {\bf 92}(12) 121101
  \urlprefix\url{https://link.aps.org/doi/10.1103/PhysRevLett.92.121101}

\bibitem{2003Natur.425..374B}
{Bertotti} B, {Iess} L and {Tortora} P 2003 {\em Nature\/} {\bf 425} 374--376
  \urlprefix\url{http://adsabs.harvard.edu/abs/2003Natur.425..374B}

\bibitem{Koyama:2015vza}
Koyama K 2016 {\em Reports on Progress in Physics\/} {\bf 79} 046902
  \urlprefix\url{https://doi.org/10.1088/0034-4885/79/4/046902}

\bibitem{Joyce:2014kja}
Joyce A, Jain B, Khoury J and Trodden M 2015 {\em Physics Reports\/} {\bf 568}
  1--98 \urlprefix\url{https://doi.org/10.1016/j.physrep.2014.12.002}

\bibitem{Khoury:2003aq}
Khoury J and Weltman A 2004 {\em Physical Reviow Letters\/} {\bf 93} 171104
  \urlprefix\url{https://doi.org/10.1103/PhysRevLett.93.171104}

\bibitem{PhysRevD.69.044026}
Khoury J and Weltman A 2004 {\em Physical Review D\/} {\bf 69}(4) 044026
  \urlprefix\url{https://link.aps.org/doi/10.1103/PhysRevD.69.044026}

\bibitem{Brax:2004qh}
Brax P, van~de Bruck C, Davis A~C, Khoury J and Weltman A 2004 {\em Physical
  Review D\/} {\bf D70} 123518 (\textit{Preprint} \eprint{astro-ph/0408415})

\bibitem{giovannetti2006quantum}
Giovannetti V, Lloyd S and Maccone L 2006 {\em Physical Review Letters\/} {\bf
  96} 010401 \urlprefix\url{https://doi.org/10.1103/PhysRevLett.96.010401}

\bibitem{upadhye2006unv}
Upadhye A, Gubser S~S and Khoury J 2006 {\em Physical Review D\/} {\bf 74}(10)
  104024 \urlprefix\url{https://link.aps.org/doi/10.1103/PhysRevD.74.104024}

\bibitem{Mota2006strongly}
Mota D~F and Shaw D~J 2006 {\em Physical Review Letters\/} {\bf 97}(15) 151102
  \urlprefix\url{https://link.aps.org/doi/10.1103/PhysRevLett.97.151102}

\bibitem{Mota2007evading}
Mota D~F and Shaw D~J 2007 {\em Physical Review D\/} {\bf 75}(6) 063501
  \urlprefix\url{https://link.aps.org/doi/10.1103/PhysRevD.75.063501}

\bibitem{Adelberger2007particle}
Adelberger E~G, Heckel B~R, Hoedl S, Hoyle C~D, Kapner D~J and Upadhye A 2007
  {\em Physical Review Letters\/} {\bf 98}(13) 131104
  \urlprefix\url{https://link.aps.org/doi/10.1103/PhysRevLett.98.131104}

\bibitem{PhysRevD.78.104021}
Brax P, van~de Bruck C, Davis A~C and Shaw D~J 2008 {\em Physical Review D\/}
  {\bf 78}(10) 104021
  \urlprefix\url{https://link.aps.org/doi/10.1103/PhysRevD.78.104021}

\bibitem{upadhye2012dark}
Upadhye A 2012 {\em Physical Review D\/} {\bf 86}(10) 102003
  \urlprefix\url{https://link.aps.org/doi/10.1103/PhysRevD.86.102003}

\bibitem{Upadhye:2012fz}
Upadhye A 2012 {Particles and forces from chameleon dark energy} {\em {8th
  Patras Workshop on Axions, WIMPs and WISPs}\/} (\textit{Preprint}
  \eprint{1211.7066})

\bibitem{PhysRevD.70.042004}
Hoyle C~D, Kapner D~J, Heckel B~R, Adelberger E~G, Gundlach J~H, Schmidt U and
  Swanson H~E 2004 {\em Physical Review D\/} {\bf 70}(4) 042004
  \urlprefix\url{https://link.aps.org/doi/10.1103/PhysRevD.70.042004}

\bibitem{PhysRevLett.98.021101}
Kapner D~J, Cook T~S, Adelberger E~G, Gundlach J~H, Heckel B~R, Hoyle C~D and
  Swanson H~E 2007 {\em Physical Review Letters\/} {\bf 98}(2) 021101
  \urlprefix\url{https://link.aps.org/doi/10.1103/PhysRevLett.98.021101}

\bibitem{Brax2007detecting}
Brax P, van~de Bruck C, Davis A~C, Mota D~F and Shaw D 2007 {\em Physical
  Review D\/} {\bf 76}(12) 124034
  \urlprefix\url{https://link.aps.org/doi/10.1103/PhysRevD.76.124034}

\bibitem{Brax2010tuning}
Brax P, van~de Bruck C, Davis A~C, Shaw D~J and Iannuzzi D 2010 {\em Physical
  Review Letters\/} {\bf 104}(24) 241101
  \urlprefix\url{https://link.aps.org/doi/10.1103/PhysRevLett.104.241101}

\bibitem{Almasi2015force}
Almasi A, Brax P, Iannuzzi D and Sedmik R~I~P 2015 {\em Physical Review D\/}
  {\bf 91}(10) 102002
  \urlprefix\url{https://link.aps.org/doi/10.1103/PhysRevD.91.102002}

\bibitem{Brax2015casimir}
Brax P and Davis A~C 2015 {\em Physical Review D\/} {\bf 91}(6) 063503
  \urlprefix\url{https://link.aps.org/doi/10.1103/PhysRevD.91.063503}

\bibitem{peters2001high}
Peters A, Chung K~Y and Chu S 2001 {\em Metrologia\/} {\bf 38} 25
  \urlprefix\url{https://doi.org/10.1088/0026-1394/38/1/4}

\bibitem{mcguirk2002sensitive}
Mcguirk J~M, Foster G, Fixler J, Snadden M and Kasevich M 2002 {\em Physical
  Review A\/} {\bf 65} 033608
  \urlprefix\url{https://doi.org/10.1103/PhysRevA.65.033608}

\bibitem{bidel2013compact}
Bidel Y, Carraz O, Charriere R, Cadoret M, Zahzam N and Bresson A 2013 {\em
  Applied Physics Letters\/} {\bf 102} 144107
  \urlprefix\url{https://doi.org/10.1063/1.4801756}

\bibitem{hu2013demonstration}
Hu Z~K, Sun B~L, Duan X~C, Zhou M~K, Chen L~L, Zhan S, Zhang Q~Z and Luo J 2013
  {\em Physical Review A\/} {\bf 88} 043610
  \urlprefix\url{https://doi.org/10.1103/PhysRevA.88.043610}

\bibitem{Burrage:2014oza}
Burrage C, Copeland E~J and Hinds E~A 2015 {\em Journal of Cosmology and
  Astroparticle Physics\/} {\bf 1503} 042
  \urlprefix\url{https://doi.org/10.1088/1475-7516/2015/03/042}

\bibitem{Burrage:2015lya}
Burrage C and Copeland E~J 2016 {\em Contemporary Physics\/} {\bf 57} 164--176
  \urlprefix\url{https://doi.org/10.1080/00107514.2015.1060058s}

\bibitem{elder2016cham}
Elder B, Khoury J, Haslinger P, Jaffe M, M\"uller H and Hamilton P 2016 {\em
  Physical Review D\/} {\bf 94}(4) 044051
  \urlprefix\url{https://link.aps.org/doi/10.1103/PhysRevD.94.044051}

\bibitem{schloegel2016prob}
Schl\"ogel S, Clesse S and F\"uzfa A 2016 {\em Physical Review D\/} {\bf
  93}(10) 104036
  \urlprefix\url{https://link.aps.org/doi/10.1103/PhysRevD.93.104036}

\bibitem{burrage2016proposed}
Burrage C, Copeland E~J and Stevenson J~A 2016 {\em Journal of Cosmology and
  Astroparticle Physics\/} {\bf 2016} 070
  \urlprefix\url{https://doi.org/10.1088/1475-7516/2016/08/070}

\bibitem{chiow2018multiloop}
Chiow S~w and Yu N 2018 {\em Physical Review D\/} {\bf 97} 044043
  \urlprefix\url{https://doi.org/10.1103/PhysRevD.97.044043}

\bibitem{Hartley:2019wzu}
Hartley D, K\"ading C, Howl R and Fuentes I 2019 {\em arXiv preprint
  arXiv:1909.02272\/} (\textit{Preprint} \eprint{1909.02272})

\bibitem{hamilton2015atom}
Hamilton P, Jaffe M, Haslinger P, Simmons Q, M{\"u}ller H and Khoury J 2015
  {\em Science\/} {\bf 349} 849--851
  \urlprefix\url{https://doi.org/10.1126/science.aaa8883}

\bibitem{sabulsky2019exp}
Sabulsky D~O, Dutta I, Hinds E~A, Elder B, Burrage C and Copeland E~J 2019 {\em
  Physical Review Letters\/} {\bf 123}(6) 061102
  \urlprefix\url{https://link.aps.org/doi/10.1103/PhysRevLett.123.061102}

\bibitem{Brax:2014zta}
Brax P and Davis A~C 2015 {\em Physical Review D\/} {\bf D91} 063503
  \urlprefix\url{https://doi.org/10.1103/PhysRevD.91.063503}

\bibitem{Serebrov2011search}
Serebrov A~P, Zherebtsov O~M, Sbitnev S~V, Varlamov V~E, Vassiljev A~V and
  Lasakov M~S 2011 {\em Physical Review C\/} {\bf 84}(4) 044001
  \urlprefix\url{https://link.aps.org/doi/10.1103/PhysRevC.84.044001}

\bibitem{Brax2011strongly}
Brax P and Pignol G 2011 {\em Physical Review Letters\/} {\bf 107}(11) 111301
  \urlprefix\url{https://link.aps.org/doi/10.1103/PhysRevLett.107.111301}

\bibitem{Serebrov2014experimental}
Serebrov A~P, Geltenbort P, Zherebtsov O~M, Sbitnev S~V, Varlamov V~E,
  Vassiljev A~V, Lasakov M~S, Krasnoschekova I~A, Ivanov S~N and Pushin D 2014
  {\em Physical Review C\/} {\bf 89}(4) 044002
  \urlprefix\url{https://link.aps.org/doi/10.1103/PhysRevC.89.044002}

\bibitem{brax2014testing}
Brax P 2014 {\em Physics Procedia\/} {\bf 51} 73--77
  \urlprefix\url{https://doi.org/10.1016/j.phpro.2013.12.017}

\bibitem{Jenke2014gravity}
Jenke T, Cronenberg G, Burgd\"orfer J, Chizhova L~A, Geltenbort P, Ivanov A~N,
  Lauer T, Lins T, Rotter S, Saul H, Schmidt U and Abele H 2014 {\em Physical
  Review Letters\/} {\bf 112}(15) 151105
  \urlprefix\url{https://link.aps.org/doi/10.1103/PhysRevLett.112.151105}

\bibitem{Cronenberg:2016Pt}
Cronenberg G, Filter H, Thalhammer M, Jenke T, Abele H and Geltenbort P 2016
  {\em PoS\/} {\bf EPS-HEP2015} 408 \urlprefix\url{10.22323/1.234.0408}

\bibitem{Brax2013probing}
Brax P, Pignol G and Roulier D 2013 {\em Physical Review D\/} {\bf 88}(8)
  083004 \urlprefix\url{https://link.aps.org/doi/10.1103/PhysRevD.88.083004}

\bibitem{pokotilovski2013strongly}
Pokotilovski Y~N 2013 {\em Physics Letters B\/} {\bf 719} 341--345
  \urlprefix\url{https://doi.org/10.1016/j.physletb.2013.01.022}

\bibitem{lemmel2015neutron}
Lemmel H, Brax P, Ivanov A, Jenke T, Pignol G, Pitschmann M, Potocar T,
  Wellenzohn M, Zawisky M and Abele H 2015 {\em Physics Letters B\/} {\bf 743}
  310--314 \urlprefix\url{https://doi.org/10.1016/j.physletb.2015.02.063}

\bibitem{Li2016neutron}
Li K, Arif M, Cory D~G, Haun R, Heacock B, Huber M~G, Nsofini J, Pushin D~A,
  Saggu P, Sarenac D, Shahi C~B, Skavysh V, Snow W~M and Young A~R (The INDEX
  Collaboration) 2016 {\em Physical Review D\/} {\bf 93}(6) 062001
  \urlprefix\url{https://link.aps.org/doi/10.1103/PhysRevD.93.062001}

\bibitem{Brax2011atomic}
Brax P and Burrage C 2011 {\em Physical Review D\/} {\bf 83}(3) 035020
  \urlprefix\url{https://link.aps.org/doi/10.1103/PhysRevD.83.035020}

\bibitem{Frugiuele2017constraining}
Frugiuele C, Fuchs E, Perez G and Schlaffer M 2017 {\em Physical Review D\/}
  {\bf 96}(1) 015011
  \urlprefix\url{https://link.aps.org/doi/10.1103/PhysRevD.96.015011}

\bibitem{Rybka2010search}
Rybka G, Hotz M, Rosenberg L~J, Asztalos S~J, Carosi G, Hagmann C, Kinion D,
  van Bibber K, Hoskins J, Martin C, Sikivie P, Tanner D~B, Bradley R and
  Clarke J 2010 {\em Physical Review Letters\/} {\bf 105}(5) 051801
  \urlprefix\url{https://link.aps.org/doi/10.1103/PhysRevLett.105.051801}

\bibitem{upadhye2012design}
Upadhye A, Steffen J~H and Chou A~S 2012 {\em Physical Review D\/} {\bf 86}(3)
  035006 \urlprefix\url{https://link.aps.org/doi/10.1103/PhysRevD.86.035006}

\bibitem{PhysRevD.97.084050}
Brax P, Davis A~C, Elder B and Wong L~K 2018 {\em Physical Review D\/} {\bf
  97}(8) 084050
  \urlprefix\url{https://link.aps.org/doi/10.1103/PhysRevD.97.084050}

\bibitem{bowenbook}
Bowen W~P and Milburn G~J 2015 {\em Quantum Optomechanics\/} (CRC Press)

\bibitem{aspelmeyer2014cavity}
Aspelmeyer M, Kippenberg T~J and Marquardt F 2014 {\em Reviews of Modern
  Physics\/} {\bf 86} 1391
  \urlprefix\url{https://doi.org/10.1103/RevModPhys.86.1391}

\bibitem{favero2009optomechanics}
Favero I and Karrai K 2009 {\em Nature Photonics\/} {\bf 3} 201
  \urlprefix\url{https://doi.org/10.1038/nphoton.2009.42}

\bibitem{barker2010cavity}
Barker P and Shneider M 2010 {\em Physical Review A\/} {\bf 81} 023826
  \urlprefix\url{https://doi.org/10.1103/PhysRevA.81.023826}

\bibitem{tsaturyan2017ultracoherent}
Tsaturyan Y, Barg A, Polzik E~S and Schliesser A 2017 {\em Nature
  Nanotechnology\/} {\bf 12} 776
  \urlprefix\url{http://dx.doi.org/10.1038/nnano.2017.101}

\bibitem{shkarin2019quantum}
Shkarin A, Kashkanova A, Brown C, Garcia S, Ott K, Reichel J and Harris J 2019
  {\em Physical Review Letters\/} {\bf 122} 153601
  \urlprefix\url{https://doi.org/10.1103/PhysRevLett.122.153601}

\bibitem{purdy2010tunable}
Purdy T~P, Brooks D, Botter T, Brahms N, Ma Z~Y and Stamper-Kurn D~M 2010 {\em
  Physical Review Letters\/} {\bf 105} 133602
  \urlprefix\url{https://doi.org/10.1103/PhysRevLett.105.133602}

\bibitem{chan2011laser}
Chan J, Alegre T~P~M, Safavi-Naeini A~H, Hill J~T, Krause A, Gr{\"o}blacher S,
  Aspelmeyer M and Painter O 2011 {\em Nature\/} {\bf 478} 89--92
  \urlprefix\url{https://doi.org/10.1038/nature10461}

\bibitem{teufel2011sideband}
Teufel J~D, Donner T, Li D, Harlow J~W, Allman M~S, Cicak K, Sirois A~J,
  Whittaker J~D, Lehnert K~W and Simmonds R~W 2011 {\em Nature\/} {\bf 475}
  359--363 \urlprefix\url{https://doi.org/10.1038/nature10261}

\bibitem{Delic892}
Deli{\'c} U, Reisenbauer M, Dare K, Grass D, Vuleti{\'c} V, Kiesel N and
  Aspelmeyer M 2020 {\em Science\/} {\bf 367} 892--895 ISSN 0036-8075
  \urlprefix\url{https://science.sciencemag.org/content/367/6480/892}

\bibitem{arcizet2006high}
Arcizet O, Cohadon P~F, Briant T, Pinard M, Heidmann A, Mackowski J~M, Michel
  C, Pinard L, Fran{\c{c}}ais O and Rousseau L 2006 {\em Physical Review
  Letters\/} {\bf 97} 133601
  \urlprefix\url{https://doi.org/10.1103/PhysRevLett.97.133601}

\bibitem{geraci2010short}
Geraci A~A, Papp S~B and Kitching J 2010 {\em Physical Review Letters\/} {\bf
  105} 101101 \urlprefix\url{https://doi.org/10.1103/PhysRevLett.105.101101}

\bibitem{hempston2017force}
Hempston D, Vovrosh J, Toro{\v{s}} M, Winstone G, Rashid M and Ulbricht H 2017
  {\em Applied Physics Letters\/} {\bf 111} 133111
  \urlprefix\url{https://doi.org/10.1063/1.4993555}

\bibitem{qvarfort2018gravimetry}
Qvarfort S, Serafini A, Barker P~F and Bose S 2018 {\em Nature
  Communications\/} {\bf 9} 3690
  \urlprefix\url{https://doi.org/10.1038/s41467-018-06037-z}

\bibitem{armata2017quantum}
Armata F, Latmiral L, Plato A and Kim M 2017 {\em Physical Review A\/} {\bf 96}
  043824
  \urlprefix\url{https://journals.aps.org/pra/abstract/10.1103/PhysRevA.96.043824}

\bibitem{schneiter2020optimal}
Schneiter F, Qvarfort S, Serafini A, Xuereb A, Braun D, R{\"a}tzel D and
  Bruschi D~E 2020 {\em Physical Review A\/} {\bf 101} 033834
  \urlprefix\url{https://doi.org/10.1103/PhysRevA.101.033834}

\bibitem{qvarfort2020optimal}
Qvarfort S, Plato A~D~K, Bruschi D~E, Schneiter F, Braun D, Serafini A and
  R{\"a}tzel D 2021 {\em Physical Review Research\/} {\bf 3} 013159
  \urlprefix\url{https://doi.org/10.1103/PhysRevResearch.3.013159}

\bibitem{PhysRevLett.117.101101}
Rider A~D, Moore D~C, Blakemore C~P, Louis M, Lu M and Gratta G 2016 {\em
  Physical Review Letters\/} {\bf 117}(10) 101101
  \urlprefix\url{https://link.aps.org/doi/10.1103/PhysRevLett.117.101101}

\bibitem{PhysRevD.96.104002}
Antoniou I and Perivolaropoulos L 2017 {\em Physical Review D\/} {\bf 96}(10)
  104002 \urlprefix\url{https://link.aps.org/doi/10.1103/PhysRevD.96.104002}

\bibitem{moore2020searching}
Moore D~C and Geraci A~A 2021 {\em Quantum Sci. Technol.\/} {\bf 6} 014008
  (\textit{Preprint} \eprint{2008.13197})

\bibitem{pernot2019general}
Pernot-Borr{\`a}s M, Berg{\'e} J, Brax P and Uzan J~P 2019 {\em Physical Review
  D\/} {\bf 100} 084006

\bibitem{pernot2021constraints}
Pernot-Borr{\`a}s M, Berg{\'e} J, Brax P, Uzan J~P, M{\'e}tris G, Rodrigues M
  and Touboul P 2021 {\em Physical Review D\/} {\bf 103} 064070

\bibitem{westphal2020measurement}
Westphal T, Hepach H, Pfaff J and Aspelmeyer M 2020 {\em arXiv preprint
  arXiv:2009.09546\/} \urlprefix\url{https://arxiv.org/abs/2009.09546}

\bibitem{schmole2017development}
Schm{\"o}le J 2017 {\em Development of a micromechanical proof-of-principle
  experiment for measuring the gravitational force of milligram masses\/} Ph.D.
  thesis uniwien

\bibitem{tufarelli2014coherently}
Tufarelli T, Ferraro A, Serafini A, Bose S and Kim M~S 2014 {\em Physical
  Review Letters\/} {\bf 112}(13) 133605
  \urlprefix\url{https://link.aps.org/doi/10.1103/PhysRevLett.112.133605}

\bibitem{rugar1991mechanical}
Rugar D and Gr{\"u}tter P 1991 {\em Physical Review Letters\/} {\bf 67} 699
  \urlprefix\url{https://doi.org/10.1103/PhysRevLett.67.699}

\bibitem{szorkovszky2013strong}
Szorkovszky A, Brawley G~A, Doherty A~C and Bowen W~P 2013 {\em Physical Review
  Letters\/} {\bf 110} 184301
  \urlprefix\url{https://doi.org/10.1103/PhysRevLett.110.184301}

\bibitem{Millen2015cavity}
Millen J, Fonseca P~Z~G, Mavrogordatos T, Monteiro T~S and Barker P~F 2015 {\em
  Physical Review Letters\/} {\bf 114}(12) 123602
  \urlprefix\url{https://link.aps.org/doi/10.1103/PhysRevLett.114.123602}

\bibitem{aranas2016split}
Aranas E~B, Fonseca P~Z~G, Barker P~F and Monteiro T~S 2016 {\em New Journal of
  Physics\/} {\bf 18} 113021
  \urlprefix\url{https://doi.org/10.1088/1367-2630/18/11/113021}

\bibitem{fonseca2016nonlinear}
Fonseca P~Z~G, Aranas E~B, Millen J, Monteiro T~S and Barker P~F 2016 {\em
  Physical Review Letters\/} {\bf 117} 173602
  \urlprefix\url{https://doi.org/10.1103/PhysRevLett.117.173602}

\bibitem{cirio2012quantum}
Cirio M, Brennen G~K and Twamley J 2012 {\em Physical Review Letters\/} {\bf
  109}(14) 147206
  \urlprefix\url{https://link.aps.org/doi/10.1103/PhysRevLett.109.147206}

\bibitem{pontin2020ultranarrow}
Pontin A, Bullier N, Toro{\v{s}} M and Barker P 2020 {\em Physical Review
  Research\/} {\bf 2} 023349
  \urlprefix\url{https://doi.org/10.1103/PhysRevResearch.2.023349}

\bibitem{pitkin2011gravitational}
Pitkin M, Reid S, Rowan S and Hough J 2011 {\em Living Reviews in Relativity\/}
  {\bf 14} 1--75 \urlprefix\url{https://doi.org/10.12942/lrr-2011-5}

\bibitem{penn2001high}
Penn S~D, Harry G~M, Gretarsson A~M, Kittelberger S~E, Saulson P~R, Schiller
  J~J, Smith J~R and Swords S~O 2001 {\em Review of Scientific Instruments\/}
  {\bf 72} 3670--3673 \urlprefix\url{https://doi.org/10.1063/1.1394183}

\bibitem{cumming2020lowest}
Cumming A, Sorazu B, Daw E, Hammond G, Hough J, Jones R, Martin I, Rowan S,
  Strain K and Williams D 2020 {\em Classical and Quantum Gravity\/} {\bf 37}
  195019 \urlprefix\url{https://doi.org/10.1088/1361-6382/abac42}

\bibitem{mehmet2011squeezed}
Mehmet M, Ast S, Eberle T, Steinlechner S, Vahlbruch H and Schnabel R 2011 {\em
  Optics express\/} {\bf 19} 25763--25772
  \urlprefix\url{https://doi.org/10.1364/OE.19.025763}

\bibitem{slusher1985observation}
Slusher R, Hollberg L~W, Yurke B, Mertz J~C and Valley J~F 1985 {\em Physical
  Review Letters\/} {\bf 55} 2409
  \urlprefix\url{https://doi.org/10.1103/PhysRevLett.55.2409}

\bibitem{wu1986generation}
Wu L~A, Kimble H~J, Hall J~L and Wu H 1986 {\em Physical Review Letters\/} {\bf
  57} 2520 \urlprefix\url{https://doi.org/10.1103/PhysRevLett.57.2520}

\bibitem{andersen201630}
Andersen U~L, Gehring T, Marquardt C and Leuchs G 2016 {\em Physica Scripta\/}
  {\bf 91} 053001 \urlprefix\url{https://doi.org/10.1088/0031-8949/91/5/053001}

\bibitem{qvarfort2020time}
Qvarfort S, Serafini A, Xuereb A, Braun D, R{\"a}tzel D and Bruschi D~E 2020
  {\em Journal of Physics A\/} {\bf 53} 075304
  \urlprefix\url{https://doi.org/10.1088/1751-8121/ab64d5}

\bibitem{qvarfort2019enhanced}
Qvarfort S, Serafini A, Xuereb A, R{\"a}tzel D and Bruschi D~E 2019 {\em New
  Journal of Physics\/}
  \urlprefix\url{https://doi.org/10.1088/1367-2630/ab1b9e}

\bibitem{doi:10.1063/1.1665613}
Lovelock D 1971 {\em Journal of Mathematical Physics\/} {\bf 12} 498--501
  \urlprefix\url{https://doi.org/10.1063/1.1665613}

\bibitem{Burrage:2017shh}
Burrage C, Copeland E~J, Moss A and Stevenson J~A 2018 {\em Journal of
  Cosmology and Astroparticle Physics\/} {\bf 01} 056
  \urlprefix\url{https://doi.org/10.1088/1475-7516/2018/01/056}

\bibitem{ellis2007causality}
Ellis G~F, Maartens R and MacCallum M~A 2007 {\em General Relativity and
  Gravitation\/} {\bf 39} 1651--1660
  \urlprefix\url{https://doi.org/10.1007/s10714-007-0479-2}

\bibitem{Khoury:2013yya}
Khoury J 2013 {\em Classical and Quantum Gravity\/} {\bf 30} 214004

\bibitem{meyer2021fisher}
Meyer J~J 2021 {\em arXiv preprint arXiv:2103.15191\/}
  \urlprefix\url{https://arxiv.org/abs/2103.15191}

\bibitem{cramer1946contribution}
Cram{\'e}r H 1946 {\em Scandinavian Actuarial Journal\/} {\bf 1946} 85--94
  \urlprefix\url{http://www.tandfonline.com/doi/pdf/10.1080/03461238.1946.10419631}

\bibitem{rao1992information}
Rao C~R 1992 Information and the accuracy attainable in the estimation of
  statistical parameters {\em Breakthroughs in statistics\/} (Springer) pp
  235--247
  \urlprefix\url{https://link.springer.com/chapter/10.1007/978-1-4612-0919-5_16}

\bibitem{pang2014}
Pang S and Brun T~A 2014 {\em Physical Review A\/} {\bf 90} 022117
  \urlprefix\url{https://doi.org/10.1103/PhysRevA.90.022117}

\bibitem{jing2014}
Jing L, Xiao-Xing J, Wei Z and Xiao-Guang W 2014 {\em Communications in
  Theoretical Physics\/} {\bf 61} 45
  \urlprefix\url{https://doi.org/10.1088/0253-6102/61/1/08}

\bibitem{wei1963lie}
Wei J and Norman E 1963 {\em J. Math. Phys.\/} {\bf 4} 575--581
  \urlprefix\url{https://doi.org/10.1063/1.1703993}

\bibitem{vahlbruch2016detection}
Vahlbruch H, Mehmet M, Danzmann K and Schnabel R 2016 {\em Physical Review
  Letters\/} {\bf 117} 110801
  \urlprefix\url{https://doi.org/10.1103/PhysRevLett.117.110801}

\bibitem{cole2011phonon}
Cole G~D, Wilson-Rae I, Werbach K, Vanner M~R and Aspelmeyer M 2011 {\em Nature
  Communications\/} {\bf 2} 1--8
  \urlprefix\url{https://doi.org/10.1038/ncomms1212}

\bibitem{delic2020levitated}
Deli{\'c} U, Grass D, Reisenbauer M, Damm T, Weitz M, Kiesel N and Aspelmeyer M
  2020 {\em Quantum Science and Technology\/} {\bf 5} 025006
  \urlprefix\url{https://doi.org/10.1088/2058-9565/ab7989}

\bibitem{magrini2020optimal}
Magrini L, Rosenzweig P, Bach C, Deutschmann-Olek A, Hofer S~G, Hong S, Kiesel
  N, Kugi A and Aspelmeyer M 2020 {\em arXiv preprint arXiv:2012.15188\/}
  \urlprefix\url{https://arxiv.org/abs/2012.15188}

\bibitem{Murata_2015}
Murata J and Tanaka S 2015 {\em Classical and Quantum Gravity\/} {\bf 32}
  033001 \urlprefix\url{https://doi.org/10.1088%2F0264-9381%2F32%2F3%2F033001}

\bibitem{PhysRevLett.124.051301}
Tan W~H, Du A~B, Dong W~C, Yang S~Q, Shao C~G, Guan S~G, Wang Q~L, Zhan B~F,
  Luo P~S, Tu L~C and Luo J 2020 {\em Physical Review Letters\/} {\bf 124}(5)
  051301
  \urlprefix\url{https://link.aps.org/doi/10.1103/PhysRevLett.124.051301}

\bibitem{burrage2018tests}
Burrage C and Sakstein J 2018 {\em Living reviews in relativity\/} {\bf 21} 1
  \urlprefix\url{https://doi.org/10.1007/s41114-018-0011-x}

\bibitem{bruschi2020time}
Bruschi D~E 2020 {\em Journal of Mathematical Physics\/} {\bf 61} 032102
  \urlprefix\url{https://doi.org/10.1063/1.5121397}

\bibitem{law1995interaction}
Law C 1995 {\em Physical Review A\/} {\bf 51} 2537
  \urlprefix\url{https://doi.org/10.1103/PhysRevA.51.2537}

\bibitem{casimir1948influence}
Casimir H~B~G and Polder D 1948 {\em Physical Review\/} {\bf 73} 360
  \urlprefix\url{https://journals.aps.org/pr/abstract/10.1103/PhysRev.73.360}

\bibitem{Bimonte2018beyond}
Bimonte G 2018 {\em Physical Review D\/} {\bf 98}(10) 105004
  \urlprefix\url{https://link.aps.org/doi/10.1103/PhysRevD.98.105004}

\bibitem{Bimonte2012exact}
Bimonte G and Emig T 2012 {\em Physical Review Letters\/} {\bf 109}(16) 160403
  \urlprefix\url{https://link.aps.org/doi/10.1103/PhysRevLett.109.160403}

\bibitem{chiaverini2003new}
Chiaverini J, Smullin S~J, Geraci A~A, Weld D~M and Kapitulnik A 2003 {\em
  Physical Review Letters\/} {\bf 90} 151101
  \urlprefix\url{https://doi.org/10.1103/PhysRevLett.90.151101}

\bibitem{munday_measured_2009}
Munday J~N, Capasso F and Parsegian V~A 2009 {\em Nature\/} {\bf 457} 170--173
  ISSN 0028-0836
  \urlprefix\url{https://www.ncbi.nlm.nih.gov/pmc/articles/PMC4169270/}

\bibitem{schmole2016micromechanical}
Schm{\"o}le J, Dragosits M, Hepach H and Aspelmeyer M 2016 {\em Classical and
  Quantum Gravity\/} {\bf 33} 125031
  \urlprefix\url{https://doi.org/10.1088/0264-9381/33/12/125031}

\bibitem{banishev_modulation_2012}
Banishev A~A, Chang C~C, Zandi R and Mohideen U 2012 {\em Applied Physics
  Letters\/} {\bf 100} 033112 ISSN 0003-6951
  \urlprefix\url{https://aip.scitation.org/doi/abs/10.1063/1.3678189}

\bibitem{intravaia2013strong}
Intravaia F, Koev S, Jung I~W, Talin A~A, Davids P~S, Decca R~S, Aksyuk V~A,
  Dalvit D~A~R and L{\'o}pez D 2013 {\em Nature Communications\/} {\bf 4} 1--8
  \urlprefix\url{https://doi.org/10.1038/ncomms3515}

\bibitem{chen_control_2007}
Chen F, Klimchitskaya G~L, Mostepanenko V~M and Mohideen U 2007 {\em Physical
  Review B\/} {\bf 76} 035338
  \urlprefix\url{https://link.aps.org/doi/10.1103/PhysRevB.76.035338}

\bibitem{kuhn2017optically}
Kuhn S, Stickler B~A, Kosloff A, Patolsky F, Hornberger K, Arndt M and Millen J
  2017 {\em Nature Communications\/} {\bf 8} 1--5
  \urlprefix\url{https://doi.org/10.1038/s41467-017-01902-9}

\bibitem{tsang2010cavity}
Tsang M 2010 {\em Physical Review A\/} {\bf 81} 063837
  \urlprefix\url{https://doi.org/10.1103/PhysRevA.81.063837}

\bibitem{serafini2017quantum}
Serafini A 2017 {\em Quantum continuous variables: a primer of theoretical
  methods\/} (CRC press)

\bibitem{motazedifard2021ultraprecision}
Motazedifard A, Dalafi A and Naderi M 2021 {\em AVS Quantum Science\/} {\bf 3}
  024701 \urlprefix\url{https://doi.org/10.1116/5.0035952}

\bibitem{rabl2011photon}
Rabl P 2011 {\em Physical Review Letters\/} {\bf 107} 063601
  \urlprefix\url{https://doi.org/10.1103/PhysRevLett.107.063601}

\bibitem{nunnenkamp2011single}
Nunnenkamp A, B{\o}rkje K and Girvin S~M 2011 {\em Physical Review Letters\/}
  {\bf 107} 063602

\bibitem{qvarfort2020master}
Qvarfort S, Vanner M~R, Barker P~F and Bruschi D~E 2020 {\em arXiv preprint
  arXiv:2009.02295\/} \urlprefix\url{https://arxiv.org/abs/2009.02295}

\bibitem{bassi2005towards}
Bassi A, Ippoliti E and Adler S~L 2005 {\em Physical Review Letters\/} {\bf 94}
  030401 \urlprefix\url{https://doi.org/10.1103/PhysRevLett.94.030401}

\bibitem{bernad2006quest}
Bern{\'a}d J~Z, Di{\'o}si L and Geszti T 2006 {\em Physical Review Letters\/}
  {\bf 97} 250404 \urlprefix\url{https://doi.org/10.1103/PhysRevLett.97.250404}

\bibitem{hu2015quantum}
Hu D, Huang S~Y, Liao J~Q, Tian L and Goan H~S 2015 {\em Physical Review A\/}
  {\bf 91} 013812 \urlprefix\url{https://doi.org/10.1103/PhysRevA.91.013812}

\bibitem{betzholz2020breakdown}
Betzholz R, Taketani B~G and Torres J~M 2020 {\em Quantum Science and
  Technology\/} {\bf 6} 015005
  \urlprefix\url{https://doi.org/10.1088/2058-9565/abc39d}

\bibitem{bender2013advanced}
Bender C~M and Orszag S~A 2013 {\em Advanced mathematical methods for
  scientists and engineers I: Asymptotic methods and perturbation theory\/}
  (Springer Science \& Business Media)

\bibitem{scharf2014finite}
Scharf G 2014 {\em Finite quantum electrodynamics: the causal approach\/}
  (Courier Corporation)

\end{thebibliography}

\end{document}